\documentclass{aa}  
\usepackage{dblfloatfix}
\usepackage{graphicx}
\usepackage{txfonts}
\usepackage{natbib}
\usepackage{capt-of}
\usepackage{textcomp}
\usepackage{epstopdf}
\usepackage{rotating}
\usepackage{siunitx}
\usepackage{multirow}
\usepackage{hhline}
\usepackage{float}
\floatstyle{plaintop}
\restylefloat{table}
\usepackage{threeparttable}
\usepackage{longtable}
\usepackage{import}
\usepackage{array}
\usepackage{amsmath}
\usepackage{mwe}
\usepackage{subfig}
\usepackage[version=4]{mhchem}
\usepackage[dvipsnames]{xcolor}
\usepackage[normalem]{ulem}
\usepackage[LGRgreek]{mathastext}

\usepackage[colorlinks=true,allcolors=blue]{hyperref}

\begin{document} 

\renewcommand{\thefootnote}{\alph{footnote}}
\renewcommand{\thefootnote}{\fnsymbol{footnote}}

   \title{Ammonium hydrosulfide (NH$_{4}$SH) as a potentially significant sulfur sink in interstellar ices}
   \subtitle{}
\titlerunning{}

   \author{ K. Slavicinska\inst{1,2}\and
            A. C. A. Boogert\inst{3}\and
            \L{}. Tychoniec\inst{2}\and
            E. F. van Dishoeck\inst{2,4}\and
            M. L. van Gelder\inst{2}\and
            M. G. Navarro\inst{5}\and
            J. C. Santos\inst{1,2}\and
            P. D. Klaassen\inst{6}\and
            P. J. Kavanagh\inst{7}\and
            K. -J. Chuang\inst{1,2}
            }

   \institute{1. Laboratory for Astrophysics, Leiden Observatory, Leiden University, P.O. Box 9513, NL 2300 RA Leiden, The Netherlands.\\
   \email{slavicinska@strw.leidenuniv.nl}\\
   2. Leiden Observatory, Leiden University, P.O. Box 9513, NL 2300 RA Leiden, The Netherlands.\\
   3. Institute for Astronomy, University of Hawai’i at Manoa, 2680 Woodlawn Drive, Honolulu, HI 96822, USA.\\
   4. Max Planck Institut f\"ur Extraterrestrische Physik (MPE), Giessenbachstrasse 1, 85748 Garching, Germany.\\
   5. INAF -- Osservatorio Astronomico di Roma, Via di Frascati 33, 00040, Monteporzio Catone, Italy.\\
   6. UK Astronomy Technology Centre, Royal Observatory Edinburgh, Blackford Hill, Edinburgh EH9 3HJ, UK.\\
   7. School of Cosmic Physics, Dublin Institute for Advanced Studies, 31 Fitzwilliam Place, Dublin 2, Ireland.}

   \date{Received 4 July 2024; accepted 16 September 2024}

  \abstract
   {Sulfur is depleted with respect to its cosmic standard abundance in dense star-forming regions. It has been suggested that this depletion is caused by the freeze-out of sulfur on interstellar dust grains, but the observed abundances and upper limits of sulfur-bearing ices remain too low to account for all of the missing sulfur. Toward the same environments, a strong absorption feature at $\sim$6.85 $\mu$m is observed, but its long-standing assignment to the NH$_{4}$$^{+}$ cation remains tentative.}
   {We aim to spectroscopically investigate the plausibility of NH$_{4}$SH salt serving as a sulfur reservoir and a carrier of the 6.85 $\mu$m band in interstellar ices by characterizing its IR signatures and apparent band strengths in water-rich laboratory ice mixtures. We then use this laboratory data to constrain NH$_{4}$SH abundances in observations of interstellar ices.}
   {Laboratory transmission IR spectra of NH$_{3}$:H$_{2}$S ice mixtures both with and without H$_{2}$O were collected. The apparent band strengths of the NH$_{4}$$^{+}$ asymmetric bending ($\nu_{4}$) mode and the SH$^{-}$ stretching mode in H$_{2}$O-containing mixtures were calculated with Beer's law plots. The IR features of the laboratory salts were compared to those observed toward a sample of four protostars (two low-mass, two high-mass) and two cold dense clouds without star formation.}
   {Apparent band strengths ranging from 3.2($\pm$0.3)-3.6($\pm$0.4)$\times$10$^{-17}$ cm molec$^{-1}$ and 3.1($\pm$0.4)-3.7($\pm$0.5)$\times$10$^{-19}$ cm molec$^{-1}$ are calculated for the NH$_{4}$$^{+}$ $\nu_{4}$ mode at $\sim$6.8 $\mu$m/1470 cm$^{-1}$ and the SH$^{-}$ stretching mode at $\sim$3.9 $\mu$m/2560 cm$^{-1}$, respectively, in NH$_{4}$SH:H$_{2}$O mixtures. The peak position of the NH$_{4}$$^{+}$ $\nu_{4}$ mode redshifts with increasing temperature but also with increasing salt concentration with respect to matrix species H$_{2}$O and NH$_{3}$. The observed 6.85 $\mu$m feature is fit well with the laboratory NH$_{4}$SH:H$_{2}$O ice spectra. NH$_{4}$$^{+}$ column densities obtained from the 6.85 $\mu$m band range from 8-23\% with respect to H$_{2}$O toward the sample of protostars and dense clouds. These column densities are consistent with the optical depths observed at 3.9 $\mu$m (the SH$^{-}$ stretching mode spectral region). A weak and broad feature observed at $\sim$5.3 $\mu$m/1890 cm$^{-1}$ is tentatively assigned to the combination mode of the NH$_{4}$$^{+}$ $\nu_{4}$ mode and the SH$^{-}$ libration. The combined upper limits of four other counter-anion candidates, OCN$^{-}$, CN$^{-}$, HCOO$^{-}$, and Cl$^{-}$, are determined to be $\lesssim$15-20\% of the total NH$_{4}$$^{+}$ column densities toward three of the protostars.}
   {The redshift of the 6.85 $\mu$m feature correlates with higher abundances of NH$_{4}$$^{+}$ with respect to H$_{2}$O in both the laboratory data presented here and observational data of dense clouds and protostars. The apparent band strength of the SH$^{-}$ feature is likely too low for the feature to be detectable in the spectrally busy 3.9 $\mu$m region, but the 5.3 $\mu$m NH$_{4}$$^{+}$ $\nu_{4}$ + SH$^{-}$ R combination mode may be an alternative means of detection. Its tentative assignment adds to mounting evidence supporting the presence of NH$_{4}$$^{+}$ salts in ices and is the first tentative observation of the SH$^{-}$ anion toward interstellar ices. If the majority ($\gtrsim$80-85\%) of the NH$_{4}$$^{+}$ cations quantified toward the investigated sources in this work are bound to SH$^{-}$ anions, then NH$_{4}$SH salts could account for up to 17-18\% of their sulfur budgets.}


   \maketitle
%
\section{Introduction}
\label{txt:intro}

The observed depletion of sulfur toward star-forming regions is a persistent topic of debate in the field of astrochemistry. In the diffuse interstellar medium, sulfur is mostly present in its singly ionized form (S$^{+}$) with an observed abundance of $\sim$1.7$\times$10$^{-5}$ with respect to hydrogen \citep{esteban2004reappraisal}, similar to its observed solar abundance as well as the abundance measured in CI chondrite meteorites \citep{lodders2003solar,asplund2009chemical}. However, in the higher-density cold interstellar environments like dense clouds, prestellar cores, and protostellar envelopes from which stellar systems eventually form, the abundance of observed gas-phase sulfur species can be two or three orders of magnitude lower \citep{joseph1986interstellar,tieftrunk1994so}.

Many chemical models suggest that in such cold and dense environments, most of the sulfur should freeze out and form H$_{2}$S ice via hydrogenation on the surfaces of interstellar grains \citep{garrod2007non,druard2012polysulphanes,vidal2017reservoir,vidal2018new}, but to date, no convincing detection of this ice in the interstellar medium exists. This is in part due to the fact that the strongest IR absorption of H$_{2}$S ice, its stretching mode at 3.93 $\mu$m (2547 cm$^{-1}$), has a broad profile that overlaps with combination modes of CH$_{3}$OH ice, a major constituent of interstellar ices, in the same spectral region. Additionally, several other species with S-H bonds have been shown to have absorption profiles similar to H$_{2}$S that peak between 3.93-3.96 $\mu$m, making a secure assignment of any absorption detected in this spectral region specifically to H$_{2}$S ice difficult \citep{hudson2018infrared}. Nevertheless, H$_{2}$S ice upper limits measured toward prestellar cores and protostars are typically $\lesssim$1\% with respect to H$_{2}$O, indicating that H$_{2}$S ice cannot be the major sulfur carrier in these dense environments \citep{smith1991search,jimenez2011sulfur,mcclure2023ice}. Two other sulfur-bearing ice species, OCS and SO$_{2}$, have been detected more securely but at very low abundances ($<$1\% with respect to H$_{2}$O, \citealt{boogert1997infrared,oberg2008c2d,boogert2022survey,mcclure2023ice,rocha2024jwst}); therefore, these ices can also only account for at maximum a couple of percent of the cosmic sulfur budget. 

This observed sulfur depletion, sometimes dubbed ``the missing sulfur problem,'' has spurred a variety of questions in the astrochemical community over the past couple of decades. These include why H$_{2}$S ice is not detected despite what models predict, whether H$_{2}$S reacts on the ice grain surfaces and forms other icy species that are not easily detected (e.g., \citealt{garozzo2010fate,jimenez2011sulfur,jimenez2014sulphur,chen2014formation,woods2015new,martin2016sulfur,laas2019modeling}), whether H$_{2}$S ice is just quick to chemisorb into the gas phase from the grain surface following its formation \citep{oba2018infrared,furuya2022quantifying,santos2023interaction}, or if most of the sulfur is actually hidden in minerals like sulfides (e.g., \citealt{keller2002identification,kama2019abundant,perrero2024theoretical}).

An essential clue may lie in the recently published analysis of in situ mass spectrometric data of the dust grains of comet 67P collected by the ROSINA Double Focusing Mass Spectrometer \citep{altwegg2022abundant}. This work shows that the cometary grains contain NH$_{4}$SH salt (NH$_{4}$$^{+}$SH$^{-}$) in exceptionally high abundance. In fact, it is estimated that approximately 90\% of the ammonium (NH$_{4}$$^{+}$) cations in the analyzed dust grains could be bound to SH$^{-}$ anions. As it is thought that comets may be chemically linked to pre- and protostellar environments \citep{bockelee2000new,drozdovskaya2019ingredients,altwegg2019cometary}, it is no stretch to hypothesize that NH$_{4}$SH salts could be similarly abundant in primitive interstellar ices.

Experiments show that at the cold temperatures expected in the environments where sulfur depletion is observed, NH$_{4}$SH forms readily via an acid-base reaction between NH$_{3}$ and H$_{2}$S \citep{loeffler2015giant,hudson2015activation,vitorino2024sulphur}. Both molecules are expected to accumulate on interstellar grains around the same time, during the prestellar epoch in which most of the H$_{2}$O ice forms, based on their deuterium ratios measured in the gas phase \citep{vastel2003first,caselli2012our}. Indeed, NH$_{3}$ ice has been securely detected toward both dense clouds and protostellar ice envelopes via its 9 $\mu$m umbrella mode, with abundances ranging from 3-10\% with respect to H$_{2}$O \citep{boogert2015observations}, and its peak profile and narrow abundance distribution with respect to H$_{2}$O indicate that it likely exists in a water-rich environment \citep{bottinelli2010c2d,oberg2011spitzer}.

Furthermore, the NH$_{4}$$^{+}$ cation has long been suspected to be the primary carrier of the ubiquitous 6.85 $\mu$m feature that is present in the majority of ice-containing lines of sight \citep{knacke1982observation,grim1989infrared,grim1989ions}. The peak position of this feature has also been proposed as a tracer of thermal processing due to its correlation with dust and ice temperatures in protostellar envelopes \citep{keane2001ice,boogert2008c2d}, a trend that is experimentally corroborated by a redshift in the NH$_{4}$$^{+}$ $\nu_{4}$ feature observed during the warm-up of laboratory ammonium salt ices \citep{schutte2003origin}. NH$_{4}$$^{+}$ pre- and protostellar ice abundances calculated from the 6.85 $\mu$m feature (using a band strength of 4.4$\times$10$^{-17}$ cm molec$^{-1}$ measured for the asymmetric bending mode $\nu_{4}$ of NH$_{4}$$^{+}$ in laboratory mixtures of NH$_{4}$$^{+}$HCOO$^{-}$ salts diluted in H$_{2}$O, \citealt{schutte2003origin}) range from 4-34\% with respect to H$_{2}$O, on par with abundances of the major ice constituents CO, CO$_{2}$, and CH$_{3}$OH \citep{boogert2015observations}.

However, the assignment remains controversial for multiple reasons. First, no laboratory data of NH$_{4}$$^{+}$-containing ices have provided a ``convincing'' fit to the observed feature \citep{boogert2015observations}. In particular, currently available laboratory spectra of ammonium salts diluted in H$_{2}$O-rich ice matrices are too broad and weak to fit the observed feature \citep{mate2009water,galvez2010ammonium} despite column density correlations indicating that the primary carrier of the 6.85 $\mu$m feature is chemically linked to H$_{2}$O ice \citep{tielens1987evolution,pontoppidan2006spatial,boogert2011ice}. Additionally, the combined abundances with respect to H$_{2}$O ice of anionic species identified so far in interstellar ices, OCN$^{-}$ and HCOO$^{-}$, are usually over an order of magnitude lower than those calculated for NH$_{4}$$^{+}$ (e.g., \citealt{schutte1999weak,van2004quantitative,van20053,boogert2022survey,mcclure2023ice,rocha2024jwst}), so if the 6.85 $\mu$m band indeed consists primarily of the $\nu$$_{4}$ mode of NH$_{4}$$^{+}$, other anionic species must be identified to fully counterbalance its net positive charge. Furthermore, no other bands have been unambiguously assigned to NH$_{4}$$^{+}$; although the 3.26 and 3.48 $\mu$m features observed toward some sources on the red wing of the 3 $\mu$m H$_{2}$O feature have been proposed to be NH$_{4}$$^{+}$ combination modes \citep{schutte2003origin}, absorptions at these wavelengths could also be attributed to PAHs \citep{sellgren1995new} and NH$_{3}$ hydrates \citep{dartois2001search}, respectively.

With this work, we seek to investigate whether NH$_{4}$SH salts formed in H$_{2}$O-rich ices could serve, metaphorically, as a ``stone'' that simultaneously kills two astrochemical ``birds'': the sink of the missing sulfur and the identity of the 6.85 $\mu$m band. Band strengths of polycrystalline NH$_{4}$SH ice at 160 K were previously derived by \cite{ferraro1980infrared}, who reported a band strength of $\sim$1.2$\times$10$^{-19}$ cm molec$^{-1}$ for the SH$^{-}$ anion stretching mode at 3.89 $\mu$m (2570 cm$^{-1}$). In comparison, \cite{yarnall2022new} reported a band strength that is over two orders of magnitude higher, 1.66$\times$10$^{-17}$ cm molec$^{-1}$, for the 3.93 $\mu$m stretching mode of neutral H$_{2}$S in a water-rich ice at 10 K . Such a dramatic difference in the reported band strengths of the H$_{2}$S and SH$^{-}$ stretching features at nearly the same wavelengths highlights the possibility that the abundance of sulfur in interstellar ices may be significantly higher than what the current ice abundances and upper limits predict if some H$_{2}$S ice has been converted into its anionic form. \cite{ferraro1980infrared} also reported a band strength of $\sim$2.6$\times$10$^{-17}$ cm molec$^{-1}$ for the NH$_{4}$$^{+}$ asymmetric bending mode ($\nu_{4}$) at 7.05 $\mu$m, almost a factor of 2 lower than the NH$_{4}$$^{+}$ band strength of 4.4$\times$10$^{-17}$ measured by \cite{schutte2003origin} that is used the most frequently in astronomical literature to quantify NH$_{4}$$^{+}$ cations in interstellar ices (e.g., \citealt{boogert2008c2d,boogert2011ice,mcclure2023ice}). Therefore, if NH$_{4}$SH is the main carrier of the 6.85 $\mu$m band, its abundance could be almost a factor of 2 higher than indicated by the current NH$_{4}$$^{+}$ column density calculations. However, it is important to note that these band strengths were calculated for pure polycrystalline NH$_{4}$SH at a relatively high temperature (160 K), and the band strengths of NH$_{4}$SH may differ significantly if the salt is primarily amorphous and in H$_{2}$O-dominated ice mixtures, as would be expected in most icy lines of sight. To our knowledge, the present literature lacks laboratory spectra and band strengths of NH$_{4}$SH salts in such astrophysically relevant conditions as well as comparisons of NH$_{4}$SH laboratory spectra to observed spectra of interstellar ices.

To that end, we have measured infrared spectra of NH$_{4}$SH formed via the acid-base reaction of co-deposited NH$_{3}$ and H$_{2}$S ice, both with and without H$_{2}$O ice, at temperatures relevant to pre- and protostellar ices. These spectra are made publicly available to the community via the Leiden Ice Database (LIDA, \citealt{rocha2022lida}) for use in fitting to observed ice spectra. We characterize how the peak profiles of the salt change as functions of temperature and dilution with H$_{2}$O. We then derive apparent band strengths of the NH$_{4}$$^{+}$ $\nu_{4}$ mode and SH$^{-}$ stretch in H$_{2}$O-rich NH$_{4}$SH ices for use in deriving column densities and upper limits from IR observations. Finally, we compare our laboratory spectra to IR spectra of pre- and protostellar ices to qualitatively evaluate whether NH$_{4}$SH is a feasible candidate carrier of the 6.85 $\mu$m band and quantify plausible column densities of NH$_{4}$SH ice in the dense star-forming environments where sulfur is depleted.

\section{Methodology}

\subsection{Experimental}

All experiments were performed on the InfraRed Absorption Setup for Ice Spectroscopy (IRASIS) in the Laboratory for Astrophysics in the Leiden Observatory. A schematic of the setup is presented in \cite{rachid2021infrared}, and recent upgrades are described in \cite{slavicinska2023hunt}. Briefly, the setup consists of a KBr substrate cooled to a minimum temperature of 15 K via a closed-cycle helium cryostat within an ultrahigh vacuum chamber (base pressure $<$1$\times$10$^{-9}$ mbar). Ices are grown on the substrate via background deposition through three independent leak valves, and each leak valve is connected to an independent reservoir on a gas manifold. Ice thickness throughout deposition is monitored via the interference patterns of a 633 nm laser that reflects off both sides of the substrate at 45$^{\circ}$ incidence, and transmission infrared spectra of the ice on the substrate are collected continuously during deposition and warm-up with a Varian 670 Fourier-transform infrared spectrometer (FTIR). The IR spectra here are collected with a resolution of  0.5 cm$^{-1}$ using 128 scans per spectrum. A Spectra Microvision Plus quadrupole mass spectrometer (QMS) is used to calibrate each leak valve to specific deposition rates for pure molecules, monitor the ice deposition, and, in some experiments, collect temperature-programmed desorption (TPD) mass spectra during warm-up. Following deposition, the substrate can be warmed until all ices have desorbed from the substrate using a LakeShore Model 335 Temperature Controller to heat the substrate at a specific rate. Most of the experiments in this paper utilized a 1 K min$^{-1}$ heating rate, resulting in a temperature difference of $\sim$3.6 K between each spectrum collected during the warm-up and corresponding to a temperature uncertainty of $\sim$$\pm$2 K for each warm-up spectrum.

The liquids and gases used in this work were ammonia (PraxAir, $\geq$99.96\%), hydrogen sulfide (Linde Gas, $\geq$99.5\%), and water (Milli-Q, Type I).

\subsubsection{Ice growth rate calibration}
\label{txt:methods_cal}

The three independent reservoirs of NH$_{3}$ gas, H$_{2}$S gas, and H$_{2}$O vapor each have their own leak valve and inlet into the main chamber. Each leak valve position is calibrated to a specific ice growth rate of a given molecule via dosing the molecule into the chamber at a constant rate and monitoring the ice growth rate with laser interference. More details on the calibration procedure can be found in Appendix~\ref{app:cal}. The benefits of using independent dosing lines compared to dosing a gas mixture from a single dosing line to create an ice mixture have been previously described in \cite{yarnall2022new} and \cite{slavicinska2023hunt}.

\subsubsection{NH$_{3}$:H$_{2}$S experiments}

\begin{table}[ht!]
\footnotesize
\caption{Summary of NH$_{3}$:H$_{2}$S experiments performed.}
\begin{center}
\begin{tabular}{|c c|c c c|c|c|}
\hline
        \multicolumn{2}{|c|}{Dosing ratios} & \multicolumn{3}{|c|}{Final ratios$^{(a)}$} & Lim. & Dep. time\\
        NH$_{3}$ & H$_{2}$S & NH$_{4}$SH & NH$_{3}$ & H$_{2}$S & react. & (min)\\
        \hline
        10 & 11 & 10 & 0 & 1 & NH$_{3}$ & 20, 40, 60 \\
        3 & 2 & 2 & 1 & 0 & H$_{2}$S & 20, 40, 60 \\
        2 & 3 & 2 & 0 & 1 & NH$_{3}$ & 20, 40, 60 \\
    \hline
\end{tabular}

\begin{tablenotes}
$^{(a)}$ once 100\% of the limiting reactant is consumed in the acid-base reaction.
\end{tablenotes}
\label{tab:experiments_pure}
\end{center}
\end{table}

\begin{table*}[ht!]

\caption{Summary of H$_{2}$O:NH$_{3}$:H$_{2}$S experiments performed.}
\begin{center}
\begin{tabular}{|c c c|c c c c|c|c|c|}
\hline
        \multicolumn{3}{|c|}{Dosing ratios} & \multicolumn{4}{|c|}{Final ratios$^{(a)}$} & \% NH$_{4}$SH$^{(a)}$ & Lim. & Dep. time\\
        H$_{2}$O & NH$_{3}$ & H$_{2}$S & H$_{2}$O & NH$_{4}$SH & NH$_{3}$ & H$_{2}$S & with respect to H$_{2}$O & react. & (min)\\
        \hline
        10 & 2 & 1 & 10 & 1 & 1 & 0 & 10 & H$_{2}$S & 10, 20, 30, 40, 50 \\
        8 & 2 & 1 & 10 & 8 & 1 & 0 & 12.5 & H$_{2}$S & 20 \\
        6 & 2 & 1 & 10 & 6 & 1 & 0 & 16.7 & H$_{2}$S & 20 \\
        5 & 2 & 1 & 5 & 1 & 1 & 0 & 20 & H$_{2}$S & 20 \\
        \hline
        12 & 3 & 2 & 12 & 2 & 1 & 0 & 16.7 & H$_{2}$S & 30 \\
        10 & 3 & 2 & 10 & 2 & 1 & 0 & 20 & H$_{2}$S & 10, 20, 30, 40, 50 \\
        6 & 3 & 2 & 6 & 2 & 1 & 0 & 33.3 & H$_{2}$S & 30 \\
        4 & 3 & 2 & 4 & 2 & 1 & 0 & 50 & H$_{2}$S & 30 \\
        3 & 3 & 2 & 3 & 2 & 1 & 0 & 66.7 & H$_{2}$S & 30 \\
        \hline
        8 & 5 & 4 & 8 & 4 & 1 & 0 & 50 & H$_{2}$S & 10, 20, 30, 40, 50 \\
        \hline
        5 & 1 & 1 & 5 & 1 & 0 & 0 & 20 & - & 30 \\
        2 & 1 & 1 & 2 & 1 & 0 & 0 & 50 & - & 40 \\
        1 & 1 & 1 & 1 & 1 & 0 & 0 & 100 & - & 40 \\
        \hline
        10 & 2 & 3 & 10 & 2 & 0 & 1 & 20 & NH$_{3}$ & 40 \\
    \hline
\end{tabular}

\begin{tablenotes}
$^{(a)}$ once 100\% of the limiting reactant is consumed by the acid-base reaction.
\end{tablenotes}
\label{tab:experiments_h2o}
\end{center}
\end{table*}

Previously, \cite{loeffler2015giant} formed NH$_{4}$SH ice in situ on a cryo-cooled substrate via the acid-base reaction of co-deposited NH$_{3}$ and H$_{2}$S at 10 K. Here, we partially replicated the methodology for the sake of having pure NH$_{4}$SH spectra to compare to those of NH$_{4}$SH mixed with H$_{2}$O ice. In our experiments, NH$_{3}$ and H$_{2}$S were co-deposited at 15 K in a solid-state ratio of 1:1.1. After deposition, the samples were held at 15 K for $\sim$1 hr and then heated at a rate of 1 K min$^{-1}$ until the salt desorbed. We repeated this methodology with 3:2 and 2:3 NH$_{3}$:H$_{2}$S mixtures to aid feature assignment in the 15 K pure salt spectrum, which consists of a complex blend of NH$_{3}$, H$_{2}$S, and NH$_{4}$SH features (Figure~\ref{fig:all_assignments}). Three experiments were performed for each ratio: a short deposition (20 min) for simultaneous TPD collection, a long deposition (60 min) for accurate characterization of weak features, and a medium-length deposition (40 min) for the sake of collecting spectra with high S/N whose strongest peaks did not eventually saturate during warm-up. These experiments are summarized in Table~\ref{tab:experiments_pure}.

\subsubsection{H$_{2}$O:NH$_{3}$:H$_{2}$S experiments}
\label{txt:methods_mixture_bs}

Table~\ref{tab:experiments_h2o} provides an overview of the NH$_{4}$SH:H$_{2}$O mixtures we characterized in our experiments. In these experiments, H$_{2}$O, NH$_{3}$, and H$_{2}$S were co-deposited at 15 K in the ratios specified in the table. As with the pure mixtures, the samples were held at 15 K for $\sim$1 hr and then heated at a rate of 1 K min$^{-1}$ until all ice species desorbed. For three mixtures, we deposited 5 samples of different thicknesses to calculate the band strengths of the NH$_{4}$$^{+}$ $\nu_{4}$ mode and the SH$^{-}$ stretching mode. In these experiments, we interrupted the warm-up ramp with a 40 min hold at 120 K (further details are provided in Section~\ref{txt:band_strengths_methods}). For the remainder of the mixtures, we deposited one sample with the goal of achieving a thickness that provided sufficient S/N for most of the ice features of interest without any of the major peaks saturating upon warm-up. Because the stretching modes of H$_{2}$S and SH$^{-}$ overlap, we aimed to minimize the unreacted H$_{2}$S present at the end of warm-up by setting H$_{2}$S as the limiting reactant in most of the mixtures.

\subsubsection{Band strength calculations}
\label{txt:band_strengths_methods}

Measuring band strengths of species that form via reactions in ices can be a complicated endeavor because it is oftentimes only possible to quantify them indirectly. Previous works that reported band strengths of salt species in ices often performed such an indirect measurement by quantifying the loss of acid or base precursor \citep{schutte1999weak,schutte2003origin,van2004quantitative,bergner2016kinetics}. However, this approach usually relies on the use of literature band strengths of the precursors that were measured in physical and chemical conditions that differ (sometimes significantly) from the ices in which the salt formation was performed, leading to inaccuracies stemming from the assumption that the precursor band strengths are identical in the two different experiments.

\begin{table*}[h!]
\caption{Summary of sources whose ice spectra are compared to laboratory NH$_{4}$SH spectra in this work.}
\begin{center}
\begin{tabular}{c c c c c}
\hline
        Source & Telescope & Source Type & Environment & Thermal processing$^{(a)}$\\
        \hline
        Elias 16 & \textit{Spitzer} & dense cloud & - & no processing \\
        CK2 & \textit{Spitzer} & dense cloud & - & no processing \\
        B1c & JWST & LYSO & clustered & little/no processing \\
        L1527 & JWST & LYSO & isolated & highly processed \\
        W33A & ISO & MYSO & - & little/no processing \\
        S140 IRS 1 & ISO & MYSO & - & highly processed \\
    \hline
     \label{tab:sources}
\end{tabular}

\begin{tablenotes}
$^{(a)}$ as is indicated by presence of a pure CO$_{2}$ component in the 4.39 $\mu$m $^{13}$CO$_{2}$ ice feature, a crystalline 3 $\mu$m H$_{2}$O ice feature, and/or a low optical depth of the 4.67 $\mu$m CO ice feature relative to the optical depth of the 4.27 $\mu$m CO$_{2}$ feature.
\end{tablenotes}

\end{center}
\end{table*}

Recently, \cite{gerakines2024sublimation} directly measured the refractive index, density, and apparent band strengths of a pure ammonium salt, NH$_{4}$CN, by dosing NH$_{3}$ and HCN vapor onto substrates at 125 K and monitoring the ice growth with laser interferometry, a quartz crystal microbalance, and transmission IR. The authors claim that 125 K is too high for the neutral species to stick to their substrates, resulting in in situ formation of pure NH$_{4}$CN salt absent of unreacted neutrals. Unfortunately, this method cannot be reliably used to find the band strengths of salts in mixtures with other ice species because calculating band strengths of species in mixtures requires knowledge of the deposition rate or deposited column density of the species of interest \citep{yarnall2022new}, and the deposition rate of a salt on a substrate may change substantially if another ice species is simultaneously accumulating on the substrate and changing the quantity of acid and base that react with each other on the substrate before they can desorb (relative to the amount that react with each other in a pure salt deposition).

With our method, we aim to eliminate the inaccuracies associated with utilizing previously published band strengths of precursors when deriving the apparent band strengths of our salts. Instead, we have relied on previously determined refractive index and density measurements of the pure limiting reactant in the acid-base reaction, H$_{2}$S, to calibrate our leak valves to a specific ice growth rate and subsequently have made two assumptions: 1) the deposition rate of H$_{2}$S on the substrate is the same when it is pure and when it is co-deposited with other species; and 2) 100\% of the deposited H$_{2}$S is converted to its ionic form at the point in the experiment when the integrals of the bands of interest are taken (135 K). We estimate from the IR and TPD spectra of our salt mixtures that the consumption of H$_{2}$S in our H$_{2}$O mixture experiments is likely $\geq$96\% at 135 K (see Section~\ref{txt:results_nh4sh_h2o}). This would result in a liberal estimate of $\sim$4\% error in our derived band strengths caused by the uncertainty in the reaction completion fraction. To remain conservative in our reported uncertainties, we opt to use a reaction completion uncertainty of twice this value, 8\%, in our apparent band strength error calculations (see Appendix~\ref{app:bs_err}).

The column density of our limiting reactant, H$_{2}$S, deposited in each experiment was calculated using a deposition rate obtained via a series of control experiments. In this series, we deposited pure H$_{2}$S ices with the same leak valve position and substrate temperature used during deposition in the mixture experiments. We then derived a deposition rate from the laser interference measured during the deposition using the previously published H$_{2}$S refractive index of 1.407 and density of 0.944 g cm$^{-3}$ \citep{yarnall2022new}. We repeated this control measurement five times to obtain an average deposition rate and a standard deviation, and we multiplied the average control deposition rate by the deposition time to obtain the total deposited H$_{2}$S column density for each mixture experiment. The standard deviation of the five control experiments yields an experimental error of 0.36\%, which represents the day-to-day variations between depositions performed in different experiments as well as noise and fitting errors in the laser interference data.

We then formed our salt-H$_{2}$O ice mixtures by depositing H$_{2}$S with the same experimental parameters as those used in the control experiments, except that we also simultaneously co-deposited NH$_{3}$ and H$_{2}$O via their independent leak valves calibrated to deposition rates that corresponded to our desired H$_{2}$O:NH$_{3}$:H$_{2}$S ratios. Following the deposition and a brief hold period at 15 K, we warmed the mixtures at 1 K min$^{-1}$ until reaching 120 K, where we held them for 40 min to maximize products formed by the beginning of ice desorption. After this second hold, we resumed the 1 K min$^{-1}$ warm-up until all ice species desorbed. Finally, we created Beer's law plots for each IR feature of interest by plotting its optical depth at 135 K as a function of total deposited H$_{2}$S column density for five different deposition times and obtained apparent band strengths from the slopes of the linear least-square fits to the data. The data at 135 K were used to simultaneously maximize the quantity of salt formed (which is greatest at high temperatures) and minimize the loss of salt, matrix, or unreacted precursor via desorption from the analyzed ice (which becomes significant $>$135 K in the TPD spectra, Figure~\ref{fig:nh4sh_10-3-2_tpd}). The linear fits were obtained using the York method for fitting data with errors on both X and Y variables \citep{york1966least}, and the apparent band strengths uncertainties calculated via this procedure range from $\sim$9-14\%. Details regarding the uncertainties of the integrated optical depths and ice column densities used in the Beer's law plots can be found in Appendix~\ref{app:bs_err}.

It should be noted that the band strengths calculated in this paper are referred to as ``apparent band strengths'' and denoted as A' because they are derived from the infrared spectra of the ices of interest rather than their optical constants. A more detailed explanation of the difference between apparent band strengths and absolute band strengths (typically denoted as A) can be found in Section 3 in \cite{hudson2014infrared}.

\subsection{Observational}
\label{txt:investigated_sources}

\begin{figure*}[h!]
\centering
\includegraphics[width=0.87\linewidth]{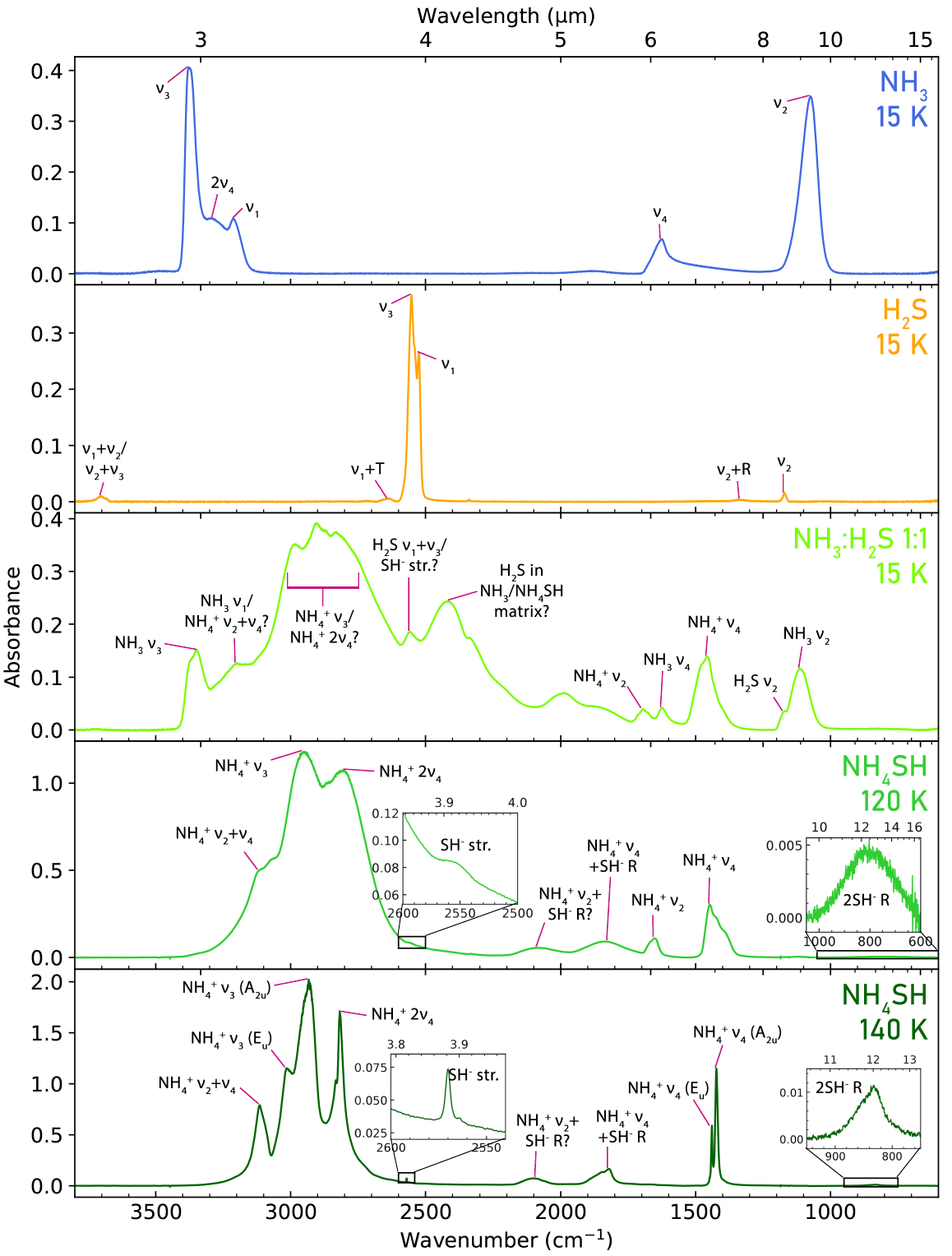}
\caption{Comparison of the infrared spectra of pure and mixed NH$_{3}$ and H$_{2}$S ices with peak assignments labeled (see Table~\ref{tab:nh4sh_pure_assignments}). Libration modes are indicated with an R. Several features in the spectrum of mixed NH$_{3}$ and H$_{2}$S ices at 15 K indicate the formation of NH$_{4}$SH salt, although peaks of nonreacted NH$_{3}$ and H$_{2}$S are still present. By 120 K, only signatures of the salt remain. The narrowing of the salt features by 140 K is characteristic of the salt undergoing a phase change (i.e., crystallization).}
\label{fig:all_assignments}
\end{figure*}

The selected observed sources for comparison to our laboratory data in this work are presented in Table~\ref{tab:sources}. These sources span a range of masses, environments, and degrees of thermal processing. The spectra of the investigated low-mass young stellar objects (LYSOs) were collected by the Near Infrared Spectrograph Integral Field Unit (NIRSpec IFU, \citealt{jakobsen2022near,boker2022near}) and Mid-Infrared Instrument Medium-Resolution Spectrometer (MIRI-MRS, \citealt{wright2023mid,argyriou2023jwst}) on the \textit{James Webb} Space Telescope (JWST, \citealt{gardner2023james}) as part of the JWST Observations of Young protoStars (JOYS+) collaboration (Program IDs 1290 and 1960, PI Ewine van Dishoeck).\footnote{\href{https://miri.strw.leidenuniv.nl}{https://miri.strw.leidenuniv.nl}} The massive young stellar objects (MYSOs) W33A and S140 IRS1 were observed by the Short Wavelength Spectrometer on the Infrared Space Observatory (ISO/SWS) \citep{gibb2004interstellar}. The dense cloud spectra were observed by the InfraRed Spectrograph (IRS) on the \textit{Spitzer} Space Telescope \citep{knez2005spitzer}.

The NIRSpec IFU data of B1c was collected with the G395M grating, resulting in continuous medium-resolution (R=$\lambda$/$\Delta\lambda$$\sim$700-1300) spectral coverage from 2.85-5.29 $\mu$m. The NRSIRS2RAPID readout pattern and a 4-point dither were used, with a total integration time of 1342 s. The instrument parameters used for the B1c MIRI data are described in \cite{chen2024joys}.

Spectra were extracted from the B1c NIRSpec and MIRI datacubes with apertures centered on the central brightest pixel over the full wavelength range (NIRSpec: RA 3:33:17.885, Dec +31:09:31.952; MIRI: RA 3:33:17.898, Dec +31:09:31.852). Differences in coordinates of the extraction apertures are a result of a known issue with WCS calibration in JWST pipeline.\footnote{\href{https://jwst-docs.stsci.edu/known-issues-with-jwst-data}{https://jwst-docs.stsci.edu/known-issues-with-jwst-data}} A wavelength-dependent aperture of seven times the diffraction-limited radius was used to extract 1D spectra from both datacubes, resulting in two spectra with overlapping data points between 4.90-5.29 $\mu$m. A scaling factor of 0.856 was used to scale the NIRSpec data to the MIRI data in the spectral overlap region.

The NIRSpec IFU data of L1527 was collected with the PRISM and G395H grating, resulting in low-resolution (R=$\lambda$/$\Delta\lambda$$\sim$30-200) spectral coverage from 0.60-2.87 and 4.06-4.18 $\mu$m and high-resolution (R=$\lambda$/$\Delta\lambda$$\sim$1900-3700) spectral coverage from 2.87-4.06 and 4.18-5.27 $\mu$m. The NRSIRS2RAPID readout pattern and a 4-point dither were used for both observations, with a total integration time of 2451 s for the G395H grating and 700 s for the PRISM. The MIRI data of L1527 spans the full MIRI range of 4.90-27.90 $\mu$m (R=$\lambda$/$\Delta\lambda$$\sim$1300-3700) and was collected using the FASTR1 readout pattern, a 2-point dither, and a total integration time of 2997 s divided equally between the three MIRI-MRS gratings.

Spectra were extracted from the L1527 NIRSpec and MIRI datacubes with apertures centered on the pixel providing the highest average signal over the full wavelength range (NIRSpec: RA 4:39:53.916, Dec +26:03:09.639; MIRI: RA 4:39:53.886, Dec +26:03:10.1248). A constant aperture with a 1" diameter was used to extract 1D spectra from both datacubes due to the source being extended and the majority of the detected flux being scattered light. The extraction resulted in two spectra with overlapping data points between 4.90-5.27 $\mu$m. A scaling factor of 0.783 was used to scale the NIRSpec data to the MIRI data in the spectral overlap region.

\section{Spectroscopic characterization}

\subsection{NH$_{3}$:H$_{2}$S}

Spectra of the NH$_{3}$:H$_{2}$S 1:1 mixture at various temperatures, along with spectra of pure NH$_{3}$ and H$_{2}$S at 15 K, are plotted with band assignments labeled in Figure~\ref{fig:all_assignments}. The peak positions for each assignment are listed in Table~\ref{tab:nh4sh_pure_assignments}. Assignments were made based on the work of \cite{bragin1977vibrational}, \cite{ferraro1980infrared} and, in the case of the NH$_{3}$:H$_{2}$S mixture at 15 K, changes in the relative intensities of features between the 3:2, 1:1, and 2:3 NH$_{3}$:H$_{2}$S mixtures (Figure~\ref{fig:nh3_h2s_conc}). For example, the relative intensity of the broad feature at 2417 cm$^{-1}$ (middle panel) increases with increasing fraction of H$_{2}$S with respect to NH$_{3}$ and decreases during warm-up as the neutral ices are consumed by the acid-base reaction, leading to a tentative assignment of the feature to H$_{2}$S in a matrix with NH$_{3}$ or NH$_{4}$SH. Some features are too blended to assign to one specific mode (e.g., the broad feature $\sim$2900 cm$^{-1}$ in the NH$_{3}$:H$_{2}$S mixture at 15 K tentatively assigned to two blended NH$_{4}$$^{+}$ modes).

\begin{figure}[h!]
\centering
\includegraphics[width=\linewidth]{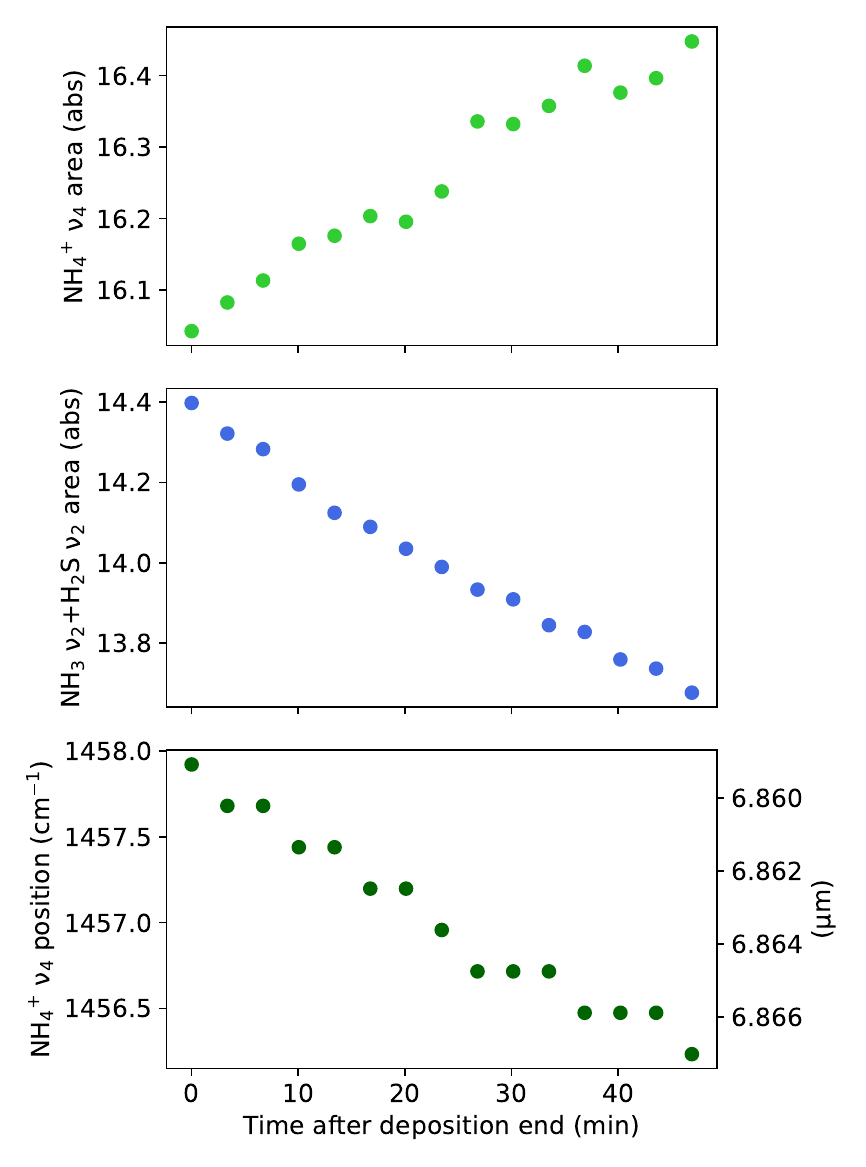}
\caption{Characterization of selected features in the NH$_{3}$:H$_{2}$S 1:1 spectra during the isothermal hold at 15 K after deposition finished. Top: peak area of the NH$_{4}$$^{+}$ $\nu_{4}$ feature. Middle: peak area of the blended NH$_{3}$ umbrella bending and H$_{2}$S bending features. Bottom: peak position of the NH$_{4}$$^{+}$ $\nu_{4}$ feature.}
\label{fig:15k_hold_pure}
\end{figure}

\subsubsection{Reaction at 15 K}

It is clear from the prominent NH$_{4}$$^{+}$ $\nu_{4}$ feature at $\sim$6.85 $\mu$m/1460 cm$^{-1}$ in the NH$_{3}$:H$_{2}$S mixtures that a significant amount of NH$_{4}$SH already forms upon co-deposition at 15 K. The quantity of reactants consumed at 15 K can be estimated via two methods: 1) by calculating the ratio between the peak area of the NH$_{3}$ umbrella mode in the mixture (with the contribution of the H$_{2}$S bending mode subtracted out via a two-Gaussian deconvolution) and the peak area in a pure NH$_{3}$ control sample deposited with identical experimental parameters, or 2) by calculating the ratio between the peak area of the NH$_{4}$$^{+}$ $\nu_{4}$ mode at 15 K and the maximum peak area achieved during warm-up. Both methods result in similar estimates of $\sim$50-60\% of the deposited NH$_{3}$ being consumed immediately upon deposition in all three of the mixing ratios investigated here. However, it is important to note that this is only a rough estimate of the reaction progression as these methods do not account for band strength variations caused by changes in temperature or chemical environment.

The in situ formation of ammonium salts below 20 K has been observed in previous works investigating the acid-base reaction between ammonia and H$_{2}$S \citep{loeffler2015giant} and other acids such as HNCO \citep{van2004quantitative}, HCN \citep{gerakines2004ultraviolet,noble2013thermal}, and HCOOH \citep{galvez2010ammonium,bergner2016kinetics}. \cite{van2004quantitative} and \cite{gerakines2004ultraviolet} suggested that the reaction may be driven at such low temperatures by the freed kinetic energy or the heat of condensation, respectively, of the room-temperature gas-phase molecules when they deposit on the cold substrate. In our experiments, we held the ice samples at 15 K for approximately 1 hr after deposition finished before beginning warm-up to monitor the reaction process at cold temperatures without energy input from the warm gas deposition. During this time, the integrated peak area of the NH$_{4}$$^{+}$ $\nu_{4}$ mode continued to increase while the integrated peak area of the blended NH$_{3}$ umbrella mode and H$_{2}$S bending mode decreased (Figure~\ref{fig:15k_hold_pure}), demonstrating that the NH$_{3}$ and H$_{2}$S ice continue to react with each other even after the deposition is complete and the reactants are completely thermalized by the actively cryo-cooled substrate.

\subsubsection{Reaction during warm-up, crystallization, and desorption}

As the ice mixtures are heated, their NH$_{4}$$^{+}$ $\nu_{4}$ features grow at a much faster rate than during the isothermal hold at 15 K, indicating an increase in the reaction rate. In the 1:1 mixture, the NH$_{4}$$^{+}$ $\nu_{4}$ feature reaches its maximum peak area at $\sim$131 K. At this temperature, the integrated peak area of the NH$_{3}$ umbrella mode is 3\% relative to its integrated peak area at the beginning of the warm-up (at which point $\sim$50\% of the NH$_{3}$ had already reacted). Moreover, the TPD spectra show that thermal desorption of NH$_{3}$ between 15-131 K only accounts for 0.4\% of the total 17 m/z mass signal (which traces the desorption of both neutral ammonia and the ammonium cation) integrated from 15-250 K. This leads us to estimate that $\sim$98\% of the deposited NH$_{3}$ is converted to NH$_{4}$SH by 131 K, with very little of the originally deposited NH$_{3}$ remaining neutral or desorbing by this temperature. 

During the warm-up from 15 to 131 K, the broadness of the NH$_{4}$SH IR features and the presence of the NH$_{4}$$^{+}$ symmetric bending mode ($\nu_{2}$) at 5.90 $\mu$m/1696 cm$^{-1}$, which is IR inactive in crystalline NH$_{4}$SH \citep{hudson2015activation}, shows that the ice remains amorphous at these temperatures. Peak sharpening and/or splitting of all of the NH$_{4}$SH IR features (except the NH$_{4}$$^{+}$ $\nu_{2}$ feature, which disappears), are observed at $\sim$135 K, indicative of the salt crystallizing. This event also coincides with an increase in the desorption of both NH$_{3}$ and H$_{2}$S in the TPD spectrum (Figure~\ref{fig:nh4sh_pure_tpd}) and the beginning of the gradual decrease in the peak area of the NH$_{4}$$^{+}$ $\nu_{4}$ feature, demonstrating that the crystallization and desorption temperatures of this salt are very close to each other. Peak desorption is observed at 164 K in the TPD, and all of the salt features disappear from the IR spectrum by $\sim$174 K. In the NH$_{3}$:H$_{2}$S 2:3 and 3:2 mixtures, the maximum NH$_{4}$$^{+}$ $\nu_{4}$ peak area (i.e., maximum reaction completion assuming the band strength does not change significantly with temperature) is achieved at lower temperatures, and the samples crystallize at higher temperatures (see Figure~\ref{fig:nh4_pure_char}).

\subsubsection{NH$_{4}$$^{+}$ $\nu_{4}$ feature}

The growth of the NH$_{4}$$^{+}$ $\nu_{4}$ feature during warm-up from 15 to 131 K is accompanied by a change in the feature's profile, which becomes more asymmetric, and a redshift in the feature's peak position. Such a redshift of this feature with increasing temperature has been observed in other laboratory ammonium salts \citep{grim1989infrared,schutte2003origin,raunier2003reactivity}. These spectral changes during the warm-up are likely a consequence of the change in the chemical environment between 15 K (when the ice matrix is still dominated by a relatively high quantity of unreacted neutral NH$_{3}$ and H$_{2}$S) and 131 K (when very little NH$_{3}$ and H$_{2}$S remain in the ice matrix, resulting in a matrix that now contains nearly exclusively ionic species). Thermal effects may also play a role in the redshift of the feature during warm-up, but the fact that we also observe a redshift in the feature during the isothermal holds at 15 K, during which the concentration of NH$_{4}$SH with respect to NH$_{3}$ and H$_{2}$S in the ice matrix is increasing (see Figure~\ref{fig:15k_hold_pure}), demonstrates that increasing salt concentration in the ice matrix must play a role in the redshift as well. The redshift of this feature has significant astronomical ramifications that are further discussed in Section~\ref{txt:astro_comparison}. The peak positions, FWHMs, and relative areas of this feature during warm-up are presented in Table~\ref{tab:nh4_pure_char} and Figure~\ref{fig:nh4_pure_char}.

\begin{figure*}[ht!]
\centering
\includegraphics[width=0.9\linewidth]{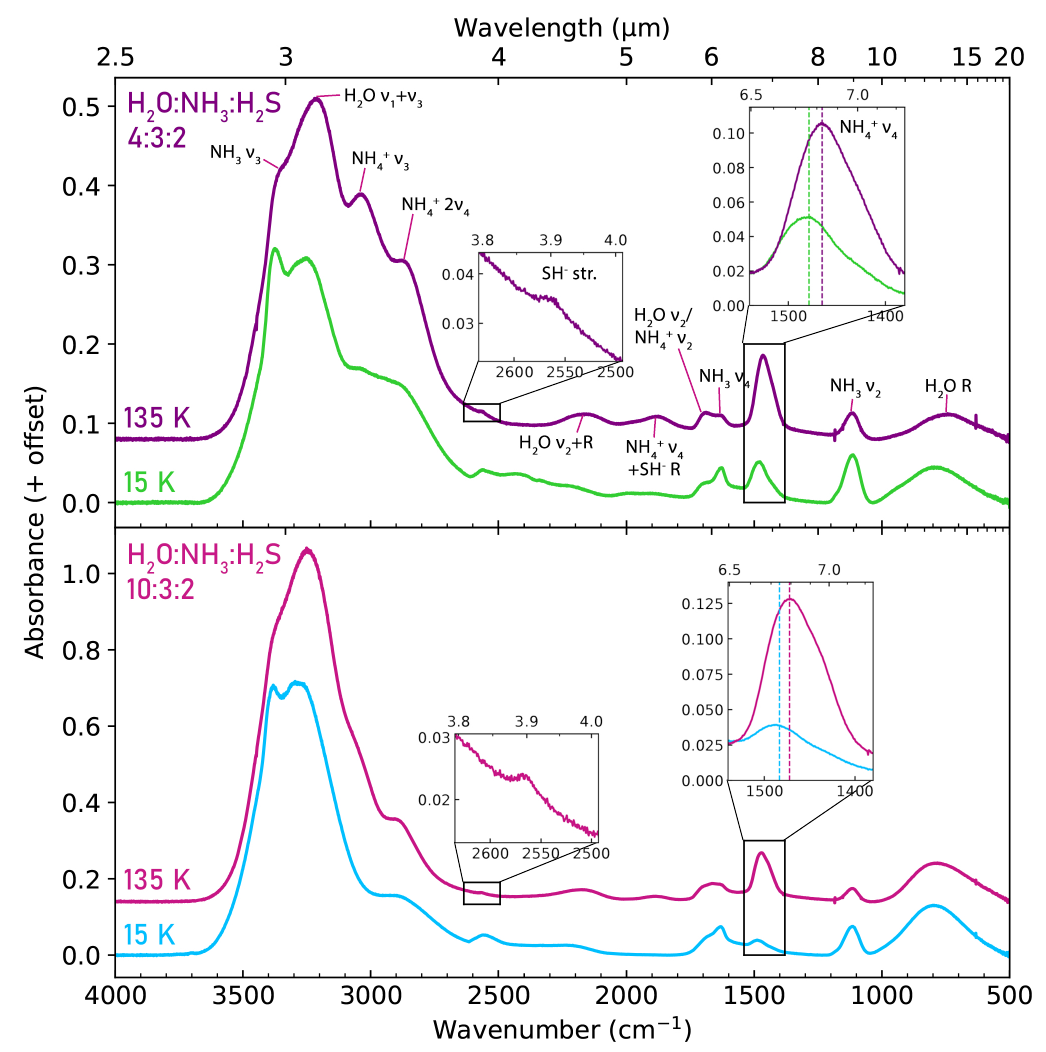}
\caption{Spectra of two H$_{2}$O:NH$_{3}$:H$_{2}$S mixtures at 15 and 135 K. The plot insets zoom in on two features of interest in this work, the SH$^{-}$ stretching mode and the NH$_{4}$$^{+}$ asymmetric bending ($\nu_{4}$) mode. The dashed vertical lines within the right inset indicate the peak positions of the NH$_{4}$$^{+}$ $\nu_{4}$ mode at the two plotted temperatures, showing how the feature redshifts as the ice is heated. Vibrational mode assignments are provided for the H$_{2}$O:NH$_{3}$:H$_{2}$S mixture at 135 K.}
\label{fig:h2o_nh3_h2s_spectra}
\end{figure*}

\subsubsection{SH$^{-}$ stretching feature}

It is difficult to similarly characterize the SH$^{-}$ stretching mode at $\sim$2555 cm$^{-1}$ throughout sample warm-up because the feature directly overlaps with the H$_{2}$S asymmetric and symmetric stretching modes (Figure~\ref{fig:all_assignments}). At 15 K, when salt concentrations are the lowest in the sample, the feature sits at 2560 cm$^{-1}$, its peak area is at its maximum, and it is blended with neighboring strong, broad features. As the sample is warmed and the acid-base reaction is facilitated, the combined H$_{2}$S and SH$^{-}$ peak redshifts and decreases in area, demonstrating that the band strength of the SH$^{-}$ stretching mode is considerably lower than that of the the H$_{2}$S stretching modes. At this point in the warm-up, the contribution of unreacted H$_{2}$S to this feature may not be negligible in the mixtures where H$_{2}$S is in excess, so the feature is only characterized further in the NH$_{3}$:H$_{2}$S 3:2 mixture. During crystallization, the feature sharpens and blueshifts to 2570 cm$^{-1}$, but its area remains small. The peak positions, FWHMs, and relative areas of the SH$^{-}$ feature at temperatures at and above 102 K (the temperature at which the reaction is complete) in the 3:2 mixture are reported in Table~\ref{tab:sh_pure_char}.

\subsection{H$_{2}$O:NH$_{3}$:H$_{2}$S}
\label{txt:results_nh4sh_h2o}

Spectra of two selected H$_{2}$O:NH$_{3}$:H$_{2}$S mixtures at 15 and 135 K are presented in Figure~\ref{fig:h2o_nh3_h2s_spectra}. Several of the features that are clear in the NH$_{3}$:H$_{2}$S mixtures are much less prominent once H$_{2}$O is added to the mixtures, in particular the various NH$_{4}$$^{+}$ peaks between $\sim$3.2-3.6 $\mu$m/3100-2800 cm$^{-1}$, which present as broad shoulders on the red wing of the H$_{2}$O OH stretching mode, the NH$_{4}$$^{+}$ scissor (symmetric) bending mode at $\sim$5.9 $\mu$m/1690 cm$^{-1}$, which blends with the H$_{2}$O bending mode, and the weak feature we tentatively assign to the combination of the NH$_{4}$$^{+}$ symmetric bend and the SH$^{-}$ libration at $\sim$4.8/2090 cm$^{-1}$, which blends with the broad H$_{2}$O combination mode. However, the  NH$_{4}$$^{+}$ $\nu_{4}$ mode at 6.80 $\mu$m/1470 cm$^{-1}$ and its combination band with the SH$^{-}$ libration at 5.3 $\mu$m/1880 cm$^{-1}$ remain distinct from the water ice bands. The SH$^{-}$ stretching mode at 3.9 $\mu$m/2560 cm$^{-1}$ is also clearly distinguishable from its local continuum but remains extremely small relative to the other salt features.

\subsubsection{Reaction at 15 K}

\begin{figure}[h!]
\centering
\includegraphics[width=\linewidth]{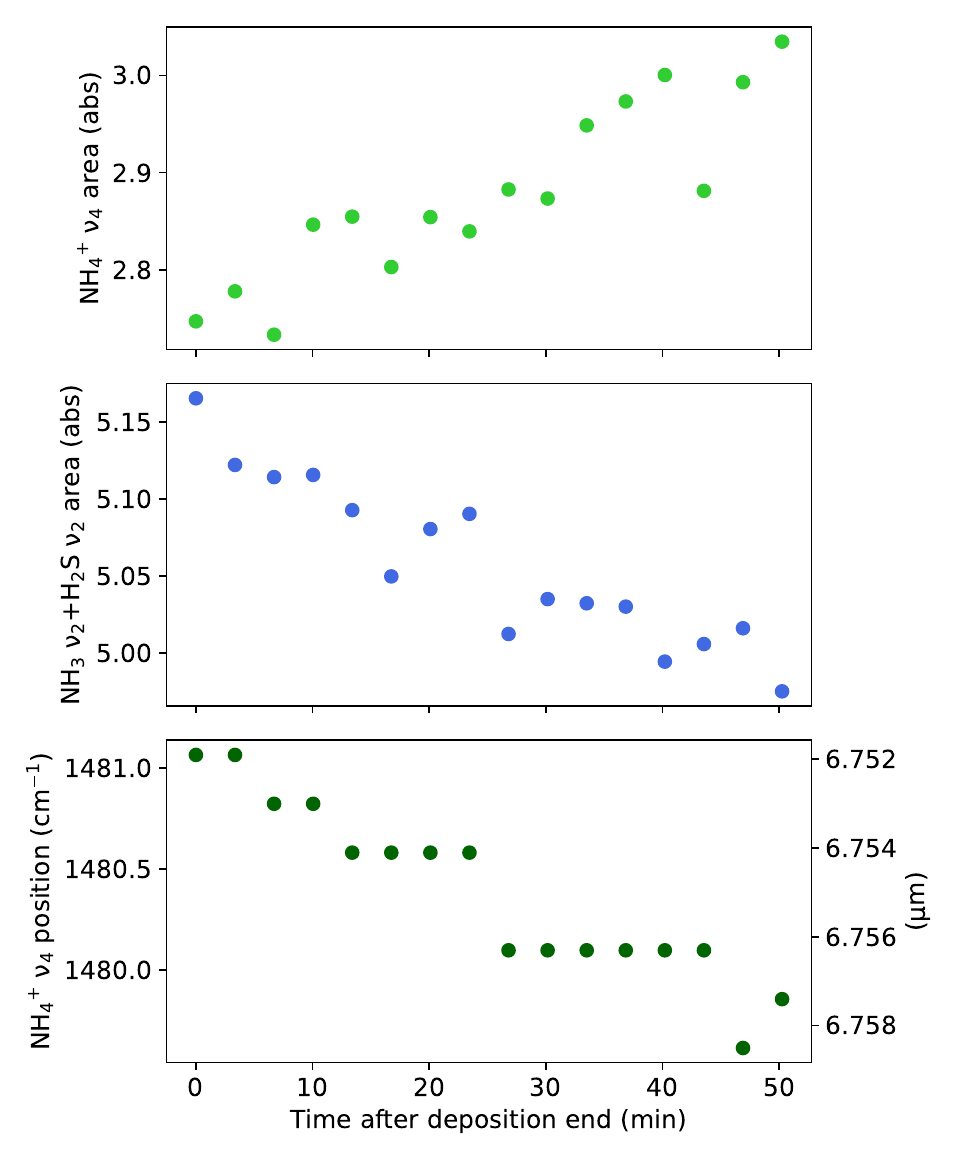}
\caption{Characterization of selected features in the H$_{2}$O:NH$_{3}$:H$_{2}$S 4:3:2 spectra in the period between the end of the ice deposition and the start of the ice warm-up. During this time, the sample was held at 15 K. Top: peak area of the NH$_{4}$$^{+}$ $\nu_{4}$ feature. Middle: peak area of the blended NH$_{3}$ umbrella bending and H$_{2}$S bending features. Bottom: peak position of the NH$_{4}$$^{+}$ $\nu_{4}$ feature.}
\label{fig:15k_hold_water}
\end{figure}

\begin{figure}[h!]
\centering
\includegraphics[width=\linewidth]{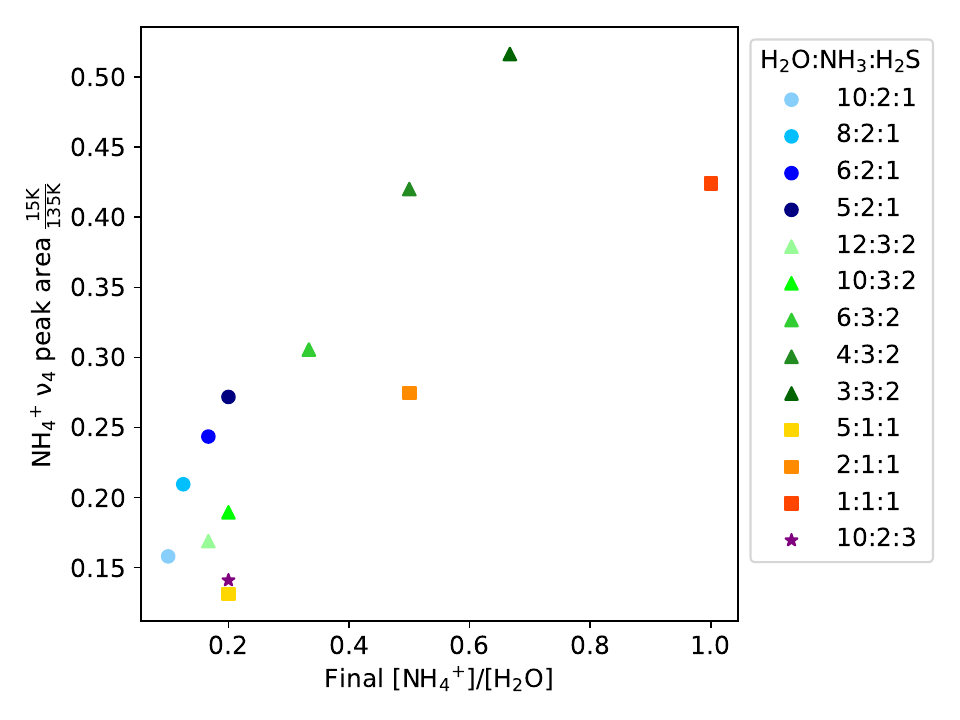}
\caption{The ratio of the integrated absorbance of the NH$_{4}$$^{+}$ $\nu_{4}$ peak at 15 K versus 135 K plotted as a function of the concentration of NH$_{4}$$^{+}$ with respect to H$_{2}$O (assuming 100\% reaction completion) for various H$_{2}$O:NH$_{3}$:H$_{2}$S mixtures.}
\label{fig:nh4_peak_area}
\end{figure}

The presence of the NH$_{4}$$^{+}$ $\nu_{4}$ mode already at 15 K in all of the water-containing mixtures demonstrates that NH$_{4}$SH can form at such low temperatures even when NH$_{3}$ and H$_{2}$S are diluted in water. Furthermore, just as in the water-free mixtures, acid-base reaction progression during an isothermal hold at 15 K after deposition has ended can be observed via the simultaneous increase in the peak area of the NH$_{4}$$^{+}$ $\nu_{4}$ mode and decrease in the peak area of the blended NH$_{3}$ umbrella bending mode and H$_{2}$S bending mode (Figure~\ref{fig:15k_hold_water}). This reaction progression is again accompanied by a redshift in the peak position of the NH$_{4}$$^{+}$ $\nu_{4}$ mode. However, the more dilute the initial concentrations of NH$_{3}$ and H$_{2}$S with respect to H$_{2}$O, the lower the integrated absorbance of the NH$_{4}$$^{+}$ $\nu_{4}$ mode at 15 K with respect to the final integrated absorbance of the NH$_{4}$$^{+}$ $\nu_{4}$ mode at 135 K -- that is, the reaction at 15 K is slowed, but not completely prevented, by the presence of water at the investigated concentrations. The ratio of the integrated absorbance of the NH$_{4}$$^{+}$ $\nu_{4}$ mode at 15 K versus 135 K is plotted as a function of final NH$_{4}$$^{+}$ abundance with respect to H$_{2}$O for all the investigated mixtures in Figure~\ref{fig:nh4_peak_area} to demonstrate this point.

\subsubsection{Reaction during warm-up, crystallization, and desorption}

During warm-up of the water-rich mixtures, the NH$_{4}$$^{+}$ $\nu_{4}$ mode grows until reaching a maximum between 135-140 K while the area of the blended NH$_{3}$ umbrella bending mode and H$_{2}$S bending mode decreases, similar to what is observed in the water-free mixtures. The sharpening of the NH$_{4}$SH features characteristic of crystallization does not occur until $\sim$150-155 K, so dilution with H$_{2}$O delays the salt crystallization just as the presence of excess unreacted NH$_{3}$ and H$_{2}$S do in the water-free mixtures. As the warm-up continues after crystallization, the salt features begin to rapidly diminish and are completely absent from the spectrum between $\sim$175-180 K, around the same temperature range in which H$_{2}$O desorbs. The TPD peaks of NH$_{3}$, H$_{2}$S, and H$_{2}$O occur around the same temperatures, $\sim$171-172 K (e.g., Figure~\ref{fig:nh4sh_10-3-2_tpd}). Such an increase ($\sim$10 K) in the peak desorption temperature of NH$_{4}$SH when it is mixed with water rather than pure was also recently observed by \cite{vitorino2024sulphur}. The consequences of the similar desorption temperatures of NH$_{4}$SH and H$_{2}$O on the detection of the NH$_{4}$SH desorption products in the gas phase are discussed in Section~\ref{txt:implications_desorption}.

\subsubsection{NH$_{4}$$^{+}$ $\nu_{4}$ feature}

\begin{figure}[h!]
\centering
\includegraphics[width=\linewidth]{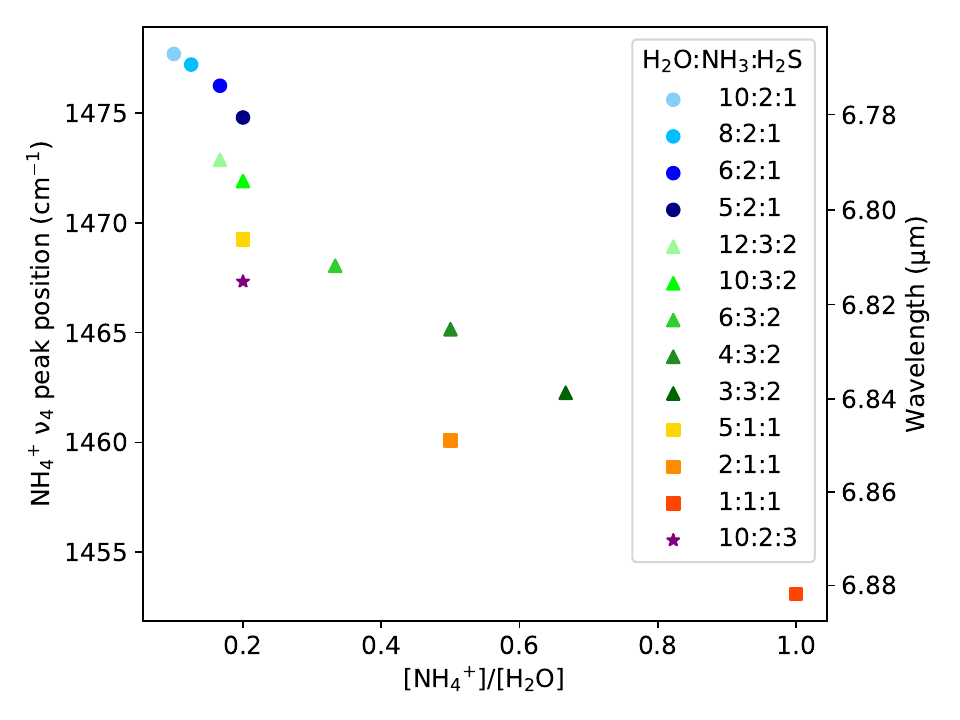}
\caption{The ratio of the peak position of the NH$_{4}$$^{+}$ $\nu_{4}$ mode at 135 K as a function of the concentration of NH$_{4}$$^{+}$ with respect to H$_{2}$O plotted for various H$_{2}$O:NH$_{3}$H$_{2}$S mixtures.}
\label{fig:nh4_peak_conc}
\end{figure}

As in the water-free spectra, the NH$_{4}$$^{+}$ $\nu_{4}$ mode redshifts during warm-up. However, the feature also redshifts with increasing salt concentration with respect to H$_{2}$O when comparing the spectra of ices at the same temperature. Figure~\ref{fig:nh4_peak_conc} visualizes this trend at 135 K. Comparison of the NH$_{4}$$^{+}$ $\nu_{4}$ mode peak position at 135 K between the H$_{2}$O:NH$_{3}$:H$_{2}$S 5:2:1, 10:3:2, 5:1:1, and 10:2:3 mixtures (where the salt concentrations with respect to H$_{2}$O are identical at 20\%, assuming reaction completion in all of the mixtures) also reveals that increasing the excess of unreacted NH$_{3}$ in the matrix causes the NH$_{4}$$^{+}$ $\nu_{4}$ mode to blueshift, while increasing the excess of unreacted H$_{2}$S in the matrix causes it to redshift.

This led us to postulate that the redshifts of this feature observed during the isothermal holds at 15 K and the warm-up are caused primarily by the increase in salt concentration with respect to the hydrogen-bonded matrix. During warm-up, the salt concentrations increase at a higher rate due to the acid and base molecules mobilizing with increasing temperature, which allows them to diffuse more easily within the ice matrix and react with each other.

In an effort to characterize the potential degeneracy between redshifts caused by increasing temperature and concentration, we conducted an annealing experiment where we deposited a H$_{2}$O:NH$_{3}$:H$_{2}$S 10:3:2 mixture, warmed it to 135 K at 1 K min$^{-1}$, and recooled it to 15 K at -5 K min$^{-1}$. Upon recooling (during which the salt concentration in the ice is not expected to change), the NH$_{4}$$^{+}$ $\nu_{4}$ mode blueshifted from 1472 cm$^{-1}$ to 1475.5 cm$^{-1}$ (Figure~\ref{fig:annealing}). This peak shift of 3.5 cm$^{-1}$ is significantly smaller than the redshift of over 11 cm$^{-1}$ (1483 to 1472 cm$^{-1}$) that occurred over the initial warm-up from 15 to 135 K during which the majority of the salt formed. This suggests that thermal changes are not the main driver of most of the feature's redshift observed during warm-up, leaving changes in the chemical environment (e.g., salt concentration) as well as irreversible thermally-driven changes in the ice morphology (e.g., ice compaction) as the remaining possible contributors to the majority of the observed redshift during warm-up.

We therefore conclude that, although some of the redshifts in the NH$_{4}$$^{+}$ $\nu_{4}$ mode associated with warm-up in our experiments as well as in some previous works may be caused by thermal effects, one of the primary physical causes of the redshift is the increase in salt concentration with respect to hydrogen-bonded matrix species as the deposited ice is warmed. We specify that the concentration increase is with respect to hydrogen-bonded matrix species because we do not observe the same trend when the salt is diluted with excess H$_{2}$S (Figure~\ref{fig:nh4_peak_conc}). Our conclusion is consistent with that drawn by \cite{mate2009water}, who stated that the redshift in the NH$_{4}$$^{+}$ $\nu_{4}$ mode observed by \cite{schutte2003origin} upon ice warm-up was likely due to ``varying interactions among all the different species in the matrix.'' Also, the NH$_{4}$$^{+}$ $\nu_{4}$ mode does not appear to vary significantly between 14 and 150 K in the NH$_{4}$HCOO:H$_{2}$O spectra formed via hyperquenching in \cite{galvez2010ammonium}, in which the salt concentration does not change during heating. The astronomical ramifications of this concentration dependence of the NH$_{4}$$^{+}$ $\nu_{4}$ mode peak position are further discussed in Section~\ref{txt:astro_nh4}. The peak positions, FWHMs, and relative areas of this feature during warm-up in all of the investigated mixtures are reported in Appendix~\ref{app:nh4sh_h2o}.

\subsubsection{SH$^{-}$ stretching feature}

As in the pure NH$_{4}$SH spectra, the blended H$_{2}$S and SH$^{-}$ stretching peak becomes smaller as the ice is heated and the NH$_{4}$$^{+}$ $\nu_{4}$ mode grows. The peak positions and FWHMs of the feature at 135 K range from 2567-2559 cm$^{-}$ and 25-32 cm$^{-}$, respectively, in the mixtures investigated here, with the highest salt concentration mixtures having the most redshifted and broadest features (see Appendix~\ref{app:nh4sh_h2o}).

\subsubsection{Band strengths}
\label{txt:band_strengths}

\begin{table}[h!]
\caption{Apparent band strengths of the NH$_{4}$$^{+}$ $\nu_{4}$ mode and the SH$^{-}$ stretching mode in H$_{2}$O:NH$_{3}$H$_{2}$S mixtures of three different concentrations.}
\begin{center}
\begin{tabular}{c c c}
\hline
        Ratio & A' NH$_{4}$$^{+}$ $\nu_{4}$$^{(a)}$ & A' SH$^{-}$ str.$^{(b)}$ \\
        H$_{2}$O:NH$_{3}$H$_{2}$S & 10$^{-17}$ cm molec$^{-1}$ & 10$^{-19}$ cm molec$^{-1}$ \\
        \hline
        10:2:1 & 3.6$\pm$0.4 & 3.7$\pm$0.5 \\
        10:3:2 & 3.6$\pm$0.4 & 3.1$\pm$0.4 \\
        8:5:4 & 3.2$\pm$0.3 & 3.2$\pm$0.4 \\
    \hline
     
\end{tabular}
\label{tab:band_strengths}

\begin{tablenotes}
$^{(a)}$ Integration range: 1560-1340 cm$^{-1}$.\\
$^{(b)}$ Integration range: 2611-2520 cm$^{-1}$.\\
\end{tablenotes}

\end{center}
\end{table}

The Beer's law plots used to calculate the apparent band strengths of the NH$_{4}$$^{+}$ $\nu_{4}$ mode and the SH$^{-}$ stretching mode in the 10:3:2 H$_{2}$O:NH$_{3}$:H$_{2}$S mixture are shown in Figures~\ref{fig:beer_nh4_20p} and~\ref{fig:beer_sh_20p} (the 10:2:1 and 8:5:4 mixture plots can be found in Appendix~\ref{app:nh4sh_h2o}). The derived values from the slopes of the linear fits in all of these plots are provided in Table~\ref{tab:band_strengths}.

Despite previous theoretical and experimental works raising concerns regarding the NH$_{4}$$^{+}$ $\nu_{4}$ mode band strength decreasing significantly with dilution in H$_{2}$O \citep{mate2009water,galvez2010ammonium}, we do not find statistically significant differences between the NH$_{4}$$^{+}$ apparent band strengths of NH$_{4}$SH:H$_{2}$O mixtures with different salt concentrations ranging from 10-50\% with respect to H$_{2}$O. Our apparent band strengths are very close to the apparent band strength of the NH$_{4}$$^{+}$ $\nu_{4}$ mode in pure NH$_{4}$CN derived recently by \cite{gerakines2024sublimation}, 3.58($\pm$0.07)$\times$10$^{-17}$ cm molec$^{-1}$. They are also a factor of$\sim$1.22-1.38 lower than the band strength of the NH$_{4}$$^{+}$ $\nu_{4}$ mode in NH$_{4}$HCOO:H$_{2}$O ice from \cite{schutte2003origin}, 4.4$\times$10$^{-17}$ cm molec$^{-1}$, which is often utilized to quantify NH$_{4}$$^{+}$ ice column densities in observational works. The band strength of the SH$^{-}$ stretching mode is almost two orders of magnitude lower than the band strength of the H$_{2}$S stretching mode at the same wavelength \citep{yarnall2022new}.

\begin{figure}[h!]
\centering
\includegraphics[width=\linewidth]{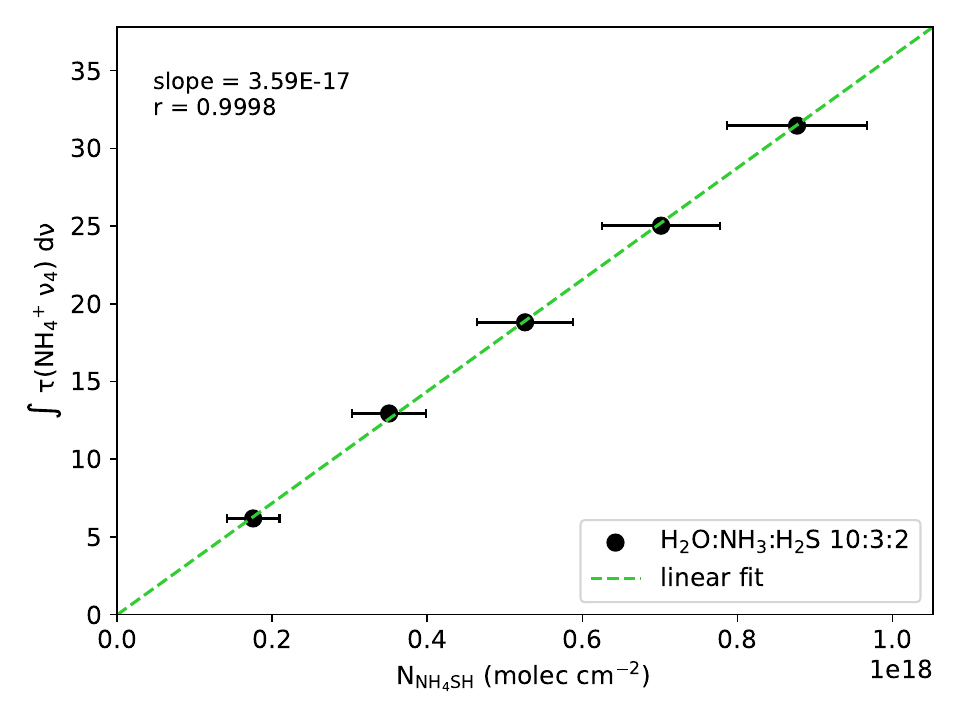}
\caption{Beer's law plot used to derive the apparent band strength of the NH$_{4}$$^{+}$ $\nu_{4}$ feature formed in the H$_{2}$O:NH$_{3}$:H$_{2}$S 10:3:2 mixture.}
\label{fig:beer_nh4_20p}
\end{figure}

\begin{figure}[h!]
\centering
\includegraphics[width=\linewidth]{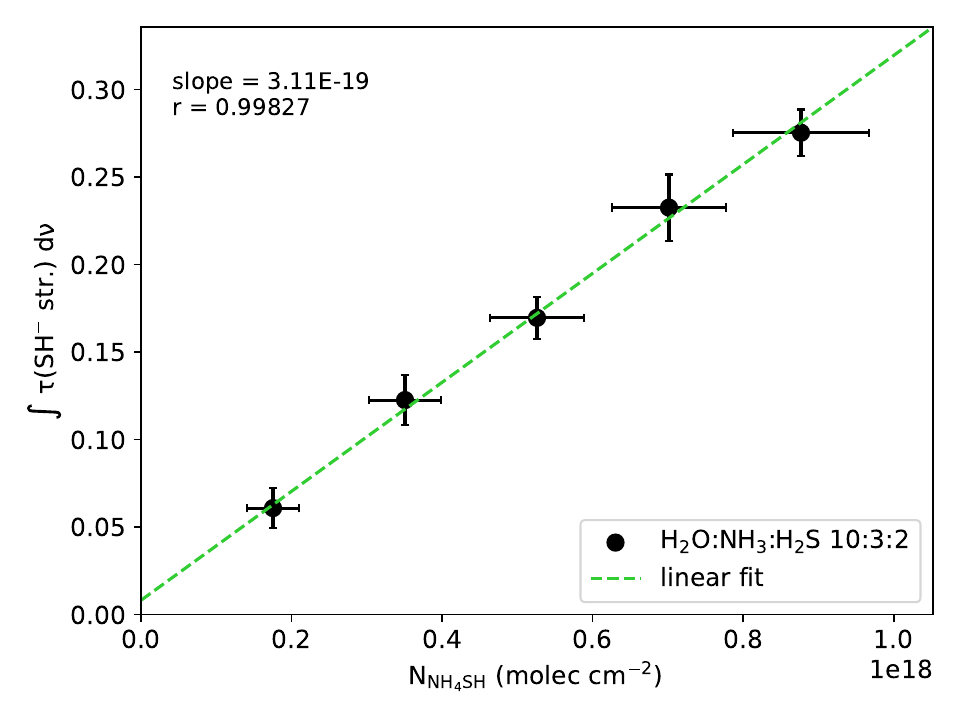}
\caption{Beer's law plot used to derive the apparent band strength of the SH$^{-}$ stretching feature formed in the H$_{2}$O:NH$_{3}$:H$_{2}$S 10:3:2 mixture.}
\label{fig:beer_sh_20p}
\end{figure}

As was mentioned previously, these band strengths assume that the limiting reactant, H$_{2}$S, was fully consumed in the reaction by 135 K. The quantity of H$_{2}$S consumed in the experiments used to calculate these band strengths cannot be estimated from the IR spectra because its strongest band, the stretching mode, overlaps closely with the strongest band of the reaction product SH$^{-}$. For this reason, we estimate instead the quantity of limiting reactant consumed by 135 K in the H$_{2}$O:NH$_{3}$:H$_{2}$S 10:2:3 mixture, where an estimate of the final abundance of the limiting reactant, NH$_{3}$, can be obtained via its umbrella mode. At 135 K, the peak area of the NH$_{3}$ umbrella mode is $\sim$4\% relative to its peak area at the beginning of the warm-up (at which point we estimate $\sim$14\% of the deposited NH$_{3}$ has already reacted based on Figure~\ref{fig:nh4_peak_area}). Assuming negligible desorption of unreacted NH$_{3}$ by 135 K due to entrapment based on QMS data, we estimate that $\sim$96\% of the deposited NH$_{3}$ has reacted with H$_{2}$S by 135 K in this experiment. Given that, in the NH$_{3}$:H$_{2}$S mixtures, the acid-base reaction appears to proceed more readily in NH$_{3}$ excess than in H$_{2}$S excess (Figure~\ref{fig:nh4_pure_char}), we extrapolate that in our H$_{2}$O-containing mixtures where H$_{2}$S is the limiting reactant, an equivalent or higher percentage of H$_{2}$S will have been consumed in the acid-base reaction by 135 K.

\section{Comparing NH$_{4}$SH:H$_{2}$O laboratory spectra to observations}
\label{txt:astro_comparison}

The assignment of the 6.85 $\mu$m band observed toward icy sightlines to the NH$_{4}$$^{+}$ cation was first suggested over 40 years ago \citep{knacke1982observation}. Since then, despite rigorous investigations in numerous observational, experimental, and theoretical works, the assignment remains tentative. As mentioned in Section~\ref{txt:intro}, previous works provided three primary arguments against a secure assignment: 1) the NH$_{4}$$^{+}$ $\nu_{4}$ mode is too broad in laboratory data of ammonium salts in water-rich matrices to match observations and, at the expected NH$_{4}$$^{+}$ concentrations, the band strength of the feature is likely so low that it may render the salt undetectable \citep{mate2009water,galvez2010ammonium}; 2) the calculated column densities of anions identified in interstellar ices are insufficient (by about an order of magnitude) to fully counterbalance the expected positive charge from the NH$_{4}$$^{+}$ cation \citep{schutte1999weak,van2004quantitative,van20053,boogert2022survey,mcclure2023ice,rocha2024jwst}; and 3) no other features that can be unambiguously assigned to NH$_{4}$$^{+}$ have been detected \citep{boogert2015observations}. In the subsequent analysis, we will address each of these points using the new laboratory data collected in this work.

\begin{figure*}[ht!]
\centering
\includegraphics[width=0.9\linewidth]{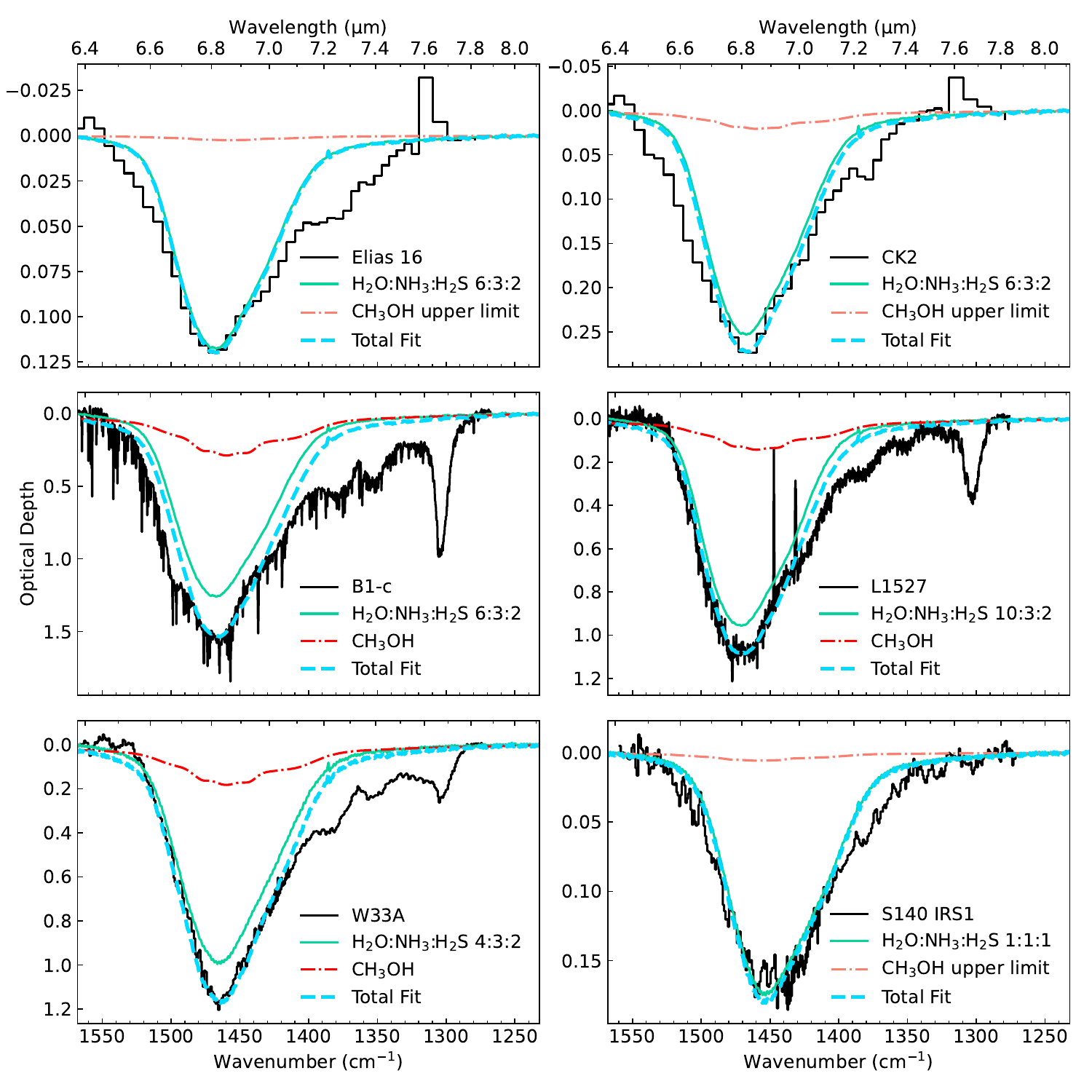}
\caption{Fits of laboratory IR spectra of NH$_{4}$SH:H$_{2}$O ice mixtures at 135 K (sea green solid trace) and pure CH$_{3}$OH ice at 15 K (dash-dotted red or pink trace) to the 6.85 $\mu$m feature observed toward four protostars and two dense clouds (see Table~\ref{tab:sources}). The total combined fits of both NH$_{4}$SH and CH$_{3}$OH are indicated with the dashed light blue trace. The best fitting mixture used in each fit is listed inside each panel. Note that many of the sharp spikes present in the JWST LYSO spectra are gas-phase lines in absorption (e.g., B1-c) or emission (e.g., L1527) rather than noise. Local continua used to extract the feature from the observational spectra can be found in Figure~\ref{fig:nh4_obs_cont}.}
\label{fig:nh4_obs}
\end{figure*}

\begin{figure}[h!]
\centering
\includegraphics[width=\linewidth]{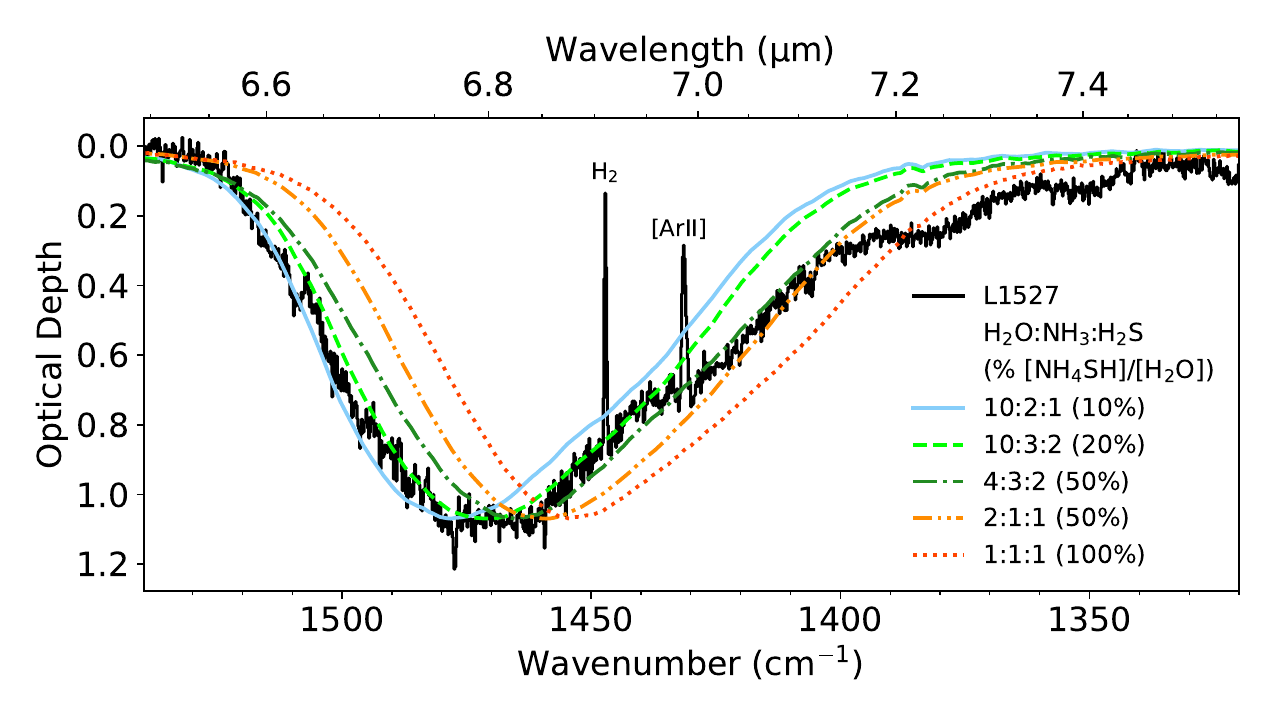}
\caption{Selected laboratory spectra of H$_{2}$O:NH$_{3}$:H$_{2}$S ice mixtures of various concentration plotted with the 6.85 $\mu$m feature observed toward L1527 with JWST.}
\label{fig:nh4_v4_labvobs}
\end{figure}

\subsection{NH$_{4}$$^{+}$ $\nu_{4}$ mode}
\label{txt:astro_nh4}

Our apparent band strength measurements of the NH$_{4}$$^{+}$ $\nu_{4}$ mode in NH$_{4}$SH diluted in H$_{2}$O ice at concentrations between 10 and 50\% with respect to H$_{2}$O range from 3.2($\pm$0.3)-3.6($\pm$0.4)$\times$10$^{-17}$ cm molec$^{-1}$. Such band strengths are comparable in order of magnitude to the band strength of the NH$_{3}$ umbrella bending mode at 9.35 $\mu$m \citep{kerkhof1999infrared,hudson2022ammonia}, which is typically detected at concentrations on the order of $\sim$5\% with respect to H$_{2}$O toward pre- and protostellar ices \citep{bottinelli2010c2d,boogert2015observations,mcclure2023ice}.

Previous laboratory experiments where ammonium salts were grown via the hyperquenching method (i.e., droplets of salts dissolved in liquid water at the desired concentration are directly dripped onto a cold IR substrate) at concentrations between 7-25\% with respect to H$_{2}$O did not formally report band strengths of the NH$_{4}$$^{+}$ $\nu_{4}$ mode, but it was noted that the feature was so weakened and broadened by the presence of water that it became implausible to securely detect assuming similar local salt concentrations with respect to H$_{2}$O in interstellar ices \citep{mate2009water,galvez2010ammonium}. \cite{galvez2010ammonium} pointed out that such dramatic weakening did not occur in salt:H$_{2}$O ice mixtures formed via vapor co-deposition, the same method used in this work, and ascribed the difference in band strengths to the fact that, in hyperquenched ice samples, the ions are expected to be densely surrounded by ``solvation spheres'' of H$_{2}$O molecules, which disturbs their ionic character and decreases their band strengths. They further argued that the water-rich ices forming in the interstellar medium via atom-addition reactions on grain surfaces are expected to be compact, mimicking more closely the morphology of their hyperquenched samples rather than the open, porous structures of water ices formed via vapor deposition in the lab.

We argue that this may not necessarily be the case, because if the salts in interstellar ice mantles form via in situ acid-base reactions, the acid and base molecules must be adjacent to each other in the ice matrix, which will result in the cations and anions formed via the reaction to be adjacent to each other as well, rather than separated and fully surrounded by solvation spheres of water. Such cation-anion pairs within the ice matrix are expected to have higher ionic character and, therefore, higher band strengths.

\begin{table}[h!]
\caption{H$_{2}$O and NH$_{4}$$^{+}$ ice column densities calculated toward the investigated sources using laboratory IR spectra and apparent band strengths of NH$_{4}$SH in H$_{2}$O-rich ices.}
\begin{center}
\begin{tabular}{c c c c}
\hline
        Source & H$_{2}$O & \multicolumn{2}{c}{NH$_{4}$$^{+}$}\\
        & 10$^{19}$ cm$^{-2}$ & 10$^{18}$ cm$^{-2}$ & \% H$_{2}$O\\
        \hline
        Elias 16 & 0.24 (0.03)$^{(a)}$ & 0.26 (0.05) & 11 (2)\\
        CK2 & 0.30 (0.03)$^{(a)}$ & 0.57 (0.11) & 19 (4) \\
        B1c & 2.5 (0.2)$^{(b)}$ & 2.8 (0.5) & 11 (2) \\
        L1527 & 2.6 (0.6)$^{(c)}$ & 2.1 (0.4) & 8 (3) \\
        W33A & 1.25 (0.3)$^{(a)}$ & 2.5 (0.5) & 20 (7) \\
        S140 IRS1 & 0.20 (0.03)$^{(a)}$ & 0.45 (0.08) & 23 (5) \\
    \hline
     
\end{tabular}
\label{tab:nh4_cd}

\begin{tablenotes}
Uncertainties are indicated in parentheses ().\\
$^{(a)}$ from \cite{boogert2013infrared}.\\
$^{(b)}$ from \cite{chen2024joys}.\\
$^{(c)}$ from the 12 $\mu$m H$_{2}$O libration using a band strength of 3.2$\times$10$^{-17}$ cm molec$^{-1}$ \citep{mastrapa2009optical,bouilloud2015bibliographic}; see Figure~\ref{fig:l1527_cont_sil_h2o}.\\
\end{tablenotes}

\end{center}
\end{table}

\begin{figure*}[h!]
\centering
\includegraphics[width=0.9\linewidth]{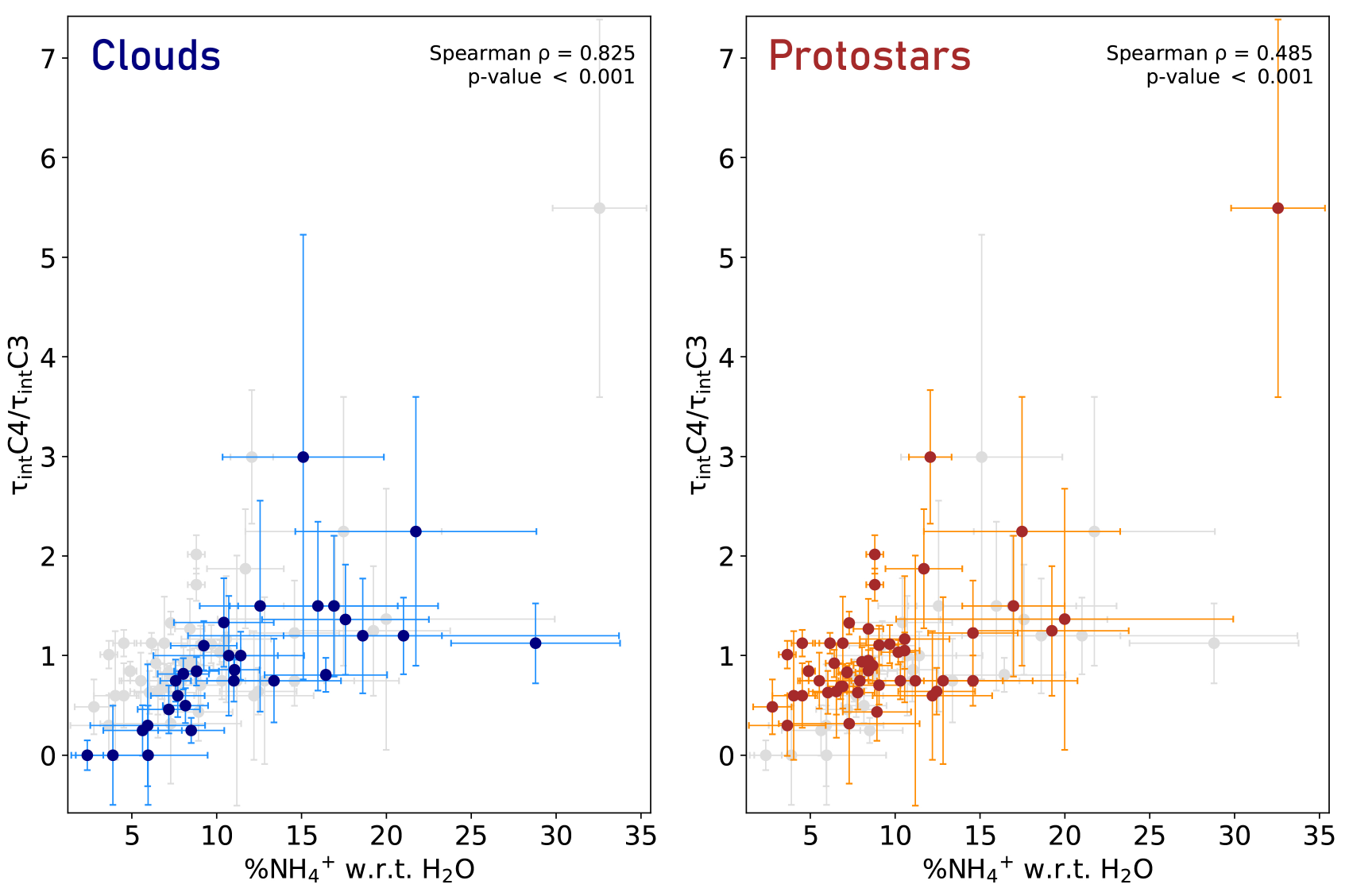}
\caption{Abundances of the ammonium cation with respect to H$_{2}$O ice plotted against the $\tau_{int}$C4/$\tau_{int}$C3 ratios of ices toward isolated dense clouds (left, blue) and protostars (right, red). To facilitate comparison, the protostellar and cloud values are also plotted in the left and right plots, respectively, in light gray. A higher $\tau_{int}$C4/$\tau_{int}$C3 is indicative of a greater redshift in the peak position of the 6.85 $\mu$m feature. The plotted values are taken from \cite{boogert2011ice} (clouds) and \cite{boogert2008c2d} (protostars). The Spearman's rank correlation coefficient $\rho$ and the p-value indicating probability of noncorrelation are provided in the top right corner. The plotted NH$_{4}$$^{+}$ abundances have been multiplied by a correction factor of 4.4/3.6 to account for the difference between NH$_{4}$$^{+}$ $\nu_{4}$ mode band strength used by \cite{boogert2008c2d} and \cite{boogert2011ice} (4.4$\times$10$^{-17}$ cm molec$^{-1}$ from \citealt{schutte2003origin}) and the band strengths calculated in this work to enable a direct comparison to our NH$_{4}$$^{+}$ ice upper limits in Table~\ref{tab:nh4_cd}. Upper limits have been excluded from this plot, as well as data from the sources 2MASS J19214480+1121203 (due to the high error of its $\tau_{int}$C4/$\tau_{int}$C3 ratio) and IRAS 03301+3111 (due to its $\tau_{int}$C4/$\tau_{int}$C3 ratio being undefined because its $\tau_{int}$C3 = 0).}
\label{fig:nh4_c4c3_corr}
\end{figure*}

In Figure~\ref{fig:nh4_obs}, we fit the 6.85 $\mu$m feature in our selected observed spectra toward dense clouds and protostars with the NH$_{4}$$^{+}$ $\nu_{4}$ mode in our NH$_{4}$SH:H$_{2}$O laboratory spectra at 135 K and use these fits to derive NH$_{4}$$^{+}$ cation column densities and abundances with respect to H$_{2}$O ice (Table~\ref{tab:nh4_cd}). We use only the laboratory spectra at 135 K because this is the temperature at which the salt concentrations are expected to be the closest to the target salt concentrations, and our annealing experiment shows that recooling the salt mixture from 135 K to 15 K has little effect on the NH$_{4}$$^{+}$ $\nu_{4}$ mode profile and position compared to the changes in the feature due to increasing NH$_{4}$SH concentration that occur during the initial warm-up. As CH$_{3}$OH ice has a weak absorption complex of O-H and C-H bending modes at 6.85 $\mu$m as well, we account for the approximate expected contribution of CH$_{3}$OH ice to our fits by scaling a laboratory spectrum of pure CH$_{3}$OH ice at 15 K using literature CH$_{3}$OH ice column densities calculated from the 3.53 or 9.74 $\mu$m features (\citealt{boogert2008c2d,boogert2011ice,chen2024joys}; Slavicinska et al. in prep.).

The observed 6.85 $\mu$m features toward these sources are fit very well by the NH$_{4}$SH:H$_{2}$O laboratory data. Specifically, the investigated dense clouds and LYSOs are best fit with laboratory spectra of more dilute NH$_{4}$SH ices (NH$_{4}$SH/H$_{2}$O $\sim$ 20-33\%) compared to the MYSOs, which require laboratory spectra of more concentrated NH$_{4}$SH ices (NH$_{4}$SH/H$_{2}$O $\sim$ 50-100\%) to fit their more redshifted features. This is consistent with previously reported trends of MYSOs, particularly thermally processed ones, generally having more redshifted 6.85 $\mu$m features than LYSOs \citep{keane2001ice,boogert2008c2d}. Given these salt concentrations of the best-fitting laboratory spectra, we used the apparent band strength calculated from the lower concentration NH$_{4}$SH:H$_{2}$O mixtures (3.6$\times$10$^{-17}$ cm molec$^{-1}$) to calculate NH$_{4}$$^{+}$ column densities in the cloud and LYSO sources, while using the apparent band strength calculated from the higher concentration NH$_{4}$SH:H$_{2}$O mixtures (3.2$\times$10$^{-17}$ cm molec$^{-1}$) to calculate NH$_{4}$$^{+}$ column densities in the MYSO sources. Figure~\ref{fig:nh4_v4_labvobs} shows the 6.85 $\mu$m feature of one of the sources, L1527, overplotted with the NH$_{4}$$^{+}$ $\nu_{4}$ feature in five selected laboratory spectra with NH$_{4}$SH/H$_{2}$O concentrations ranging from 10-100\% to help the reader visualize how NH$_{4}$SH concentration with respect to H$_{2}$O (and excess NH$_{3}$) affects the fitting of this feature.

The resulting calculated NH$_{4}$$^{+}$ ice column densities presented in Table~\ref{tab:nh4_cd} range between 8-23\% with respect to H$_{2}$O, within the previously reported ranges of NH$_{4}$$^{+}$ ice abundances with respect to H$_{2}$O toward dense clouds and protostars \citep{boogert2011ice,boogert2015observations}. Considering the typical NH$_{3}$ ice abundances in such objects (3-10\% with respect to H$_{2}$O, \citealt{boogert2015observations}) results in combined NH$_{3}$+NH$_{4}$$^{+}$ abundances on the order of 10$^{-6}$-10$^{-5}$ with respect to H$_{2}$ (see Section~\ref{txt:implications_budget} for H column densities). Such values are in line with the typical gas-phase NH$_{3}$ abundances with respect to H$_{2}$ derived from small-beam radio/cm measurements targeting highly excited rotational lines in the most inner regions of hot cores, where temperatures are high enough that most NH$_{3}$ and NH$_{4}$$^{+}$ are expected to have desorbed into the gas phase \citep{pauls1983clumping,wilson1983non,mauersberger1986hot,henkel1987multilevel,brown1988model,walmsley1994ammonia}. However, the uncertainties of these gas NH$_{3}$ abundances are very high (up to an order of magnitude, due primarily to the poorly constrained warm H$_{2}$ column densities in their denominators), so the total abundance of NH$_{3}$ in dense star-forming regions remains an open question.

Our reported bulk NH$_{4}$$^{+}$ abundances with respect to H$_{2}$O are also approximately a factor of 2 lower than the NH$_{4}$$^{+}$ abundances of the laboratory ices whose spectra best fit the observed 6.85 $\mu$m features ($\geq$20\%). One explanation for such a discrepancy could be an addition to the current layered ice model described in \cite{boogert2015observations} and references therein. This model explains why, despite H$_{2}$O being the most abundantly detected icy species, the peak profiles of molecules like CO, CH$_{3}$OH, and OCS indicate that a significant fraction of these ices are in water-poor chemical environments: there is a difference in the timescale/cloud density at which H$_{2}$O formation and CO freeze-out occur, leading to distinct ice layers rather than a homogeneous ice mixture. Although H$_{2}$S is expected to form on ice grains at timescales closer to H$_{2}$O formation than CO freeze-out, models predict that H$_{2}$S formation should be slightly delayed (by $\sim$1.4 A$_{v}$) relative to H$_{2}$O formation due to S atoms' lower abundance and heavier mass with respect to O atoms \citep{goicoechea2021bottlenecks}. Such a delay could result in a gradient within the polar ice layer where the outer layer of polar ice that formed at later times in the dense cloud has a higher NH$_{4}$SH salt concentration than the inner layer of polar ice formed at earlier times. In such a scenario, high local salt concentrations in the salt-rich layer could still be consistent with overall lower bulk salt concentrations.

However, it is also possible that the discrepancies in NH$_{4}$$^{+}$ peak positions and calculated concentrations are caused by the presence of other species in the ice matrix (e.g., CO$_{2}$), overlap with other ice features in this region, or differences in temperature between the laboratory ices and the observed ices. Furthermore, the NH$_{4}$$^{+}$ abundances reported here are calculated only using laboratory data of NH$_{4}$SH, and although contributions from other ammonium salts like NH$_{4}$OCN, NH$_{4}$CN, NH$_{4}$HCOO, and NH$_{4}$Cl cannot dominate the 6.85 $\mu$m feature given their low abundances and upper limits with respect to the total NH$_{4}$$^{+}$ column densities (see Section~\ref{txt:anions}), they could still cause some redshift of the 6.85 $\mu$m feature if their NH$_{4}$$^{+}$ $\nu_{4}$ peak positions are redder than that of NH$_{4}$SH. A systematic characterization and comparison of multiple ammonium salts with carefully controlled concentrations in astrophysically relevant ice mixtures is needed to constrain and differentiate potential contributions from other ammonium salts to this feature.

\subsection{Redshift of the 6.85 $\mu$m feature}
\label{txt:redshift}

In the past, observational studies empirically decomposed the observed 6.85 $\mu$m feature into two components, a blue component centered at $\sim$6.75 $\mu$m and a red component centered at $\sim$6.95 $\mu$m (referred to respectively as ``component 1'' and ``component 2'' in \citealt{keane2001ice} and ``C3'' and ``C4'' in \citealt{boogert2008c2d,boogert2011ice,boogert2015observations}). A combination of both components was needed to fit the 6.85 $\mu$m feature toward most observed icy lines of sight, but sight-lines with unprocessed ices typically had low $\tau_{int}$C4/$\tau_{int}$C3 ratios (i.e., a very blueshifted 6.85 $\mu$m feature, \citealt{boogert2011ice}) while high $\tau_{int}$C4/$\tau_{int}$C3 ratios (i.e., a very redshifted 6.85 $\mu$m feature) were associated with thermally processed sources with low H$_{2}$O column densities (with respect to the depth of the 9.7 $\mu$m silicate band, \citealt{boogert2008c2d}). The association of redshift in the 6.85 $\mu$m feature with thermal processing was also supported by the previously available laboratory data of NH$_{4}$$^{+}$ salts in astrophysically relevant ice mixtures, in which only ices at very high temperatures ($>$200 K, \citealt{schutte2003origin}) could reproduce the redshifted peak position of the C4 component.

\begin{figure}[h!]
\centering
\includegraphics[width=\linewidth]{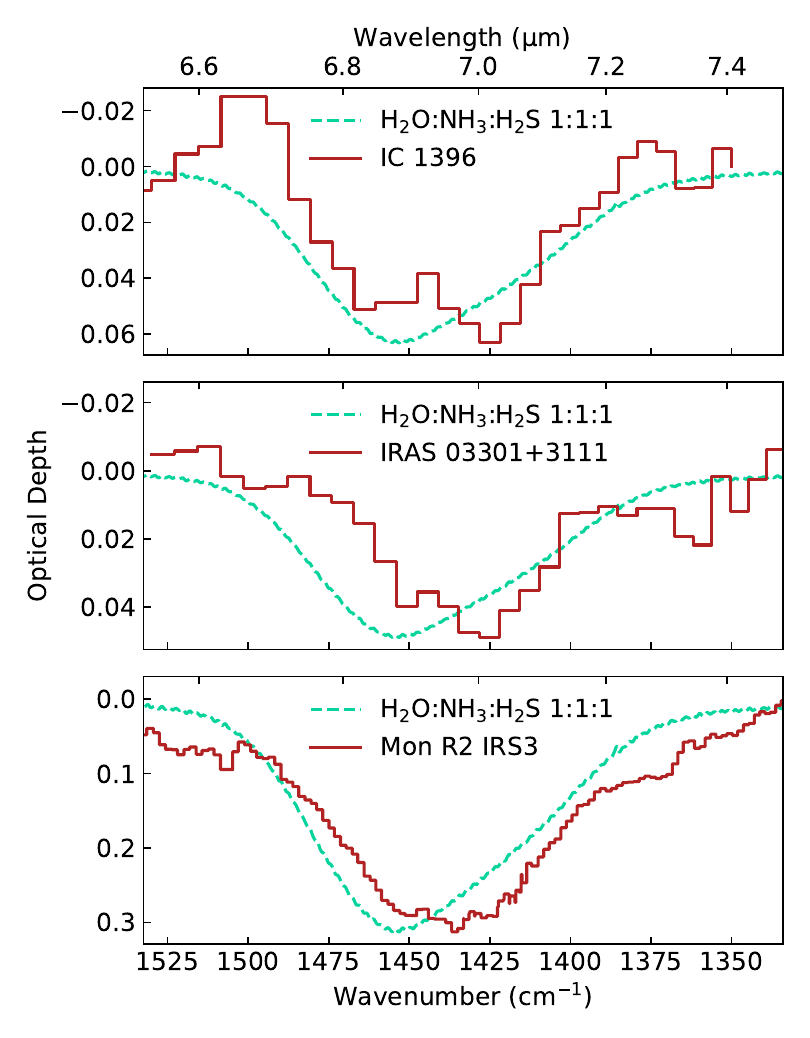}
\caption{The highly redshifted 6.85 $\mu$m band of three unusual protostellar sources (red solid traces) which are not sufficiently fit with any of our laboratory NH$_{4}$SH:H$_{2}$O mixtures. The laboratory spectrum with the greatest redshift out of our full lab sample, the H$_{2}$O:NH$_{3}$:H$_{2}$S 1:1:1 ice heated to 135 K, is overplotted in dotted sea green traces for comparison.}
\label{fig:redshifted_sources}
\end{figure}

Our experiments show clearly that the NH$_{4}$$^{+}$ $\nu_{4}$ mode redshifts with increasing salt concentration with respect to hydrogen-bonding matrix species. Moreover, the redshift observed when salt concentration with respect to H$_{2}$O increases from 10 to 50\% is approximately a factor of 3 larger (on the wavenumber scale) than the redshift observed when the ice temperature increases from 15 to 135 K in an annealed ice, where the salt concentration does not change upon warm-up (Figure~\ref{fig:annealing}).

Toward dense clouds without star formation, thermal processing cannot explain any variation in the peak position of the 6.85 $\mu$m feature. Thus, the small variations in the 6.85 $\mu$m feature peak position toward different prestellar sight-lines should be due purely to differences in the ice chemical composition. Indeed, we find that the published $\tau_{int}$C4/$\tau_{int}$C3 ratios of ices in the isolated dense clouds from \cite{boogert2011ice} correlate very strongly with the reported NH$_{4}$$^{+}$ abundances with respect to H$_{2}$O (Figure~\ref{fig:nh4_c4c3_corr}, left plot). The Spearman's rank correlation coefficient, which expresses the probability of two variables being related by a monotonically increasing function on a scale of 0 to 1, is very high at $\rho$=0.825. Such a correlation is consistent with the trend noted in our laboratory data. However, it is important to once again keep in mind that the redshift of the NH$_{4}$$^{+}$ $\nu_{4}$ feature with increasing concentration could also be degenerate with other chemical effects that are not investigated here (e.g., presence of CO$_{2}$ in the ice matrix, ionic bonds with other anions, etc.).

Toward protostars, the correlation between the NH$_{4}$$^{+}$ abundances and the peak position is more moderate with a Spearman's rank of $\rho$=0.485 (Figure~\ref{fig:nh4_c4c3_corr}, right plot). The redshift of the 6.85 $\mu$m band observed toward thermally processed protostellar sources could be due in part to the formation of more NH$_{4}$$^{+}$ ions from leftover unreacted neutrals driven by diffusion at warm temperatures, but it is also possible that increases in local concentrations of NH$_{4}$$^{+}$ ions caused by thermally-driven ice restructuring, such as that which occurs during the heating of CO$_{2}$:H$_{2}$O ice mixtures \citep{ehrenfreund1997infrared,boogert2000iso}, also contribute to the observed redshifts. Such structural changes would not change the overall line-of-sight averaged NH$_{4}$$^{+}$ abundance with respect to H$_{2}$O and would therefore result in a lesser correlation between the two variables.

Finally, comparisons of our laboratory spectra to unique, highly redshifted sources with extreme conditions reveal a third possible driver of the 6.85 $\mu$m feature's redshift in thermally processed sources. While our set of NH$_{4}$SH data is able to fit well the majority of the 6.85 $\mu$m features observed toward icy lines of sight, which have $\tau_{int}$C4/$\tau_{int}$C3 ratios below 3, a few exceptional sources exist whose 6.85 $\mu$m bands are too redshifted to be covered by even our most concentrated NH$_{4}$SH:H$_{2}$O $\sim$1:1 ice mixture (Figure~\ref{fig:redshifted_sources}). These sources are the LYSO IC 1396$\alpha$ \citep{reach2008properties}, the LYSO IRAS 03301+3111 \citep{boogert2008c2d}, and the MYSO Mon R2 IRS3 \citep{keane2001ice}. The 6.85 $\mu$m bands of these features can be fit almost exclusively with the empirical C4 component.

\begin{figure*}[ht!]
\centering
\includegraphics[width=0.9\linewidth]{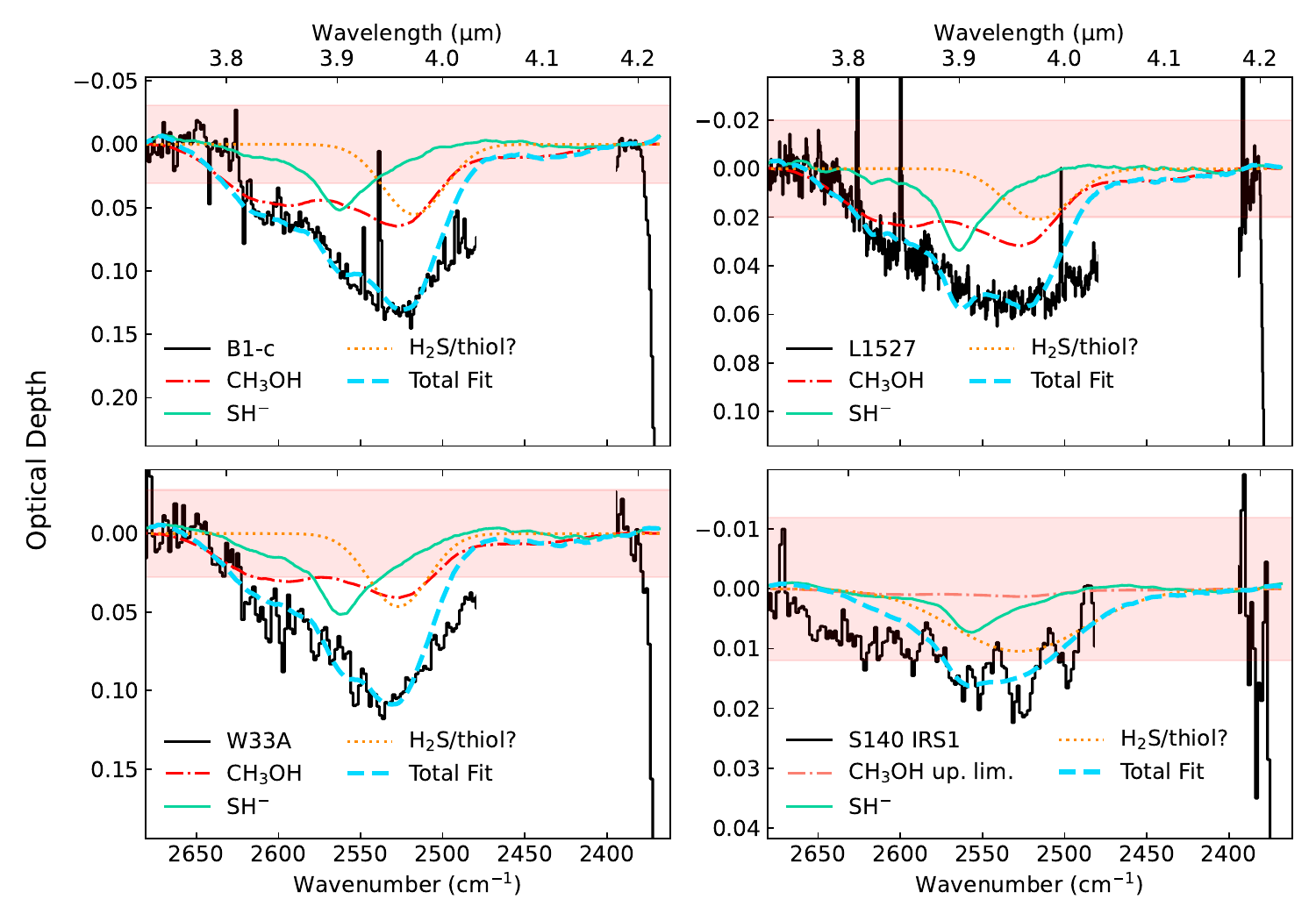}
\caption{Fits of the SH$^{-}$ stretching mode from laboratory NH$_{4}$SH:H$_{2}$O spectra (dark green solid trace) and the combination modes from laboratory CH$_{3}$OH spectra (dash-dotted red or pink trace) to the absorption complex between 3.8-4.2 $\mu$m observed toward four protostars. The fit SH$^{-}$ features are isolated from the laboratory spectra via a local continuum subtraction and scaled with the same factors as those used to fit the 6.85 $\mu$m bands in Figure~\ref{fig:nh4_obs}. The CH$_{3}$OH features are similarly scaled using literature CH$_{3}$OH ice column densities or upper limits calculated from the 3.53 or 9.74 $\mu$m features. As the SH$^{-}$ stretching mode and CH$_{3}$OH combination modes are some of the weakest spectral features of the investigated ices, both the NH$_{4}$SH and CH$_{3}$OH laboratory spectra are smoothed with a Savgol filter prior to fitting to remove experimental noise. Excess absorptions that is not fit with SH$^{-}$ and CH$_{3}$OH features are fit with Gaussians (dotted orange trace) that could be attributed to a combination of H$_{2}$S and thiol ices. The full fit is indicated with a dashed light green trace, and the horizontal shaded red bars indicate 3$\sigma$ levels calculated from RMS errors in the 3.72-3.77 $\mu$m range. Data from 4.03-4.18 $\mu$m (G395H gap in JWST data, low S/N region in ISO data) is excluded from the plot. Plots of the local continua applied to the observed spectra can be found in Appendix~\ref{app:obs_cont}.}
\label{fig:sh_obs}
\end{figure*}

IRAS 03301+3111 and Mon R2 IRS3 are unusual in that they have exceptionally low H$_{2}$O ice abundances with respect to H (a factor of 3-7 lower than typical protostellar sources like B1c, L1527, and W33A, \citealt{boogert2013infrared}), indicating that the thermal processing of their ice envelopes has been extensive enough to sublimate a significant fraction of their water ice. IC 1395$\alpha$ also has a low water ice column and additionally exists in a harsh, highly clustered and irradiated environment \citep{reach2008properties}.

We find in our experiments that NH$_{4}$SH salts have similar desorption temperatures as H$_{2}$O, so if these sources' envelopes have lost a significant fraction of their water ice to thermal desorption, we would expect them to have lost a significant fraction of their NH$_{4}$SH salts, if present, as well. This would imply that NH$_{4}$SH cannot be the sole carrier of the empirically derived C4 component (although it may still contribute to the component because some water ice remains in these lines of sight). A significant part of the C4 component must therefore spectrally consist of either a more refractory ammonium salt with a more redshifted 6.85 $\mu$m feature or an entirely different species.

The next most abundant ammonium salt after NH$_{4}$SH detected in the dust grains of comet 67P, NH$_{4}$CN, has laboratory desorption temperatures similar to NH$_{4}$SH \citep{gerakines2024sublimation}, so it is not a suitable C4 candidate. The third-, fourth-, and fifth-most abundant ammonium salts, NH$_{4}$Cl, NH$_{4}$OCN, and NH$_{4}$HCOO, are significantly more refractory with laboratory desorption temperatures above at least 200 K \citep{mate2009water,jimenez2014sulphur,ligterink2018formation,bergner2016kinetics,kruczkiewicz2021ammonia}. Perhaps one or a combination of these salts (or other refractory ammonium salts) are major contributors to the C4 component. The highly redshifted reported peak positions of NH$_{4}$Cl:H$_{2}$O ice formed via hyperquenching in \cite{galvez2010ammonium} and ammonium salts formed by photolysis experiments in \cite{schutte2003origin} point to a promising future direction for elucidating the carriers of these highly redshifted features.

\subsection{SH$^{-}$ stretching mode}
\label{txt:astro_sh}

Following the fits of the NH$_{4}$$^{+}$ $\nu_{4}$ mode to the 6.85 $\mu$m band, we aimed to investigate if the optical depths of the fit NH$_{4}$$^{+}$ features were consistent with SH$^{-}$ column densities or upper limits that could be calculated using the SH$^{-}$ stretching mode at 3.9 $\mu$m. We performed local continuum subtractions on the 3.7-4.2 $\mu$m region toward the sources in our sample whose spectra in this range are available to us (Figure~\ref{fig:sh_obs_cont}) and compared the subtracted spectra to the local continuum-subtracted SH$^{-}$ feature in the laboratory spectra, scaled with the same factor as was used to fit the NH$_{4}$$^{+}$ feature in each source in Figure~\ref{fig:nh4_obs}. Similarly to the 6.85 $\mu$m band fits, the expected contribution of CH$_{3}$OH ice to this spectral region was approximated by scaling a laboratory spectrum of CH$_{3}$OH ice at 15 K using literature CH$_{3}$OH ice column densities calculated from the 3.53 or 9.74 $\mu$m features (\citealt{boogert2008c2d,boogert2011ice,chen2024joys}; Slavicinska et al. in prep.).

\begin{figure*}[ht!]
\centering
\includegraphics[width=0.9\linewidth]{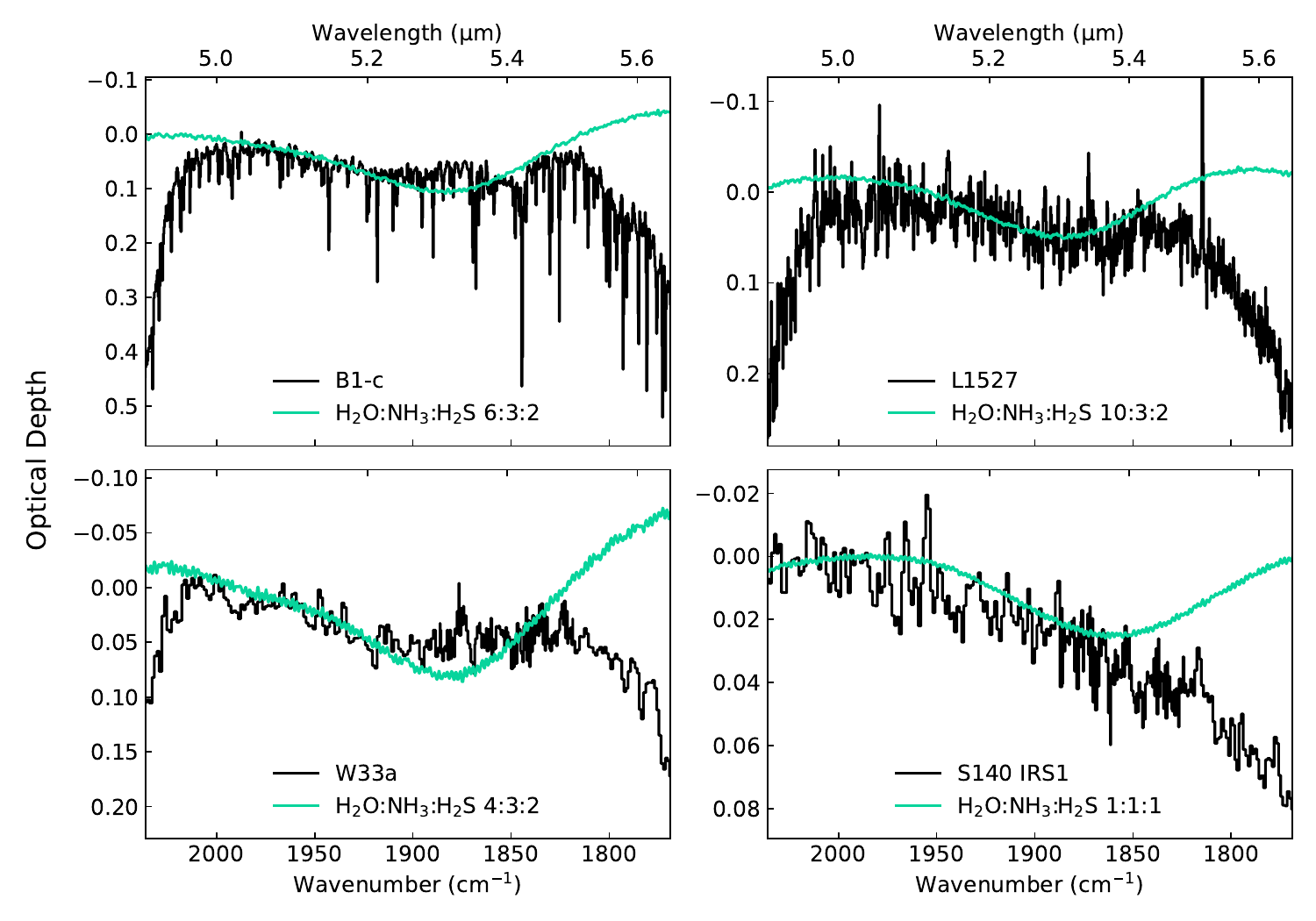}
\caption{Laboratory NH$_{4}$SH:H$_{2}$O data scaled with the same factor as that used to fit the 6.85 $\mu$m band in Figure~\ref{fig:nh4_obs} plotted with the global continuum-subtracted spectra of four protostars, zoomed in on the 4.8-5.6 $\mu$m region where a weak and broad absorption is observed. The profile of the observed absorption matches well the NH$_{4}$$^{+}$ $\nu_{4}$ + SH$^{-}$ libration combination mode in the laboratory spectra. Plots of the global continua applied to the observed spectra can be found in Appendix~\ref{app:obs_cont}.}
\label{fig:nh4_sh_comb_obs}
\end{figure*}

The excess absorptions left over toward each source after accounting for the CH$_{3}$OH and SH$^{-}$ ice features were fit with Gaussians, whose peak positions and FWHMs range from 2716-2728 cm$^{-1}$ and 20-40 cm$^{-1}$, respectively. Such profiles hint at the possible presence of thiol ices like methanethiol and ethanethiol toward these lines of sight, but secure assignment to a specific species is precluded by the broad, overlapping profiles of such molecules \citep{hudson2018infrared}. Furthermore, the fitting of a single Gaussian to this excess is merely a first-order approximation to what could be a complex of multiple Gaussian-like absorptions attributable to several different species containing S-H bonds. The upper limits obtained from assigning these Gaussians to either H$_{2}$S or thiols are on the order of $\lesssim$1\% and $\lesssim$5\% with respect to H$_{2}$O, respectively (using the band strengths of 1.66$\times$10$^{-17}$ cm molec$^{-1}$ for H$_{2}$S, \citealt{yarnall2022new}, and 3.40$\times$10$^{-18}$ cm molec$^{-1}$ for 1-propanethiol, \citealt{hudson2018infrared}). The underfit red shoulder of the absorption complex $>$4 $\mu$m may be the blue wing of an HDO ice absorption \citep{slavicinska2024jwst}.

The resulting fits are presented in Figure~\ref{fig:sh_obs}, where the green solid trace indicates the maximum possible SH$^{-}$ contribution (assuming [NH$_{4}$$^{+}$] = [SH$^{-}$]). Given the blending of multiple overlapping ice features in this spectral region as well as the extremely weak band strength of the SH$^{-}$ stretching mode, we cannot claim tentative or secure detection of the SH$^{-}$ anion in this spectral region toward these sources, and it is unlikely that this feature could serve as a reliable tracer of NH$_{4}$SH ice in other interstellar ice spectra. However, these fits by no means preclude the presence of SH$^{-}$ column densities comparable to the NH$_{4}$$^{+}$ column densities calculated from the 6.85 $\mu$m features of these sources. In fact, these fits show that sizable column densities of SH$^{-}$ anions can be masked very effectively by the absorptions of CH$_{3}$OH ice, H$_{2}$S or thiol ices, and observational noise at the wavelengths where the SH$^{-}$ stretching mode is found.

\subsection{NH$_{4}$$^{+}$ $\nu_{4}$ and SH$^{-}$ libration combination mode}

\begin{table*}[h!]
\caption{Estimated abundances and upper limits of four other counterion candidates, reported in percentages (\%) with respect to NH$_{4}$$^{+}$ column densities.}
\begin{center}
\begin{tabular}{c c c c c c c}
\hline
        Source & OCN$^{-}$$^{(a)}$ & CN$^{-}$$^{(b)}$ & HCOO$^{-}$$^{(c)}$ & Cl$^{-}$$^{(d)}$ & \multicolumn{2}{c}{Total anion budget} \\
        & & & & & w/o Cl$^{-}$ & incl. Cl$^{-}$ \\
        \hline
        B1c & 6.7 (1.3) & $\leq$0.9 & $\leq$6.4 & $\leq$8.5 & $\leq$14 & $\leq$22 \\
        L1527 & 4.5 (0.9) & $\leq$0.1 & $\leq$2.9 & $\leq$8.2 & $\leq$7.5 & $\leq$16 \\
        W33A & 10.8 (2.2) & n.d. & $\leq$2.4 & $\leq$8.1 & $\leq$13 & $\leq$21 \\
    \hline
     
\end{tabular}
\label{tab:anions}

\begin{tablenotes}
Uncertainties are indicated in parentheses ().\\
$^{(a)}$ calculated using a band strength of 1.3$\times$10$^{-16}$ cm molec$^{-1}$ \citep{van2004quantitative}.\\
$^{(b)}$ calculated using a band strength of 5.74$\times$10$^{-18}$ cm molec$^{-1}$ \citep{gerakines2024sublimation}.\\
$^{(c)}$ calculated using a band strength of 2.1$\times$10$^{-17}$ cm molec$^{-1}$ \citep{bergner2016kinetics}.\\
$^{(d)}$ calculated using the diffuse ISM Cl/H of 2.88$\times$10$^{-7}$ \citep{esteban2004reappraisal}.\\
\end{tablenotes}

\end{center}
\end{table*}

\cite{boogert2022survey} first reported the presence of a weak, broad absorption centered at approximately 5.25 $\mu$m in the IRTF spectra of W33A and a ``number'' of other MYSOs. This feature has not been assigned to any species and, prior to JWST, was not reported toward any dense clouds or LYSOs, likely due to the limited spectral coverage of the full width of this band by both ground-based NIR data and \textit{Spitzer} Space Telescope MIR data. Here, we report the presence of a similar feature centered around $\sim$5.3 $\mu$m/1890 cm$^{-1}$ with a FWHM of $\sim$0.26 $\mu$m/92 cm$^{-1}$ toward both of the LYSOs observed with JWST in our sample of sources and tentatively assign it to the combination mode of the NH$_{4}$$^{+}$ $\nu_{4}$ mode and the SH$^{-}$ libration. The tentative assignment has two-fold significance: it reinforces the presence of NH$_{4}$$^{+}$ in interstellar ices, and it is a tentative detection of a SH$^{-}$ signature in a spectral region that is less busy than the 3.9 $\mu$m region where the SH$^{-}$ stretching mode is found. 
The only other major ice bands present in this region are the OCS feature at 4.9 $\mu$m and the 6.0 µm band, whose wings overlap slightly with the blue and red edges, respectively, of the very broad 5.3 µm band.

In Figure~\ref{fig:nh4_sh_comb_obs}, the observed feature is fit with the same NH$_{4}$SH laboratory spectra and scaling factors as those used to fit the 6.85 $\mu$m band to each source in Figure~\ref{fig:nh4_obs}. The profile of the features in the laboratory spectra match the profile of the observed features quite well, and the observed depths are similar to the depths of the laboratory features when they are scaled with the same factor that is used to fit the 6.85 $\mu$m band. However, it must be stressed that, due to the weak and broad profile of the feature as well as its overlap with the blue wing of the large 6.0 $\mu$m ice feature, it is difficult to accurately extract its profile and depth from the source continuum, and even slight inflections in the continuum choice can significantly affect its calculated depth. Therefore, we do not consider this feature to be a reliable means of quantifying the SH$^{-}$ anion column density. Nevertheless, we conclude that the observed depth of this feature is not inconsistent with the observed depth of the NH$_{4}$$^{+}$ band, and its presence shows that SH$^{-}$ is an excellent candidate anion to counterbalance the majority of the NH$_{4}$$^{+}$ cations quantified via the 6.85 $\mu$m band.

\subsection{Other anions (OCN$^{-}$, CN$^{-}$, HCOO$^{-}$, Cl$^{-}$)}
\label{txt:anions}

After the SH$^{-}$ fitting attempt at 3.9 $\mu$m, we estimated abundances and upper limits of four other NH$_{4}$$^{+}$ counterion candidates identified in prior studies of interstellar ices and 67P dust grains to constrain how much NH$_{4}$$^{+}$ with unbalanced positive charge remains after accounting for these counterions, leaving room for other anions like SH$^{-}$.

Table~\ref{tab:anions} presents the estimated abundances of OCN$^{-}$ and the upper limits of CN$^{-}$, HCOO$^{-}$, and Cl$^{-}$ in percentages with respect to the NH$_{4}$$^{+}$ column densities (Table~\ref{tab:nh4_cd}) toward three of the four selected protostellar sources. The spectrum of S 140 IRS1 was not included in this analysis due to its relatively low S/N that prevented the detection of any known anionic features.

The OCN$^{-}$, CN$^{-}$, and HCOO$^{-}$ values were calculated via fitting their IR features at $\sim$4.62 $\mu$m/2165 cm$^{-1}$, $\sim$4.78 $\mu$m/2093 cm$^{-1}$, and $\sim$7.39 $\mu$m/1352 cm$^{-1}$ with Gaussians (Figure~\ref{fig:anions}). The calculated column densities of CN$^{-}$ and HCOO$^{-}$ are only considered upper limits due to the possible contribution of other ice species to their observed bands (e.g., $^{13}$CO for CN$^{-}$ \citep{mcclure2023ice,brunken2024joys}; CH$_{3}$CHO for HCOO$^{-}$, \citealt{rocha2024jwst,chen2024joys}). Because the column density of Cl$^{-}$ cannot be similarly constrained from IR spectra, its upper limit was constrained using the cosmic standard abundance of chlorine (i.e., assuming all Cl atoms are in the form of NH$_{4}$Cl toward these objects using Cl/H = 2.88$\times$10$^{-7}$ from \citealt{esteban2004reappraisal}).

The alternative anion candidates can only counterbalance up to $\sim$15-20\% of the ammonium cations toward these lines of sight. This upper limit is similar to the total combined abundance of the neutral acid counterparts of these anions with respect to NH$_{3}$ quantified from the 67P dust grain impact event \#4 in \cite{altwegg2022abundant}.

Therefore, a minimum of $\sim$80-85\% of the ammonium cations observed toward these lines of sight must be counterbalanced by other anions. It has been previously suggested that this counterbalance could come from a ``variety'' of multiple other anions, perhaps formed by energetic processing, whose weak and blended spectral signatures could create an undetectable ``pseudo-continuum'' \citep{schutte2003origin}. Here, we show that SH$^{-}$ column densities as high as the total measured NH$_{4}$$^{+}$ column densities can be effectively hidden in the 3.9 $\mu$m spectral region via blending with CH$_{3}$OH, H$_{2}$S, and thiol features (Section~\ref{txt:astro_sh}). Furthermore, SH$^{-}$ can form purely via low-temperature acid-base chemistry without needing to invoke energetic processing (Section~\ref{txt:results_nh4sh_h2o}).

\section{Astronomical implications}
\subsection{Sulfur budgets}
\label{txt:implications_budget}

\begin{table}[h!]
\caption{Estimated contribution of NH$_{4}$SH salt to the total S budgets of each of the investigated sources, assuming 80-85\% of the NH$_{4}$$^{+}$ cations quantified via the 6.85 $\mu$m band are bound to SH$^{-}$ anions.}
\begin{center}
\begin{tabular}{c c c}
\hline
        Source & N$_{\rm{H}}$ & S in NH$_{4}$SH \\
        & 10$^{23}$ cm$^{-2}$ & \% S cosmic std.$^{(a)}$ \\
        \hline
        Elias 16 & 0.89 (0.27)$^{(b)}$ & 14-15 \\
        CK2 & 1.9 (0.6)$^{(b)}$ & 14-15 \\
        B1c & 8.3 (2.5)$^{(b)}$ & 17-18 \\
        L1527 & 5.9 (1.8)$^{(c)}$ & 17-18 \\
        W33A & 7.1 (2.1)$^{(b)}$ & 17-18 \\
        S140 IRS1 & 2.2 (0.7)$^{(b)}$ & 10 \\
    \hline
     
\end{tabular}
\label{tab:s_budget}

\begin{tablenotes}
Uncertainties are indicated in parentheses ().\\
$^{(a)}$ relative to the diffuse ISM S/H of 1.66$\times$10$^{-5}$ from \cite{esteban2004reappraisal}.\\
$^{(b)}$ from \cite{boogert2013infrared}.\\
$^{(c)}$ from the 9.7 $\mu$m silicate feature; see Appendix~\ref{app:obs_nh}.\\
\end{tablenotes}

\end{center}
\end{table}

Estimates of the total sulfur budgets that could be accounted for by solid NH$_{4}$SH toward all of the investigated sources are provided in Table~\ref{tab:s_budget}. The ranges in this table are calculated with the assumption that 80-85\% of the NH$_{4}$$^{+}$ cations quantified via the 6.85 $\mu$m band (see Table~\ref{tab:nh4_cd}) are bound to SH$^{-}$ anions (as the relative abundance of H$_{2}$S with respect to the combined abundances of all acids in the comet 67P dust grain impact event \#4 was $\sim$82\% in \citealt{altwegg2022abundant}, and the total NH$_{4}$$^{+}$ column density left without corresponding counterions after accounting for OCN$^{-}$, CN$^{-}$, HCOO$^{-}$, and Cl$^{-}$ abundances and upper limits is $\gtrsim$80-85\%). The hydrogen column densities (N$_{\rm{H}}$) used to determine S/H were taken from \cite{boogert2013infrared} except in the case of L1527, whose N$_{\rm{H}}$ was similarly calculated from the optical depth of the 9.7 $\mu$m silicate feature in the L1527 MIRI spectrum (see Appendix~\ref{app:obs_nh}).

Our observationally constrained estimates indicate that 10-18\% of the sulfur expected toward the investigated lines of sight could be in the form of NH$_{4}$SH ice. The estimated sulfur abundance in NH$_{4}$SH is lowest toward S140 IRS1, the source whose ices have experienced the most thermal processing and which also has a very low H$_{2}$O ice abundance with respect to H \citep{boogert2013infrared}. Therefore, its envelope may have already lost a significant portion of its ices, including NH$_{4}$SH, to thermal desorption. In the case of the rest of the protostars in our sample, almost a fifth of the expected sulfur column may be locked in NH$_{4}$SH salt, much more than the previously constrained $\lesssim$5\% S locked in ices calculated from OCS column densities and H$_{2}$S, thiol, and SO$_{2}$ upper limits \citep{palumbo1997solid,boogert1997infrared,jimenez2011sulfur,boogert2015observations,boogert2022survey,mcclure2023ice,rocha2024jwst}.

These estimates demonstrate that NH$_{4}$SH is a potentially significant sulfur sink in pre- and protostellar environments. They also show that NH$_{4}$SH alone cannot solve the sulfur problem, leaving the field open to other potential sulfur sinks whose abundances in dense star-forming regions are still not spectroscopically well-constrained (e.g., sulfur allotropes and minerals). Notably, the $\sim$82-83\% of the sulfur that remains missing after accounting for the plausible abundance of NH$_{4}$SH is close to the proposed S/H abundance of highly refractory sulfur-bearing species (e.g., FeS) in protoplanetary disks quantified indirectly via the accretion-contaminated photospheres of young stars (89$\pm$8\%, \citealt{kama2019abundant}).

\subsection{NH$_{4}$SH ice desorption front}
\label{txt:implications_desorption}

In protostellar envelopes and protoplanetary disks, molecules in ices warm up via heating by the growing central source and, at high enough temperatures, thermally desorb into the gas phase, where they can be detected via their rotational transitions by millimeter and submillimeter spectroscopy. The temperature at which desorption of a given ice molecule occurs is dictated by its binding energy to its surrounding molecules (i.e., other ices or grain materials like silicates). A molecule's thermal desorption can also occur at lower or higher temperatures if it is deeply trapped in a matrix of molecules that desorb at lower or higher temperatures.

Although we do not derive binding energies of NH$_{4}$SH in this work, the TPD experiments here indicate that NH$_{4}$SH desorbs at nearly the same temperature as H$_{2}$O, both when it is pure and when it is in H$_{2}$O-rich mixtures. As ionic salts desorb as their neutral counterparts, this means that any gas-phase NH$_{3}$ or H$_{2}$S that has thermally desorbed from NH$_{4}$SH ices should have a sublimation front very similar to H$_{2}$O. While pure neutral NH$_{3}$ and H$_{2}$S ices desorb at much lower temperatures ($\sim$110 K and $\sim$90 K, respectively, in our experiments), neutral NH$_{3}$ and H$_{2}$S in interstellar ices may have sublimation fronts at temperatures closer to the sublimation front of H$_{2}$O ($\sim$170 K in our experiments) due to entrapment by a surrounding H$_{2}$O ice matrix. Whether the sublimation fronts of neutral NH$_{3}$ and H$_{2}$S mixed in H$_{2}$O ices would be different enough to distinguish them from NH$_{3}$ and H$_{2}$S originating from thermally desorbed NH$_{4}$SH salts must be determined with studies measuring the entrapment of NH$_{3}$ and H$_{2}$S in H$_{2}$O ices.

\section{Conclusions}

In this work, we present a systematic characterization of the laboratory IR spectra of NH$_{4}$SH ices, both pure and mixed with H$_{2}$O. From this characterization and a subsequent comparison of the laboratory spectra to spectra observed toward dense clouds and protostellar ice envelopes, we draw the following conclusions:

\begin{enumerate}

\item NH$_{4}$SH salts form via the in situ reaction of NH$_{3}$ and H$_{2}$S ices already at 15 K even when diluted in H$_{2}$O ice in mixing ratios relevant to interstellar ices.

\item The apparent band strength of the NH$_{4}$$^{+}$ $\nu_{4}$ feature ($\sim$6.8 $\mu$m/1470 cm$^{-1}$) in 10-50\% NH$_{4}$SH:H$_{2}$O mixtures ranges from 3.2($\pm$0.3)-3.6($\pm$0.4)$\times$10$^{-17}$, which does not preclude the detection of ammonium cations at similar concentrations with respect to H$_{2}$O in interstellar ice spectra. These apparent band strengths are a factor of $\sim$1.22-1.38 lower than the NH$_{4}$$^{+}$ $\nu_{4}$ band strength in NH$_{4}$HCOO:H$_{2}$O ice from \cite{schutte2003origin} that is frequently used to quantify NH$_{4}$$^{+}$ ice column densities in interstellar ice spectra.

\item The apparent band strength of the SH$^{-}$ stretching mode ($\sim$3.9 $\mu$m/2560 cm$^{-1}$) in 10-50\% NH$_{4}$SH:H$_{2}$O mixturesranges from 3.1($\pm$0.4)-3.7($\pm$0.5)$\times$10$^{-19}$ cm molec$^{-1}$. This is nearly two orders of magnitude lower than the apparent band strength of the H$_{2}$S stretching mode in the same spectral region. Therefore, it is significantly more challenging to detect H$_{2}$S if it is anionic.

\item The peak profile of the NH$_{4}$$^{+}$ $\nu_{4}$ feature in the laboratory NH$_{4}$SH:H$_{2}$O spectra matches well with the profile of the observed 6.85 $\mu$m feature in the pre- and protostellar sources investigated here. From these fits, we derive NH$_{4}$$^{+}$ column densities ranging from $\sim$8-23\% with respect to H$_{2}$O.

\item The NH$_{4}$$^{+}$ $\nu_{4}$ feature in laboratory NH$_{4}$SH ices redshifts not only when heated, but also when the salt's concentration increases with respect to hydrogen-bonding matrix species like H$_{2}$O and NH$_{3}$. This could explain why the ammonium ion abundance with respect to H$_{2}$O is very strongly correlated with the peak position of the 6.85 $\mu$m feature toward cold dense clouds and moderately correlated toward protostars. The redshifts in the 6.85 $\mu$m feature toward thermally processed protostars could be driven by multiple physicochemical effects, such as an increase in the local concentration of ammonium salts with respect to H$_{2}$O caused by the reaction of leftover neutrals in the ice promoted by diffusion or thermally-driven ice restructuring. In ices found toward protostars with extreme environments and exceptionally redshifted features, the thermal desorption of NH$_{4}$SH with water ice may also contribute to the observed redshift.

\item While the extremely low apparent band strength of the SH$^{-}$ stretching mode makes it very difficult to reliably detect in the spectrally busy region at 3.9 $\mu$m, the weak and broad NH$_{4}$$^{+}$ $\nu_{4}$ + SH$^{-}$ libration combination mode at 5.3 $\mu$m is an alternative means of SH$^{-}$ detection due to a lack of other major ice features at this wavelength. Here, we tentatively assign the weak and broad feature observed at 5.3 $\mu$m in the JWST spectra of our two investigated LYSOs to this combination band of NH$_{4}$SH.

\item The maximum possible combined contributions of the salts NH$_{4}$OCN, NH$_{4}$CN, NH$_{4}$HCOO, and NH$_{4}$Cl to the total ammonium salt abundance range between 15-20\% toward the three investigated protostars with clearly detected 4.62 and 7.39 $\mu$m features.

\item If the SH$^{-}$ anion is indeed the primary counterion to the observed column density of NH$_{4}$$^{+}$ cations in these spectra (as it is in the dust grains of comet 67P), then NH$_{4}$SH could account for up to 17-18\% of the S budget toward our investigated sources, making it a potentially significant sulfur sink in dense star-forming regions.

\end{enumerate}

\section{Data availability}

The appendix of this paper can be found on Zenodo at \href{https://doi.org/10.5281/zenodo.13864529}{DOI 10.5281/zenodo.13864529}.

\begin{acknowledgements}
This work is based on observations made with the NASA/ESA/CSA \textit{James Webb} Space Telescope. The data were obtained from the Mikulski Archive for Space Telescopes at the Space Telescope Science Institute, which is operated by the Association of Universities for Research in Astronomy, Inc., under NASA contract NAS 5-03127 for JWST. These observations are associated with programs \#1290 and \#1960. All the JWST data used in this paper can be found in MAST: \href{https://archive.stsci.edu/doi/resolve/resolve.html?doi=10.17909/3kky-t040}{10.17909/3kky-t040}. The following National and International Funding Agencies funded and supported the MIRI development: NASA; ESA; Belgian Science Policy Office (BELSPO); Centre Nationale d’Etudes Spatiales (CNES); Danish National Space Centre; Deutsches Zentrum fur Luft- und Raumfahrt (DLR); Enterprise Ireland; Ministerio De Economiá y Competividad; Netherlands Research School for Astronomy (NOVA); Netherlands Organisation for Scientific Research (NWO); Science and Technology Facilities Council; Swiss Space Office; Swedish National Space Agency; and UK Space Agency. Astrochemistry at Leiden is supported by funding from the European Research Council (ERC) under the European Union’s Horizon 2020 research and innovation programme (grant agreement No. 101019751 MOLDISK), the Netherlands Research School for Astronomy (NOVA), and the Danish National Research Foundation through the Center of Excellence “InterCat” (Grant agreement no.: DNRF150). We thank an anomymous reviewer for their constructive and insightful comments that we believe improved this paper.
\end{acknowledgements}


\bibliographystyle{aa}
\bibliography{biblio}

\begin{thebibliography}{101}
\expandafter\ifx\csname natexlab\endcsname\relax\def\natexlab#1{#1}\fi

\bibitem[{Altwegg {et~al.}(2019)Altwegg, Balsiger, \& Fuselier}]{altwegg2019cometary}
Altwegg, K., Balsiger, H., \& Fuselier, S.~A. 2019, ARA\&A, 57, 113

\bibitem[{Altwegg {et~al.}(2022)Altwegg, Combi, Fuselier, H{\"a}nni, De~Keyser, Mahjoub, M{\"u}ller, Pestoni, Rubin, \& Wampfler}]{altwegg2022abundant}
Altwegg, K., Combi, M., Fuselier, S., {et~al.} 2022, MNRAS, 516, 3900

\bibitem[{Argyriou {et~al.}(2023)Argyriou, Glasse, Law, Labiano, {\'A}lvarez-M{\'a}rquez, Patapis, Kavanagh, Gasman, Mueller, Larson, {et~al.}}]{argyriou2023jwst}
Argyriou, I., Glasse, A., Law, D.~R., {et~al.} 2023, A\&A, 675, A111

\bibitem[{Asplund {et~al.}(2009)Asplund, Grevesse, Sauval, \& Scott}]{asplund2009chemical}
Asplund, M., Grevesse, N., Sauval, A.~J., \& Scott, P. 2009, ARA\&A, 47, 481

\bibitem[{Bergner {et~al.}(2016)Bergner, {\"O}berg, Rajappan, \& Fayolle}]{bergner2016kinetics}
Bergner, J.~B., {\"O}berg, K.~I., Rajappan, M., \& Fayolle, E.~C. 2016, ApJ, 829, 85

\bibitem[{Bockel{\'e}e-Morvan {et~al.}(2000)Bockel{\'e}e-Morvan, Lis, Wink, Despois, Crovisier, Bachiller, Benford, Biver, Colom, Davies, {et~al.}}]{bockelee2000new}
Bockel{\'e}e-Morvan, D., Lis, D., Wink, J., {et~al.} 2000, A\&A, 353, 1101

\bibitem[{Bohlin {et~al.}(1978)Bohlin, Savage, \& Drake}]{bohlin1978survey}
Bohlin, R.~C., Savage, B.~D., \& Drake, J. 1978, ApJ, 224, 132

\bibitem[{B{\"o}ker {et~al.}(2022)B{\"o}ker, Arribas, L{\"u}tzgendorf, de~Oliveira, Beck, Birkmann, Bunker, Charlot, De~Marchi, Ferruit, {et~al.}}]{boker2022near}
B{\"o}ker, T., Arribas, S., L{\"u}tzgendorf, N., {et~al.} 2022, A\&A, 661, A82

\bibitem[{Boogert {et~al.}(2022)Boogert, Brewer, Brittain, \& Emerson}]{boogert2022survey}
Boogert, A., Brewer, K., Brittain, A., \& Emerson, K. 2022, ApJ, 941, 32

\bibitem[{Boogert {et~al.}(2013)Boogert, Chiar, Knez, {\"O}berg, Mundy, Pendleton, Tielens, \& Van~Dishoeck}]{boogert2013infrared}
Boogert, A., Chiar, J., Knez, C., {et~al.} 2013, ApJ, 777, 73

\bibitem[{Boogert {et~al.}(2000)Boogert, Ehrenfreund, Gerakines, Tielens, Whittet, Schutte, Van~Dishoeck, de~Graauw, Decin, \& Prusti}]{boogert2000iso}
Boogert, A., Ehrenfreund, P., Gerakines, P.~A., {et~al.} 2000, A\&A, 353, 349

\bibitem[{Boogert {et~al.}(2011)Boogert, Huard, Cook, Chiar, Knez, Decin, Blake, Tielens, \& Van~Dishoeck}]{boogert2011ice}
Boogert, A., Huard, T., Cook, A., {et~al.} 2011, ApJ, 729, 92

\bibitem[{Boogert {et~al.}(1997)Boogert, Schutte, Helmich, Tielens, \& Wooden}]{boogert1997infrared}
Boogert, A., Schutte, W., Helmich, F., Tielens, A., \& Wooden, D. 1997, A\&A, 317, 929

\bibitem[{Boogert {et~al.}(2015)Boogert, Gerakines, \& Whittet}]{boogert2015observations}
Boogert, A.~A., Gerakines, P.~A., \& Whittet, D.~C. 2015, ARA\&A, 53, 541

\bibitem[{Boogert {et~al.}(2008)Boogert, Pontoppidan, Knez, Lahuis, Kessler-Silacci, van Dishoeck, Blake, Augereau, Bisschop, Bottinelli, {et~al.}}]{boogert2008c2d}
Boogert, A.~C., Pontoppidan, K.~M., Knez, C., {et~al.} 2008, ApJ, 678, 985

\bibitem[{Bottinelli {et~al.}(2010)Bottinelli, Boogert, Bouwman, Beckwith, Van~Dishoeck, {\"O}berg, Pontoppidan, Linnartz, Blake, Evans, {et~al.}}]{bottinelli2010c2d}
Bottinelli, S., Boogert, A.~A., Bouwman, J., {et~al.} 2010, ApJ, 718, 1100

\bibitem[{Bouilloud {et~al.}(2015)Bouilloud, Fray, B{\'e}nilan, Cottin, Gazeau, \& Jolly}]{bouilloud2015bibliographic}
Bouilloud, M., Fray, N., B{\'e}nilan, Y., {et~al.} 2015, MNRAS, 451, 2145

\bibitem[{Bragin {et~al.}(1977)Bragin, Diem, Guthals, \& Chang}]{bragin1977vibrational}
Bragin, J., Diem, M., Guthals, D., \& Chang, S. 1977, J. Chem. Phys., 67, 1247

\bibitem[{Brown {et~al.}(1988)Brown, Charnley, \& Millar}]{brown1988model}
Brown, P., Charnley, S., \& Millar, T. 1988, MNRAS, 231, 409

\bibitem[{Brunken {et~al.}(2024)Brunken, van Dishoeck, Slavicinska, le~Gouellec, Rocha, Francis, Tychoniec, van Gelder, Navarro, Boogert, Kavanagh, Nazari, Greene, Ressler, \& Majumdar}]{brunken2024joys}
Brunken, N., van Dishoeck, E., Slavicinska, K., {et~al.} 2024, arXiv preprint arXiv:2409.17237

\bibitem[{Caselli \& Ceccarelli(2012)}]{caselli2012our}
Caselli, P. \& Ceccarelli, C. 2012, A\&AR, 20, 1

\bibitem[{Chen {et~al.}(2024)Chen, Rocha, van Dishoeck, van Gelder, Nazari, Slavicinska, Francis, Tabone, Ressler, Beuther, Klaassen, Kavanagh, Perotti, Le~Gouellec, Majumdar, G\"udel, \& Henning}]{chen2024joys}
Chen, Y., Rocha, W. R.~M., van Dishoeck, E.~F., {et~al.} 2024, A\&A

\bibitem[{Chen {et~al.}(2014)Chen, Juang, Nuevo, Jim{\'e}nez-Escobar, Caro, Qiu, Chu, Yih, Wu, Fung, {et~al.}}]{chen2014formation}
Chen, Y.-J., Juang, K.-J., Nuevo, M., {et~al.} 2014, ApJ, 798, 80

\bibitem[{Dartois \& d'Hendecourt(2001)}]{dartois2001search}
Dartois, E. \& d'Hendecourt, L. 2001, A\&A, 365, 144

\bibitem[{Drozdovskaya {et~al.}(2019)Drozdovskaya, van Dishoeck, Rubin, J{\o}rgensen, \& Altwegg}]{drozdovskaya2019ingredients}
Drozdovskaya, M.~N., van Dishoeck, E.~F., Rubin, M., J{\o}rgensen, J.~K., \& Altwegg, K. 2019, MNRAS, 490, 50

\bibitem[{Druard \& Wakelam(2012)}]{druard2012polysulphanes}
Druard, C. \& Wakelam, V. 2012, MNRAS, 426, 354

\bibitem[{Ehrenfreund {et~al.}(1997)Ehrenfreund, Boogert, Gerakines, Tielens, \& Van~Dishoeck}]{ehrenfreund1997infrared}
Ehrenfreund, P., Boogert, A., Gerakines, P., Tielens, A., \& Van~Dishoeck, E. 1997, A\&A, 328, 649

\bibitem[{Esteban {et~al.}(2004)Esteban, Peimbert, Garc{\'\i}a-Rojas, Ruiz, Peimbert, \& Rodr{\'\i}guez}]{esteban2004reappraisal}
Esteban, C., Peimbert, M., Garc{\'\i}a-Rojas, J., {et~al.} 2004, MNRAS, 355, 229

\bibitem[{Ferraro {et~al.}(1980)Ferraro, Sill, \& Fink}]{ferraro1980infrared}
Ferraro, J.~R., Sill, G., \& Fink, U. 1980, Appl. Spectrosc., 34, 525

\bibitem[{Furuya {et~al.}(2022)Furuya, Oba, \& Shimonishi}]{furuya2022quantifying}
Furuya, K., Oba, Y., \& Shimonishi, T. 2022, ApJ, 926, 171

\bibitem[{Galvez {et~al.}(2010)Galvez, Mat{\'e}, Herrero, \& Escribano}]{galvez2010ammonium}
Galvez, O., Mat{\'e}, B., Herrero, V.~J., \& Escribano, R. 2010, ApJ, 724, 539

\bibitem[{Gardner {et~al.}(2023)Gardner, Mather, Abbott, Abell, Abernathy, Abney, Abraham, Abraham, Abul-Huda, Acton, {et~al.}}]{gardner2023james}
Gardner, J.~P., Mather, J.~C., Abbott, R., {et~al.} 2023, PASP, 135, 068001

\bibitem[{Garozzo {et~al.}(2010)Garozzo, Fulvio, Kanuchova, Palumbo, \& Strazzulla}]{garozzo2010fate}
Garozzo, M., Fulvio, D., Kanuchova, Z., Palumbo, M., \& Strazzulla, G. 2010, A\&A, 509, A67

\bibitem[{Garrod {et~al.}(2007)Garrod, Wakelam, \& Herbst}]{garrod2007non}
Garrod, R.~T., Wakelam, V., \& Herbst, E. 2007, A\&A, 467, 1103

\bibitem[{Gerakines {et~al.}(2004)Gerakines, Moore, \& Hudson}]{gerakines2004ultraviolet}
Gerakines, P., Moore, M., \& Hudson, R. 2004, Icarus, 170, 202

\bibitem[{Gerakines {et~al.}(2024)Gerakines, Yarnall, \& Hudson}]{gerakines2024sublimation}
Gerakines, P.~A., Yarnall, Y.~Y., \& Hudson, R.~L. 2024, Icarus, 116007

\bibitem[{Gibb {et~al.}(2004)Gibb, Whittet, Boogert, \& Tielens}]{gibb2004interstellar}
Gibb, E., Whittet, D., Boogert, A., \& Tielens, A. 2004, ApJ supplement series, 151, 35

\bibitem[{Goicoechea {et~al.}(2021)Goicoechea, Aguado, Cuadrado, Roncero, Pety, Bron, Fuente, Riquelme, Chapillon, Herrera, {et~al.}}]{goicoechea2021bottlenecks}
Goicoechea, J.~R., Aguado, A., Cuadrado, S., {et~al.} 2021, A\&A, 647, A10

\bibitem[{Grim {et~al.}(1989{\natexlab{a}})Grim, Greenberg, De~Groot, Baas, Schutte, \& Schmitt}]{grim1989infrared}
Grim, R., Greenberg, J., De~Groot, M., {et~al.} 1989{\natexlab{a}}, A\&AS, 78, 161

\bibitem[{Grim {et~al.}(1989{\natexlab{b}})Grim, Greenberg, Schutte, \& Schmitt}]{grim1989ions}
Grim, R.~J., Greenberg, J.~M., Schutte, W.~A., \& Schmitt, B. 1989{\natexlab{b}}, ApJL, 341, L87

\bibitem[{Henkel {et~al.}(1987)Henkel, Wilson, \& Mauersberger}]{henkel1987multilevel}
Henkel, C., Wilson, T., \& Mauersberger, R. 1987, A\&A, 182, 137

\bibitem[{Hudson {et~al.}(2014)Hudson, Ferrante, \& Moore}]{hudson2014infrared}
Hudson, R., Ferrante, R., \& Moore, M. 2014, Icarus, 228, 276

\bibitem[{Hudson {et~al.}(2015)Hudson, Gerakines, \& Loeffler}]{hudson2015activation}
Hudson, R., Gerakines, P., \& Loeffler, M. 2015, PCCP, 17, 12545

\bibitem[{Hudson \& Gerakines(2018)}]{hudson2018infrared}
Hudson, R.~L. \& Gerakines, P.~A. 2018, ApJ, 867, 138

\bibitem[{Hudson {et~al.}(2022)Hudson, Gerakines, \& Yarnall}]{hudson2022ammonia}
Hudson, R.~L., Gerakines, P.~A., \& Yarnall, Y.~Y. 2022, ApJ, 925, 156

\bibitem[{Jakobsen {et~al.}(2022)Jakobsen, Ferruit, de~Oliveira, Arribas, Bagnasco, Barho, Beck, Birkmann, B{\"o}ker, Bunker, {et~al.}}]{jakobsen2022near}
Jakobsen, P., Ferruit, P., de~Oliveira, C.~A., {et~al.} 2022, A\&A, 661, A80

\bibitem[{Jim{\'e}nez-Escobar \& Caro(2011)}]{jimenez2011sulfur}
Jim{\'e}nez-Escobar, A. \& Caro, G.~M. 2011, A\&A, 536, A91

\bibitem[{Jim{\'e}nez-Escobar {et~al.}(2014)Jim{\'e}nez-Escobar, Mu{\~n}oz~Caro, \& Chen}]{jimenez2014sulphur}
Jim{\'e}nez-Escobar, A., Mu{\~n}oz~Caro, G., \& Chen, Y.-J. 2014, MNRAS, 443, 343

\bibitem[{Joseph {et~al.}(1986)Joseph, Snow~Jr, Seab, \& Crutcher}]{joseph1986interstellar}
Joseph, C.~L., Snow~Jr, T.~P., Seab, C.~G., \& Crutcher, R.~M. 1986, ApJ, 309, 771

\bibitem[{Kama {et~al.}(2019)Kama, Shorttle, Jermyn, Folsom, Furuya, Bergin, Walsh, \& Keller}]{kama2019abundant}
Kama, M., Shorttle, O., Jermyn, A.~S., {et~al.} 2019, ApJ, 885, 114

\bibitem[{Keane {et~al.}(2001)Keane, Tielens, Boogert, Schutte, \& Whittet}]{keane2001ice}
Keane, J., Tielens, A., Boogert, A., Schutte, W., \& Whittet, D. 2001, A\&A, 376, 254

\bibitem[{Keller {et~al.}(2002)Keller, Hony, Bradley, Molster, Waters, Bouwman, De~Koter, Brownlee, Flynn, Henning, {et~al.}}]{keller2002identification}
Keller, L., Hony, S., Bradley, J., {et~al.} 2002, Nature, 417, 148

\bibitem[{Kemper {et~al.}(2004)Kemper, Vriend, \& Tielens}]{kemper2004absence}
Kemper, F., Vriend, W., \& Tielens, A. 2004, ApJ, 609, 826

\bibitem[{Kerkhof {et~al.}(1999)Kerkhof, Schutte, \& Ehrenfreund}]{kerkhof1999infrared}
Kerkhof, O., Schutte, W., \& Ehrenfreund, P. 1999, A\&A, 346, 990

\bibitem[{Knacke {et~al.}(1982)Knacke, McCorkle, Puetter, Erickson, \& Kr{\"a}tschmer}]{knacke1982observation}
Knacke, R., McCorkle, S., Puetter, R., Erickson, E., \& Kr{\"a}tschmer, W. 1982, ApJ, 260, 141

\bibitem[{Knez {et~al.}(2005)Knez, Boogert, Pontoppidan, Kessler-Silacci, van Dishoeck, Evans~II, Augereau, Blake, \& Lahuis}]{knez2005spitzer}
Knez, C., Boogert, A.~A., Pontoppidan, K.~M., {et~al.} 2005, ApJ, 635, L145

\bibitem[{Kruczkiewicz {et~al.}(2021)Kruczkiewicz, Vitorino, Congiu, Theul{\'e}, \& Dulieu}]{kruczkiewicz2021ammonia}
Kruczkiewicz, F., Vitorino, J., Congiu, E., Theul{\'e}, P., \& Dulieu, F. 2021, A\&A, 652, A29

\bibitem[{Laas \& Caselli(2019)}]{laas2019modeling}
Laas, J.~C. \& Caselli, P. 2019, A\&A, 624, A108

\bibitem[{Ligterink {et~al.}(2018)Ligterink, Terwisscha~van Scheltinga, Taquet, J{\o}rgensen, Cazaux, van Dishoeck, \& Linnartz}]{ligterink2018formation}
Ligterink, N., Terwisscha~van Scheltinga, J., Taquet, V., {et~al.} 2018, MNRAS, 480, 3628

\bibitem[{Lodders(2003)}]{lodders2003solar}
Lodders, K. 2003, ApJ, 591, 1220

\bibitem[{Loeffler {et~al.}(2015)Loeffler, Hudson, Chanover, \& Simon}]{loeffler2015giant}
Loeffler, M.~J., Hudson, R.~L., Chanover, N.~J., \& Simon, A.~A. 2015, Icarus, 258, 181

\bibitem[{Mart{\'\i}n-Dom{\'e}nech {et~al.}(2016)Mart{\'\i}n-Dom{\'e}nech, Jim{\'e}nez-Serra, Caro, M{\"u}ller, Occhiogrosso, Testi, Woods, \& Viti}]{martin2016sulfur}
Mart{\'\i}n-Dom{\'e}nech, R., Jim{\'e}nez-Serra, I., Caro, G.~M., {et~al.} 2016, A\&A, 585, A112

\bibitem[{Mastrapa {et~al.}(2009)Mastrapa, Sandford, Roush, Cruikshank, \& Dalle~Ore}]{mastrapa2009optical}
Mastrapa, R., Sandford, S., Roush, T., Cruikshank, D., \& Dalle~Ore, C. 2009, ApJ, 701, 1347

\bibitem[{Mat{\'e} {et~al.}(2009)Mat{\'e}, G{\'a}lvez, Herrero, Fern{\'a}ndez-Torre, Moreno, \& Escribano}]{mate2009water}
Mat{\'e}, B., G{\'a}lvez, O., Herrero, V., {et~al.} 2009, ApJ, 703, L178

\bibitem[{Mauersberger {et~al.}(1986)Mauersberger, Henkel, Wilson, \& Walmsley}]{mauersberger1986hot}
Mauersberger, R., Henkel, C., Wilson, T., \& Walmsley, C. 1986, A\&A, 162, 199

\bibitem[{McClure {et~al.}(2023)McClure, Rocha, Pontoppidan, Crouzet, Chu, Dartois, Lamberts, Noble, Pendleton, Perotti, {et~al.}}]{mcclure2023ice}
McClure, M.~K., Rocha, W., Pontoppidan, K., {et~al.} 2023, Nat. astron., 7, 431

\bibitem[{Noble {et~al.}(2013)Noble, Theule, Borget, Danger, Chomat, Duvernay, Mispelaer, \& Chiavassa}]{noble2013thermal}
Noble, J.~A., Theule, P., Borget, F., {et~al.} 2013, MNRAS, 428, 3262

\bibitem[{Oba {et~al.}(2018)Oba, Tomaru, Lamberts, Kouchi, \& Watanabe}]{oba2018infrared}
Oba, Y., Tomaru, T., Lamberts, T., Kouchi, A., \& Watanabe, N. 2018, Nat. Astron., 2, 228

\bibitem[{{\"O}berg {et~al.}(2008){\"O}berg, Boogert, Pontoppidan, Blake, Evans, Lahuis, \& van Dishoeck}]{oberg2008c2d}
{\"O}berg, K.~I., Boogert, A.~A., Pontoppidan, K.~M., {et~al.} 2008, ApJ, 678, 1032

\bibitem[{{\"O}berg {et~al.}(2011){\"O}berg, Boogert, Pontoppidan, Van~den Broek, Van~Dishoeck, Bottinelli, Blake, \& Evans}]{oberg2011spitzer}
{\"O}berg, K.~I., Boogert, A.~A., Pontoppidan, K.~M., {et~al.} 2011, ApJ, 740, 109

\bibitem[{Palumbo {et~al.}(1997)Palumbo, Geballe, \& Tielens}]{palumbo1997solid}
Palumbo, M., Geballe, T., \& Tielens, A.~G. 1997, ApJ, 479, 839

\bibitem[{Pauls {et~al.}(1983)Pauls, Wilson, Bieging, \& Martin}]{pauls1983clumping}
Pauls, A., Wilson, T., Bieging, J., \& Martin, R. 1983, A\&A, 124, 23

\bibitem[{Perrero {et~al.}(2024)Perrero, Beitia-Antero, Fuente, Ugliengo, \& Rimola}]{perrero2024theoretical}
Perrero, J., Beitia-Antero, L., Fuente, A., Ugliengo, P., \& Rimola, A. 2024, MNRAS, 527, 10697

\bibitem[{Pontoppidan(2006)}]{pontoppidan2006spatial}
Pontoppidan, K.~M. 2006, A\&A, 453, L47

\bibitem[{Rachid {et~al.}(2021)Rachid, Brunken, De~Boe, Fedoseev, Boogert, \& Linnartz}]{rachid2021infrared}
Rachid, M., Brunken, N., De~Boe, D., {et~al.} 2021, A\&A, 653, A116

\bibitem[{Raunier {et~al.}(2003)Raunier, Chiavassa, Marinelli, Allouche, \& Aycard}]{raunier2003reactivity}
Raunier, S., Chiavassa, T., Marinelli, F., Allouche, A., \& Aycard, J. 2003, Chem. phys. lett., 368, 594

\bibitem[{Reach {et~al.}(2008)Reach, Faied, Rho, Boogert, Tappe, Jarrett, Morris, Cambr{\'e}sy, Palla, \& Valdettaro}]{reach2008properties}
Reach, W.~T., Faied, D., Rho, J., {et~al.} 2008, ApJ, 690, 683

\bibitem[{Rocha {et~al.}(2022)Rocha, Rachid, Olsthoorn, Van~Dishoeck, McClure, \& Linnartz}]{rocha2022lida}
Rocha, W., Rachid, M., Olsthoorn, B., {et~al.} 2022, A\&A, 668, A63

\bibitem[{Rocha {et~al.}(2024)Rocha, van Dishoeck, Ressler, van Gelder, Slavicinska, Brunken, Linnartz, Ray, Beuther, o~Garatti, {et~al.}}]{rocha2024jwst}
Rocha, W., van Dishoeck, E., Ressler, M., {et~al.} 2024, A\&A, 683, A124

\bibitem[{Santos {et~al.}(2023)Santos, Linnartz, \& Chuang}]{santos2023interaction}
Santos, J.~C., Linnartz, H., \& Chuang, K.-J. 2023, A\&A, 678, A112

\bibitem[{Schutte {et~al.}(1999)Schutte, Boogert, Tielens, Whittet, Gerakines, Chiar, Ehrenfreund, Greenberg, Van~Dishoeck, \& De~Graauw}]{schutte1999weak}
Schutte, W., Boogert, A., Tielens, A., {et~al.} 1999, A\&A, 343, 966

\bibitem[{Schutte \& Khanna(2003)}]{schutte2003origin}
Schutte, W. \& Khanna, R. 2003, A\&A, 398, 1049

\bibitem[{Sellgren {et~al.}(1995)Sellgren, Brooke, Smith, \& Geballe}]{sellgren1995new}
Sellgren, K., Brooke, T., Smith, R., \& Geballe, T. 1995, ApJ, 449, L69

\bibitem[{Slavicinska {et~al.}(2023)Slavicinska, Rachid, Rocha, Chuang, van Dishoeck, \& Linnartz}]{slavicinska2023hunt}
Slavicinska, K., Rachid, M.~G., Rocha, W. R.~M., {et~al.} 2023, A\&A, 677, A13

\bibitem[{Slavicinska {et~al.}(2024)Slavicinska, van Dishoeck, Tychoniec, Nazari, Rubinstein, Gutermuth, Tyagi, Chen, Brunken, Rocha, {et~al.}}]{slavicinska2024jwst}
Slavicinska, K., van Dishoeck, E.~F., Tychoniec, {\L}., {et~al.} 2024, A\&A, 688, A29

\bibitem[{Smith(1991)}]{smith1991search}
Smith, R.~G. 1991, MNRAS, 249, 172

\bibitem[{Tieftrunk {et~al.}(1994)Tieftrunk, Pineau~des Forets, Schilke, \& Walmsley}]{tieftrunk1994so}
Tieftrunk, A., Pineau~des Forets, G., Schilke, P., \& Walmsley, C. 1994, A\&A, 289, 579

\bibitem[{Tielens \& Allamandola(1987)}]{tielens1987evolution}
Tielens, A. \& Allamandola, L. 1987, Physical processes in interstellar clouds, 333

\bibitem[{van Broekhuizen {et~al.}(2004)van Broekhuizen, Keane, \& Schutte}]{van2004quantitative}
van Broekhuizen, F., Keane, J., \& Schutte, W. 2004, A\&A, 415, 425

\bibitem[{van Broekhuizen {et~al.}(2005)van Broekhuizen, Pontoppidan, Fraser, \& Van~Dishoeck}]{van20053}
van Broekhuizen, F.~A., Pontoppidan, K., Fraser, H., \& Van~Dishoeck, E. 2005, A\&A, 441, 249

\bibitem[{Vastel {et~al.}(2003)Vastel, Phillips, Ceccarelli, \& Pearson}]{vastel2003first}
Vastel, C., Phillips, T., Ceccarelli, C., \& Pearson, J. 2003, ApJ, 593, L97

\bibitem[{Vidal {et~al.}(2017)Vidal, Loison, Jaziri, Ruaud, Gratier, \& Wakelam}]{vidal2017reservoir}
Vidal, T.~H., Loison, J.-C., Jaziri, A.~Y., {et~al.} 2017, MNRAS, 469, 435

\bibitem[{Vidal \& Wakelam(2018)}]{vidal2018new}
Vidal, T.~H. \& Wakelam, V. 2018, MNRAS, 474, 5575

\bibitem[{Vitorino {et~al.}(2024)Vitorino, Loison, Wakelam, Congiu, \& Dulieu}]{vitorino2024sulphur}
Vitorino, J., Loison, J.-C., Wakelam, V., Congiu, E., \& Dulieu, F. 2024, MNRAS, 533, 52

\bibitem[{Walmsley(1994)}]{walmsley1994ammonia}
Walmsley, C. 1994, in AIP Conference Proceedings, American Institute of Physics, 463--475

\bibitem[{Whittet(2018)}]{whittet2018dust}
Whittet, D.~C. 2018, Dust in the galactic environment (CRC press)

\bibitem[{Wilson {et~al.}(1983)Wilson, Mauersberger, Walmsley, \& Batrla}]{wilson1983non}
Wilson, T., Mauersberger, R., Walmsley, C., \& Batrla, W. 1983, A\&A, 127, L19

\bibitem[{Woods {et~al.}(2015)Woods, Occhiogrosso, Viti, Kanuchov{\'a}, Palumbo, \& Price}]{woods2015new}
Woods, P.~M., Occhiogrosso, A., Viti, S., {et~al.} 2015, MNRAS, 450, 1256

\bibitem[{Wright {et~al.}(2023)Wright, Rieke, Glasse, Ressler, Mar{\'\i}n, Aguilar, Alberts, {\'A}lvarez-M{\'a}rquez, Argyriou, Banks, {et~al.}}]{wright2023mid}
Wright, G.~S., Rieke, G.~H., Glasse, A., {et~al.} 2023, PASP, 135, 048003

\bibitem[{Yarnall \& Hudson(2022)}]{yarnall2022new}
Yarnall, Y.~Y. \& Hudson, R.~L. 2022, ApJL, 931, L4

\bibitem[{York(1966)}]{york1966least}
York, D. 1966, Can. J. Phys., 44, 1079

\end{thebibliography}

\begin{appendix}

\section{Calibration procedure supplementary information}
\label{app:cal}
The thickness of a pure ice, d, can be calculated from its laser interference pattern during its deposition if its refractive index is known via the following equation:

\begin{equation}
    \indent d = \frac{m\lambda}{2\sqrt{n^{2} - sin^{2}\theta}},
\end{equation}

\noindent where m is an integer number of constructive fringes, $\lambda$ is the laser wavelength, n is the ice refractive index, and $\theta$ is the angle of incidence (45$^{\circ}$ on IRASIS). The ice thickness can be converted into an ice column density using a measured ice density value, $\rho$, allowing for the calculation of an ice deposition rate in units of column density per unit of time when the period of the laser interference oscillation is determined using a curve fitting procedure.

The refractive index and density values used to calibrate the deposition rates of our three ices of interest, H$_{2}$O, NH$_{3}$, H$_{2}$S, in our experiments are presented in Table~\ref{tab:calibration}.

\begin{table}[h]
\caption{Refractive indexes and densities used for calibration.}
\begin{center}
\begin{tabular}{c c c c}
\hline
        Molecule & n & $\rho$ (g cm$^{-3}$) & Ref. \\
        \hline
        H$_{2}$O & 1.234 & 0.719 & 1 \\
        NH$_{3}$ & 1.33 & 0.68 & 2 \\
        H$_{2}$S & 1.407 & 0.944 & 1 \\
    \hline
     \label{tab:calibration}
\end{tabular}
\begin{tablenotes}
\item 1. \citealt{yarnall2022new}
\item 2. \citealt{hudson2022ammonia}
\end{tablenotes}

\end{center}
\end{table}

\section{Beer's law plot errors}
\label{app:bs_err}
Errors of the integrated optical depths of the features used in the Beer's law plots are estimated by smoothing the integrated optical depth data throughout the entire warm-up in each individual experiment with a Savgol filter and then taking the standard deviation of five raw integrated optical depth data points collected at the end of the 120 K hold from the smoothed data (used here as expected values). This provides an individual error for each data point in the Beer's law plots.

Three sources of error contribute to the uncertainty of the ice column densities in the Beer's law plots. First, there is the experimental error caused by day-to-day fluctuations in the calibrated deposition rate, which we estimate to be 0.36\% from our repeat measurements of the laser interference patterns during pure ice depositions (Section~\ref{txt:band_strengths_methods}). Then there is the error caused by the uncertainty of laser incidence angle in our optical alignment that we estimate to be $\pm$5$^{\circ}$, resulting in an error of $\sim$4\% in the ice thickness values calculated from the laser interference patterns. The largest source of uncertainty in the ice column density is the uncertainty associated with reaction completion, for which we use a conservative estimate of 8\%. Propagating these errors results in a final error of 9\% on the ice column densities.

\section{Pure NH$_{4}$SH supplementary laboratory data}
\label{app:nh4sh_pure}

This appendix contains supplementary figures of the laboratory data from the pure NH$_{4}$SH experiments.

\begin{figure}[ht!]
\centering
\includegraphics[width=\linewidth]{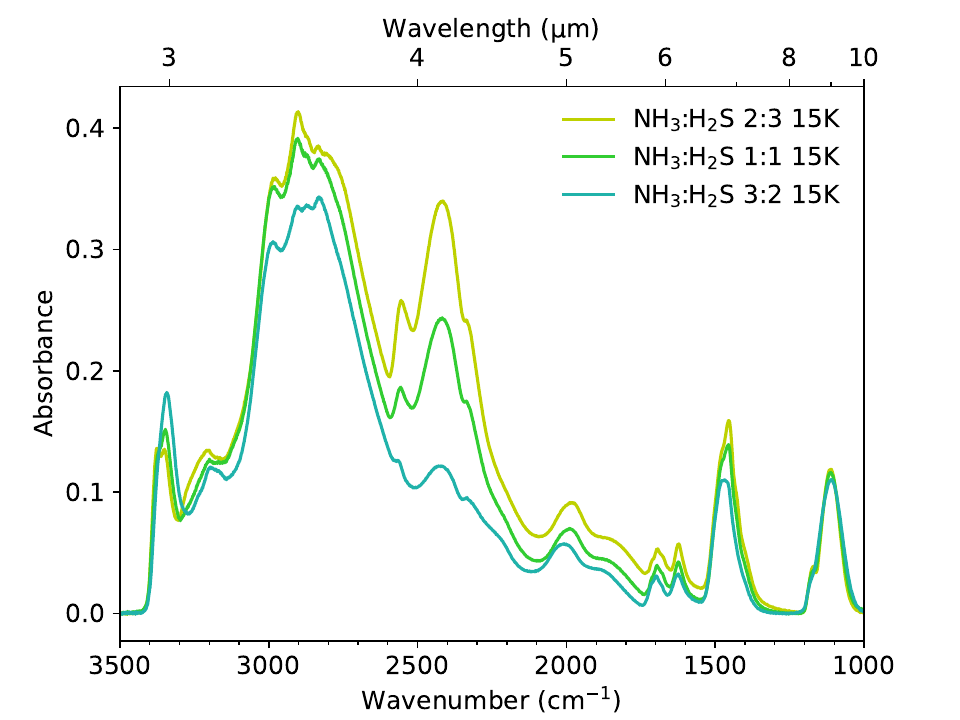}
\caption{Spectra of NH$_{3}$:H$_{2}$S mixtures with various mixing ratios at 15 K. In all ices, the same quantity of NH$_{3}$ was deposited.}
\label{fig:nh3_h2s_conc}
\end{figure}

\begin{figure}[ht!]
\centering
\includegraphics[width=\linewidth]{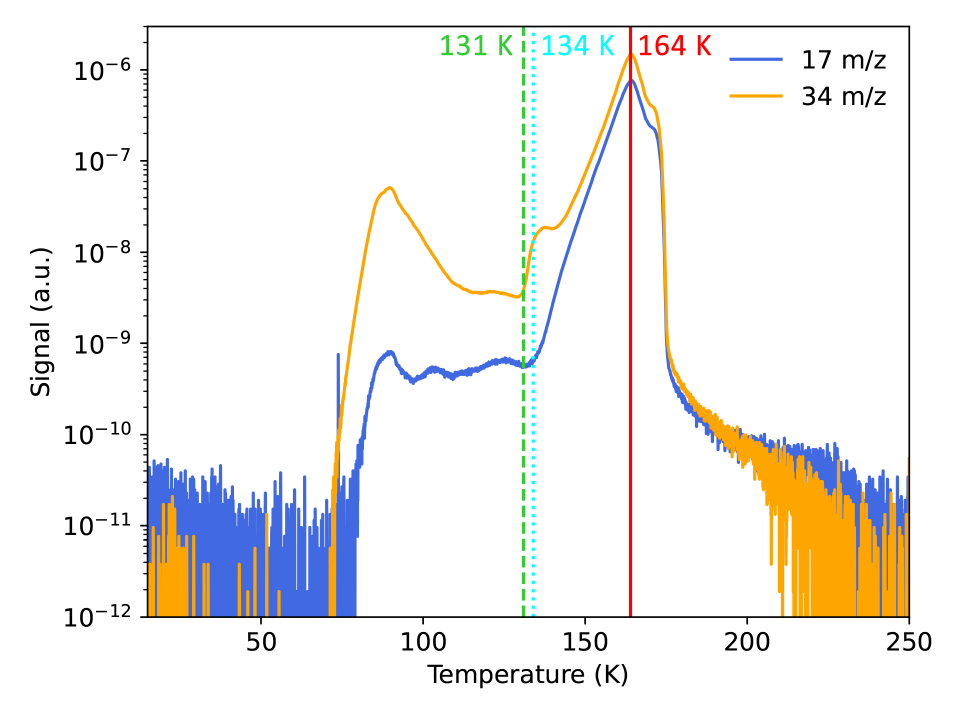}
\caption{TPD curve of 17 m/z (NH$_{3}$$^{+}$) and 34 m/z (H$_{2}$S$^{+}$) from the NH$_{3}$:H$_{2}$S 1:1 mixture. The temperature at the maximum area of the NH$_{4}$$^{+}$ $\nu_{4}$ mode (131 K) is marked with a green dashed line, the temperature when crystallization begins (134 K) is marked with a cyan dotted line, and the temperature of peak desorption (164 K) is marked with a red solid line.}
\label{fig:nh4sh_pure_tpd}
\end{figure}

\begin{figure}[ht!]
\centering
\includegraphics[width=\linewidth]{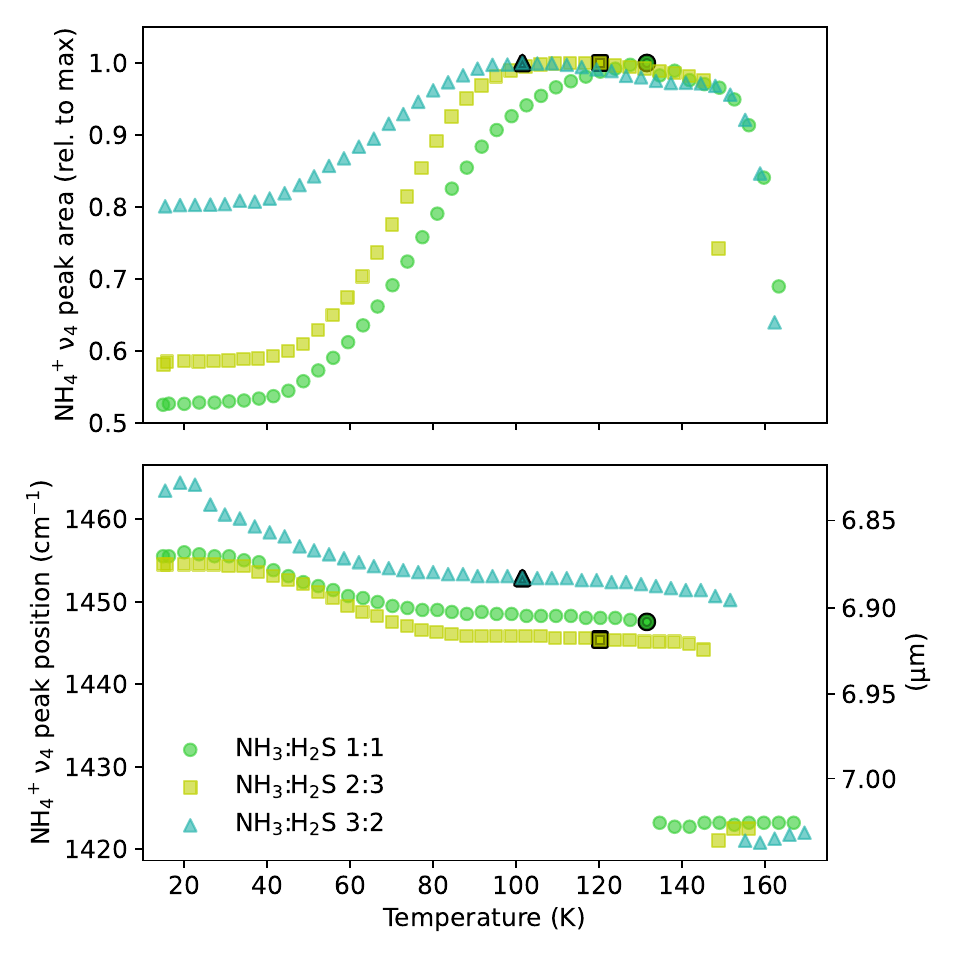}
\caption{Areas (relative to the maximum peak area for each experiment) and positions of the NH$_{4}$$^{+}$ $\nu_{4}$ mode during warm-up of the NH$_{3}$:H$_{2}$S ice mixtures. The points at the temperatures when the peak areas are at their maximum for each experiment are indicated with a black outline.}
\label{fig:nh4_pure_char}
\end{figure}

\begin{table*}[h!]
\caption{Vibrational mode assignments of the peaks observed in the IR spectra of pure NH$_{3}$ ice at 15 K, pure H$_{2}$S ice at 15 K, NH$_{3}$:H$_{2}$S ice at 15 K, NH$_{4}$SH ice at 120 K, and NH$_{4}$SH ice at 140 K.}
\begin{center}
\begin{tabular}{c c c c c}
\hline
        Ice & Temp. & Phase & Mode & Peak position \\
        & (K) &  &  & (cm$^{-1}$/$\mu$m)\\
        \hline
        \multirow{5}{*}{NH$_{3}$} & \multirow{5}{*}{15} & \multirow{5}{*}{Amorphous} & $\nu_{3}$ (asymm. stretch) & 3375/2.963 \\
        & & & 2$\nu_{4}$ & 3291/3.039 \\
        & & & $\nu_{1}$ (symm. stretch) & 3213/3.112 \\
        & & & $\nu_{4}$ (sciss. bend) & 1625/6.154 \\
        & & & $\nu_{2}$ (umb. bend) & 1074/9.311 \\
        \hline
        \multirow{6}{*}{H$_{2}$S} & \multirow{6}{*}{15} & \multirow{6}{*}{Amorphous} & $\nu_{1}$+$\nu_{2}$/$\nu_{2}$+$\nu_{3}$ & 3702/2.701 \\
        & & & $\nu_{1}$+T & 2636/3.794 \\
        & & & $\nu_{3}$ (asymm. stretch) & 2251/4.442 \\
        & & & $\nu_{1}$ (symm. stretch) & 2525/3.960 \\
        & & & $\nu_{2}$+R & 1334/7.496 \\
        & & & $\nu_{2}$ (bend) & 1170/8.547 \\
        \hline
        \multirow{10}{*}{NH$_{3}$:H$_{2}$S 1:1} & \multirow{10}{*}{15} & \multirow{10}{*}{Amorphous} & NH$_{3}$ $\nu_{3}$ & 3348/2.987 \\
        & & & NH$_{3}$ $\nu_{1}$/NH$_{4}$$^{+}$ $\nu_{2}$+$\nu_{4}$ & 3202/3.123 \\
        & & & NH$_{4}$$^{+}$ $\nu_{3}$/NH$_{4}$$^{+}$ 2$\nu_{4}$ & 2905/3.442 \\
        & & & H$_{2}$S $\nu_{1}$+$\nu_{3}$/SH$^{-}$ str. & 2559/3.908 \\
        & & & H$_{2}$S in NH$_{3}$/NH$_{4}$SH matrix? & 2418/4.136 \\
        & & & NH$_{4}$$^{+}$ $\nu_{2}$ & 1696/5.896 \\
        & & & NH$_{3}$ $\nu_{4}$ & 1624/6.158 \\
        & & & NH$_{4}$$^{+}$ $\nu_{4}$ & 1455/6.873 \\
        & & & H$_{2}$S $\nu_{2}$ & 1177/8.496 \\
        & & & NH$_{3}$ $\nu_{2}$ & 1112/8.993 \\
        \hline
        \multirow{9}{*}{NH$_{4}$SH} & \multirow{9}{*}{120 K} & \multirow{9}{*}{Amorphous} & NH$_{4}$$^{+}$ $\nu_{2}$+$\nu_{4}$ & 3121/3.204 \\
        & & & NH$_{4}$$^{+}$ $\nu_{3}$ (asymm. stretch) & 2953/3.386 \\
        & & & NH$_{4}$$^{+}$ 2$\nu_{4}$ & 2804/3.566 \\
        & & & SH$^{-}$ str. & 2553/3.917 \\
        & & & NH$_{4}$$^{+}$ $\nu_{2}$+SH$^{-}$ R? & 2089/4.787 \\
        & & & NH$_{4}$$^{+}$ $\nu_{4}$+SH$^{-}$ R & 1837/5.444 \\
        & & & NH$_{4}$$^{+}$ $\nu_{2}$ (sciss. bend) & 1651/6.057 \\
        & & & NH$_{4}$$^{+}$ $\nu_{4}$ (umb. bend) & 1448/6.906 \\
        & & & 2 SH$^{-}$ R & 803/12.453 \\
        \hline
        \multirow{10}{*}{NH$_{4}$SH} & \multirow{10}{*}{140 K} & \multirow{10}{*}{Crystalline} & NH$_{4}$$^{+}$ $\nu_{2}$+$\nu_{4}$ & 3115/3.210 \\
        & & & NH$_{4}$$^{+}$ $\nu_{3}$ E$_{u}$ & 3017/3.315 \\
        & & & NH$_{4}$$^{+}$ $\nu_{3}$ A$_{2u}$ & 2930/3.413 \\
        & & & NH$_{4}$$^{+}$ 2$\nu_{4}$ & 2817/3.550 \\
        & & & SH$^{-}$ str. & 2570/3.891 \\
        & & & NH$_{4}$$^{+}$ $\nu_{2}$+SH$^{-}$ R? & 2099/4.764 \\
        & & & NH$_{4}$$^{+}$ $\nu_{4}$+SH$^{-}$ R & 1821/5.491 \\
        & & & NH$_{4}$$^{+}$ $\nu_{4}$ E$_{u}$ & 1440/6.944 \\
        & & & NH$_{4}$$^{+}$ $\nu_{4}$ A$_{2u}$ & 1423/7.027 \\
        & & & 2 SH$^{-}$ R & 832/12.019 \\
    \hline
     \label{tab:nh4sh_pure_assignments}
\end{tabular}

\begin{tablenotes}
T = translation; R = libration
\end{tablenotes}

\end{center}
\end{table*}

\begin{table*}[h]
\caption{Peak positions, FWHMs, and integrated band areas of the NH$_{4}$$^{+}$ $\nu_{4}$ mode in NH$_{3}$:H$_{2}$S ices with various mixing ratios at temperatures $\sim$15-159 K. The uncertainties of the peak positions and FWHMs are 0.5 cm$^{-1}$. Reported band areas are provided relative to the maximum amorphous value recorded during the warm-up. The reported band areas at different temperatures should not be used to calculate relative band strengths, as the salt abundance in the laboratory ice varies during warm-up as the acid-base reaction proceeds, and the salt begins desorbing at temperatures >135 K.}
\begin{center}
\begin{tabular}{|c|c|cc|cc|c|}
\hline
    NH$_{3}$:H$_{2}$S & T & \multicolumn{2}{c|}{Peak} & \multicolumn{2}{c|}{FWHM} & \multirow{2}{*}{Rel. band area} \\
    mixing ratio & (K) & (cm$^{-1}$) & ($\mu$m) & (cm$^{-1}$) & ($\mu$m) & \\
    \hline 
    \multirow{9}{*}{1:1} & 15 & 1455.5 & 6.870 & 76.7 & 0.358 & 0.53 \\
     & 34 & 1455.0 & 6.873 & 74.0 & 0.346 & 0.53 \\
     & 56 & 1451.4 & 6.890 & 65.3 & 0.308 & 0.59 \\
     & 77 & 1449.0 & 6.901 & 49.9 & 0.240 & 0.76 \\
     & 99 & 1448.5 & 6.904 & 55.4 & 0.268 & 0.93 \\
     & 120 & 1448.0 & 6.906 & 63.9 & 0.311 & 0.99 \\
     & 132 & 1447.6 & 6.906 & 63.9 & 0.311 & 1 \\
     & 138 & 1422.7 & 7.029 & 12.1 & 0.060 & 0.99 \\
     & 156 & 1423.2 & 7.026 & 10.8 & 0.054 & 0.91 \\
    \hline
    \multirow{8}{*}{3:2} & 15 & 1463.5 & 6.833 & 78.8 & 0.366 & 0.80 \\
    & 37 & 1459.1 & 6.853 & 77.4 & 0.359 & 0.81 \\
    & 59 & 1455.3 & 6.872 & 70.1 & 0.328 & 0.87 \\
    & 80 & 1453.6 & 6.880 & 66.8 & 0.314 & 0.96 \\
    & 102 & 1452.9 & 6.883 & 66.8 & 0.315 & 1 \\
    & 123 & 1452.4 & 6.885 & 67.0 & 0.317 & 0.99 \\
    & 145 & 1451.4 & 6.890 & 65.8 & 0.313 & 0.97 \\
    & 159 & 1420.8 & 7.038 & 11.6 & 0.057 & 0.85 \\
    \hline
    \multirow{8}{*}{2:3} & 15 & 1454.5 & 6.875 & 76.7 & 0.359 & 0.58 \\
    & 34 & 1454.3 & 6.876 & 74.0 & 0.347 & 0.59 \\
    & 56 & 1450.4 & 6.894 & 65.6 & 0.310 & 0.65 \\
    & 77 & 1446.6 & 6.913 & 47.7 & 0.230 & 0.85 \\
    & 99 & 1445.9 & 6.916 & 49.2 & 0.238 & 0.99 \\
    & 120 & 1445.4 & 6.919 & 50.6 & 0.245 & 1 \\
    & 142 & 1444.9 & 6.921 & 51.8 & 0.251 & 0.98 \\
    & 152 & 1422.5 & 7.030 & 11.6 & 0.057 & 0.16 \\
    \hline
\end{tabular}
\end{center}
\label{tab:nh4_pure_char}
\end{table*}

\begin{table*}[h]
\caption{Peak positions, FWHMs, and integrated band areas of the SH$^{-}$ stretching mode in NH$_{3}$:H$_{2}$S 3:2 ice at temperatures 102-159 K. The uncertainties of the peak positions and FWHMs are 0.5 cm$^{-1}$. Reported band areas are provided relative to the SH$^{-}$ band area at 102 K, the temperature when the band area of the NH$_{4}$$^{+}$ $\nu_{4}$ mode reaches its maximum amorphous peak area in the NH$_{3}$:H$_{2}$S 3:2 ice mixture. The reported band areas at different temperatures should not be used to calculate relative band strengths, as the salt begins desorbing at temperatures >135 K.}
\begin{center}
\begin{tabular}{|c|cc|cc|c|}
\hline
    T & \multicolumn{2}{c|}{Peak} & \multicolumn{2}{c|}{FWHM} & \multirow{2}{*}{Rel. band area} \\
    (K) & (cm$^{-1}$) & ($\mu$m) & (cm$^{-1}$) & ($\mu$m) & \\
    \hline 
    102 & 2557.9 & 3.910 & 24.8 & 0.038 & 1 \\
    123 & 2556.9 & 3.911 & 25.3 & 0.039 & 1.00 \\
    145 & 2557.4 & 3.910 & 24.1 & 0.037 & 0.88 \\
    159 & 2569.9 & 3.891 & 4.1 & 0.006 & 0.63 \\
    \hline
\end{tabular}
\end{center}
\label{tab:sh_pure_char}
\end{table*}

\clearpage

\section{NH$_{4}$SH:H$_{2}$O supplementary laboratory data}
\label{app:nh4sh_h2o}

This appendix contains supplementary data for the laboratory NH$_{4}$SH:H$_{2}$O ice mixture experiments.

\begin{figure}[ht!]
\centering
\includegraphics[width=\linewidth]{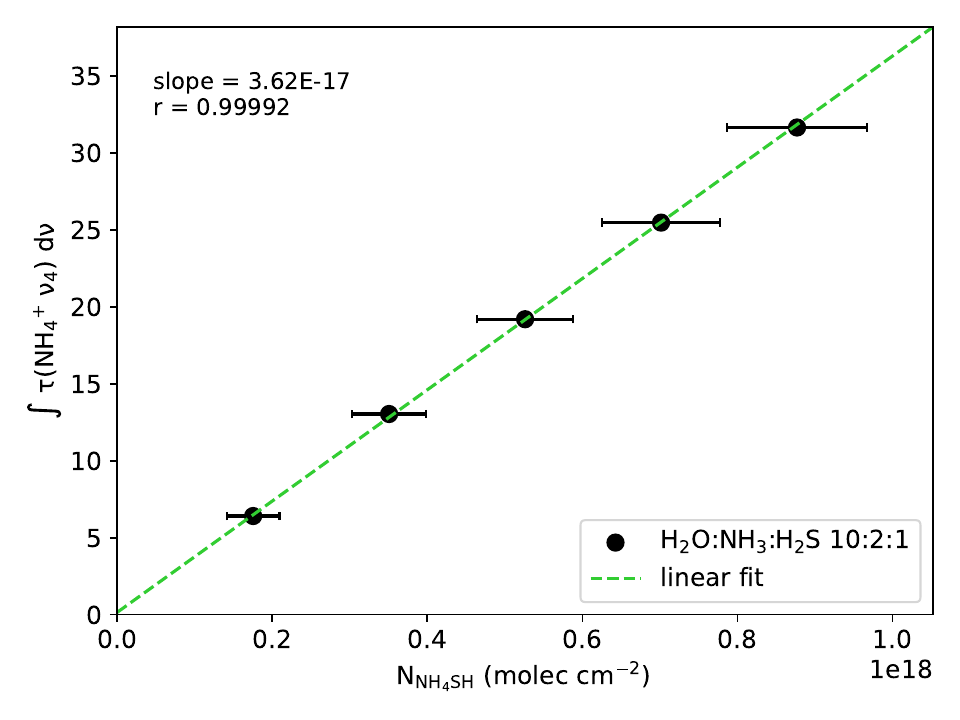}
\caption{Beer's law plot used to derive the apparent band strength of the NH$_{4}$$^{+}$ $\nu_{4}$ feature formed in the H$_{2}$O:NH$_{3}$:H$_{2}$S 10:2:1 mixture.}
\label{fig:beer_nh4_10p}
\end{figure}

\begin{figure}[ht!]
\centering
\includegraphics[width=\linewidth]{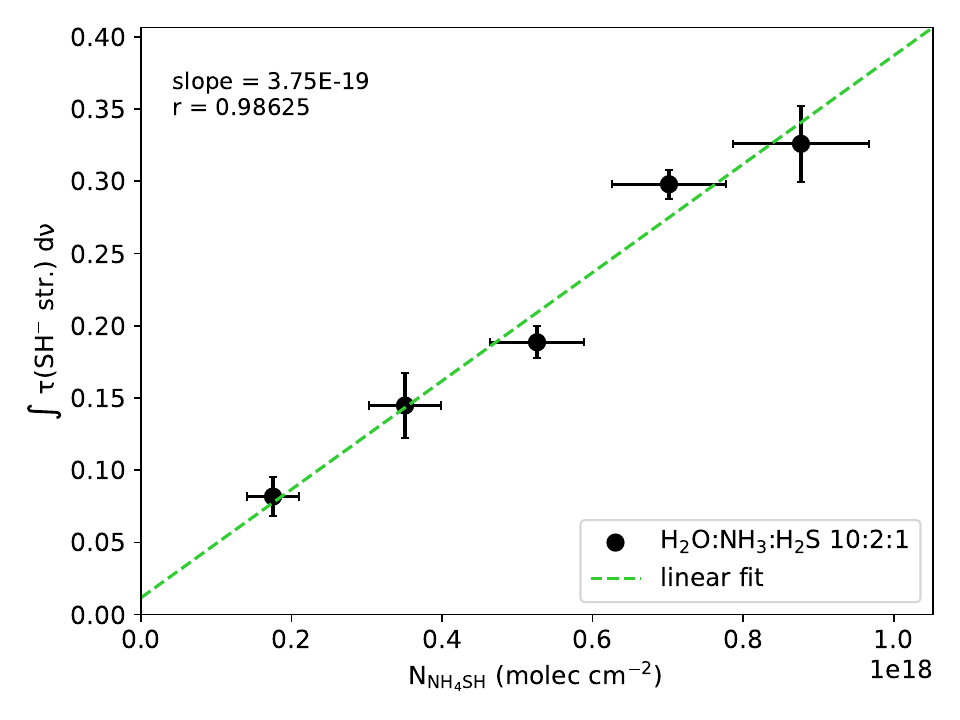}
\caption{Beer's law plot used to derive the apparent band strength of the SH$^{-}$ stretching feature formed in the H$_{2}$O:NH$_{3}$:H$_{2}$S 10:2:1 mixture.}
\label{fig:beer_sh_10p}
\end{figure}

\begin{figure}[ht!]
\centering
\includegraphics[width=\linewidth]{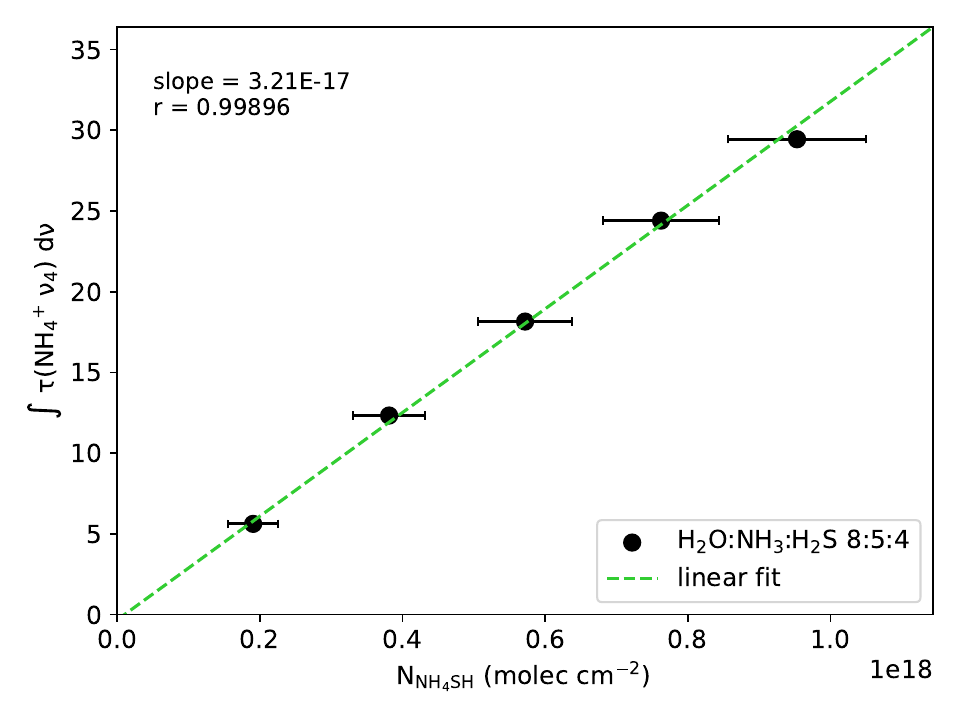}
\caption{Beer's law plot used to derive the apparent band strength of the NH$_{4}$$^{+}$ $\nu_{4}$ feature formed in the H$_{2}$O:NH$_{3}$:H$_{2}$S 8:5:4 mixture.}
\label{fig:beer_nh4_50p}
\end{figure}

\begin{figure}[ht!]
\centering
\includegraphics[width=\linewidth]{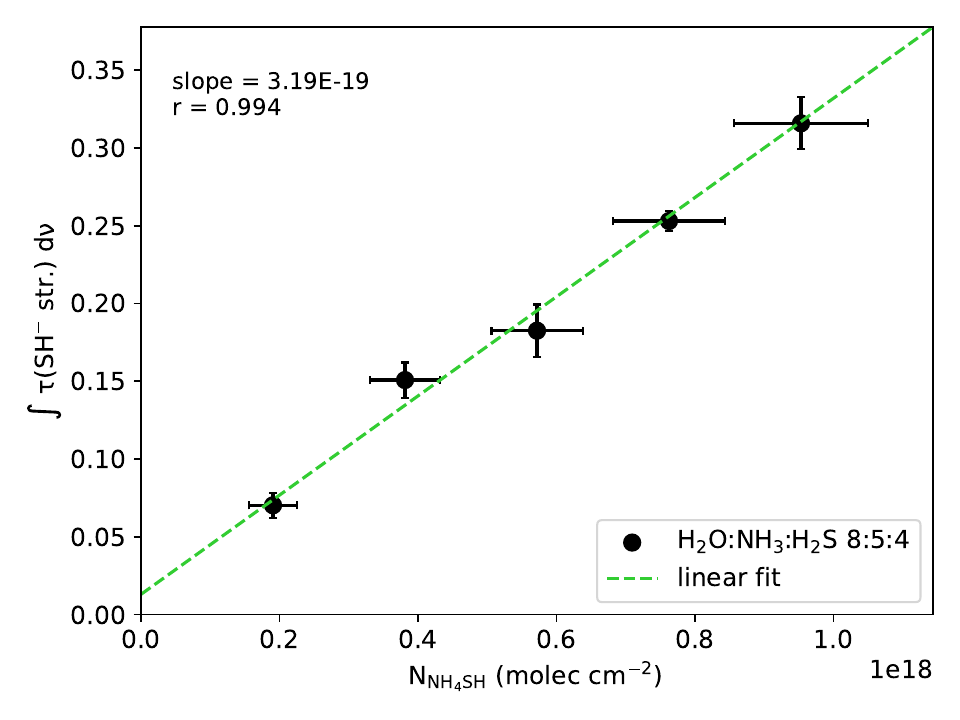}
\caption{Beer's law plot used to derive the apparent band strength of the SH$^{-}$ stretching feature formed in the H$_{2}$O:NH$_{3}$:H$_{2}$S 8:5:4 mixture.}
\label{fig:beer_sh_50p}
\end{figure}

\begin{figure}[ht!]
\centering
\includegraphics[width=\linewidth]{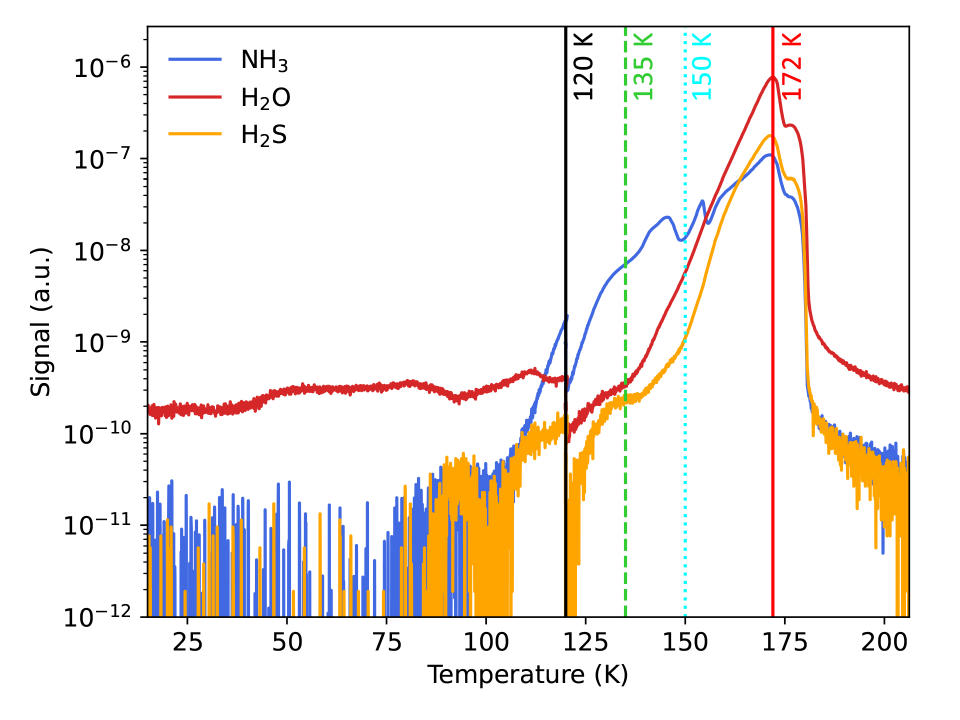}
\caption{TPD curve of NH$_{3}$, H$_{2}$O, and H$_{2}$S from the H$_{2}$O:NH$_{3}$:H$_{2}$S 10:3:2 mixture. The temperature of the 40 min hold (120 K) is marked with a solid black line, the temperature at the maximum area of the NH$_{4}$$^{+}$ $\nu_{4}$ mode (135 K) is marked with a green dashed line, the temperature when crystallization begins (150 K) is marked with a cyan dotted line, and the temperature of peak desorption is marked with a red solid line (172 K). The H$_{2}$S signal consists of the 34 m/z peak (H$_{2}$S$^{+}$). The NH$_{3}$ and H$_{2}$O signals are derived by deconvolving the 17 m/z (NH$_{3}$$^{+}$ and OH$^{+}$) and 18 m/z (H$_{2}$O$^{+}$ and $^{15}$NH$_{3}$$^{+}$) signals using systems of equations and scaling factors obtained from QMS data from pure depositions.}
\label{fig:nh4sh_10-3-2_tpd}
\end{figure}

\begin{figure}[ht!]
\centering
\includegraphics[width=\linewidth]{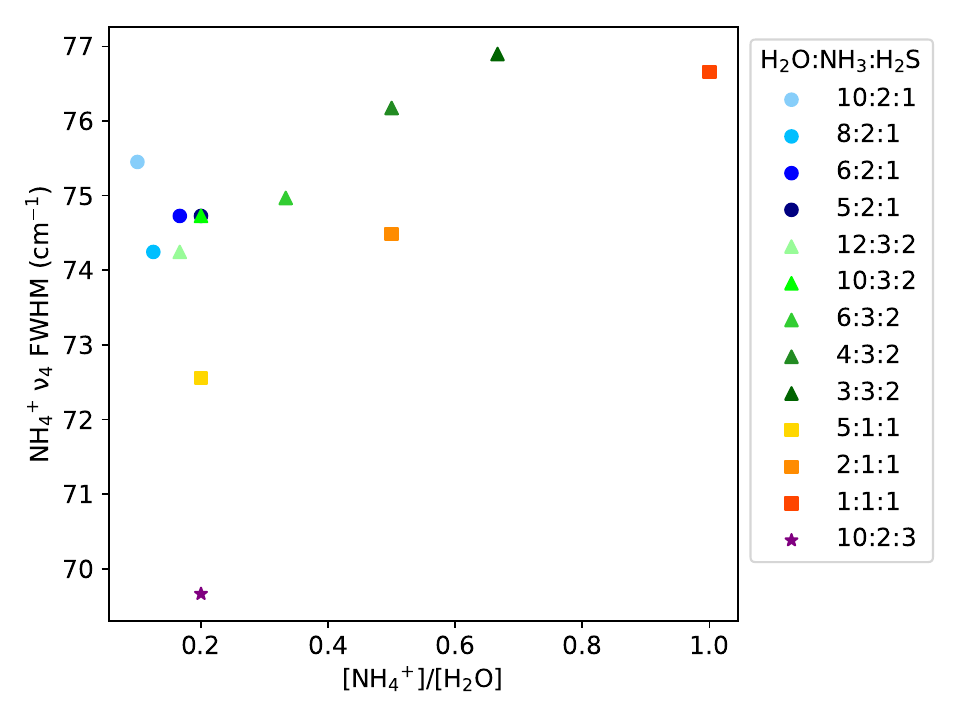}
\caption{The FWHM of the NH$_{4}$$^{+}$ $\nu_{4}$ mode at 135 K plotted as a function of the concentration of NH$_{4}$$^{+}$ with respect to H$_{2}$O (assuming 100\% reaction completion) for various H$_{2}$O:NH$_{3}$:H$_{2}$S mixtures.}
\label{fig:nh4_fwhm_conc}
\end{figure}

\begin{figure}[ht!]
\centering
\includegraphics[width=\linewidth]{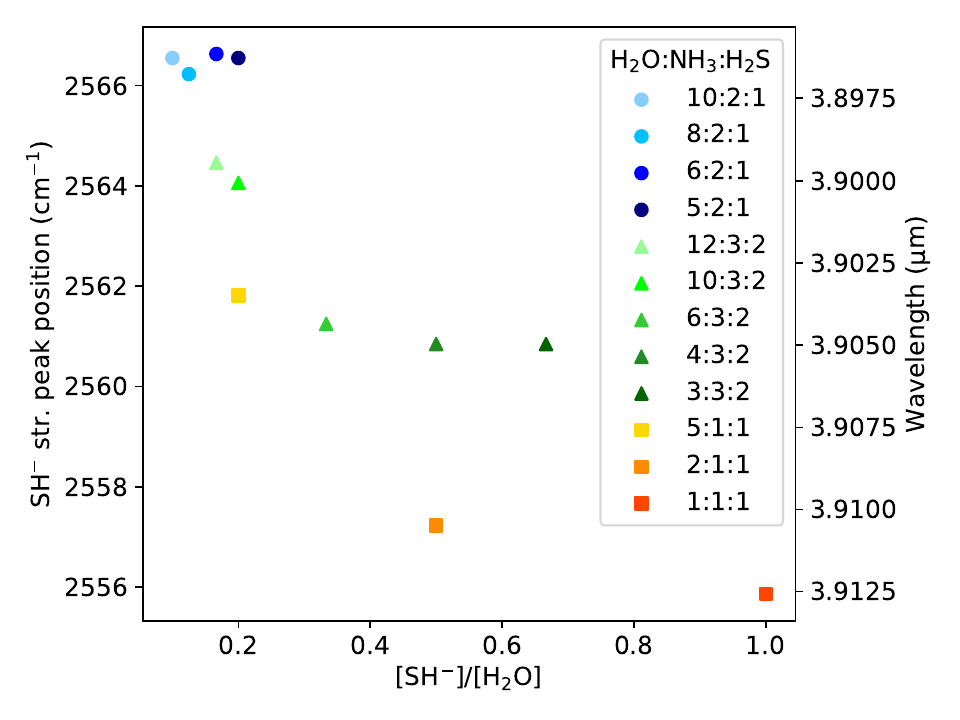}
\caption{The peak position of the SH$^{-}$ stretching mode at 135 K plotted as a function of the concentration of SH$^{-}$ with respect to H$_{2}$O (assuming 100\% reaction completion) for various H$_{2}$O:NH$_{3}$:H$_{2}$S mixtures.}
\label{fig:sh_peak_conc}
\end{figure}

\begin{figure}[ht!]
\centering
\includegraphics[width=\linewidth]{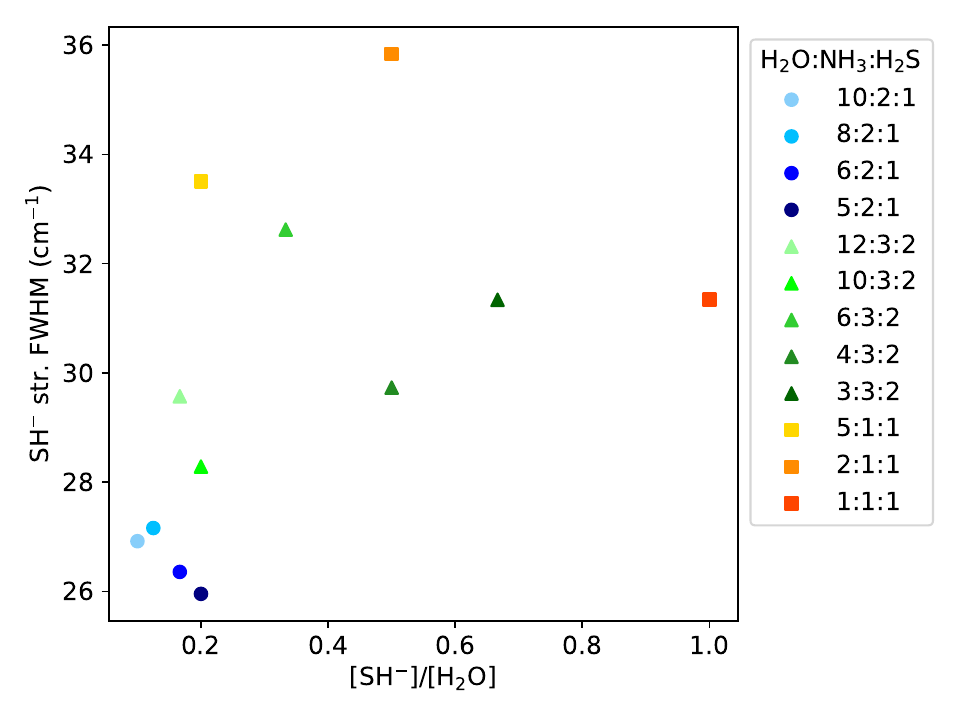}
\caption{The FWHM of the SH$^{-}$ stretching mode at 135 K plotted as a function of the concentration of SH$^{-}$ with respect to H$_{2}$O (assuming 100\% reaction completion) for various H$_{2}$O:NH$_{3}$:H$_{2}$S mixtures.}
\label{fig:sh_fwhm_conc}
\end{figure}

\begin{figure}[ht!]
\centering
\includegraphics[width=\linewidth]{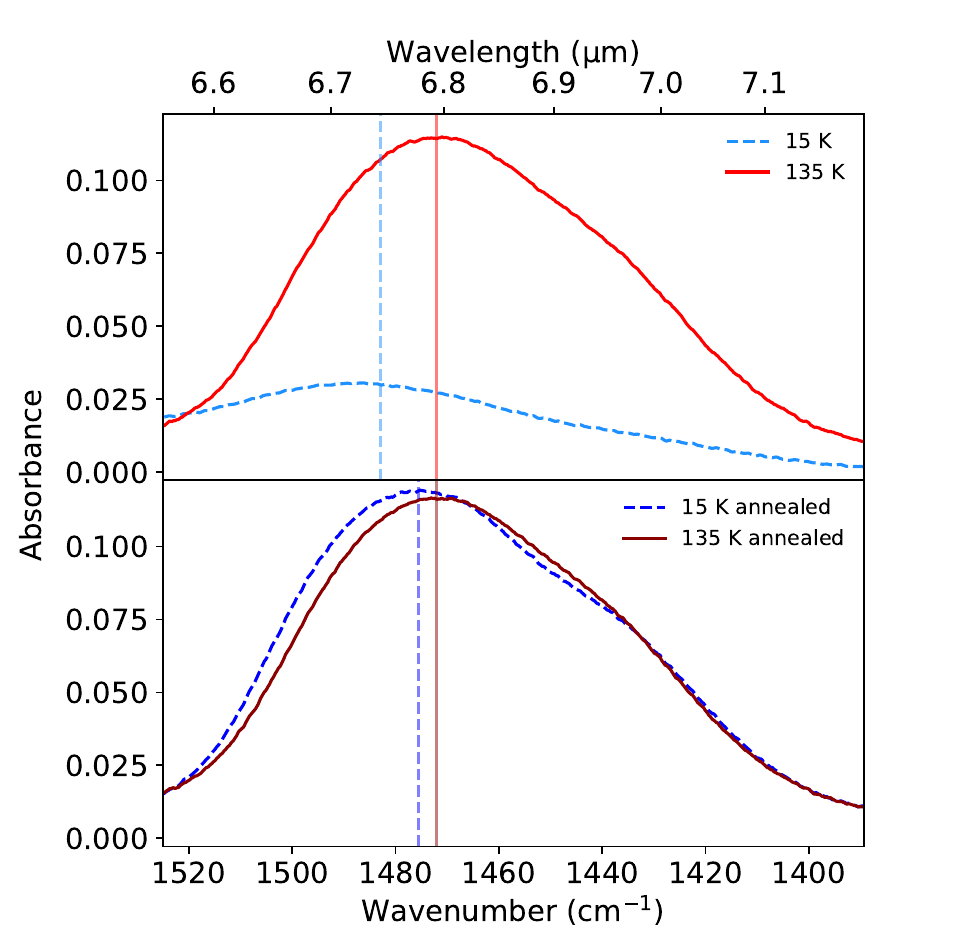}
\caption{Comparison of the NH$_{4}$$^{+}$ $\nu_{4}$ features in the spectra of the H$_{2}$O:NH$_{3}$:H$_{2}$S 10:3:2 ice mixtures before and after annealing. Top: NH$_{4}$$^{+}$ $\nu_{4}$ mode after deposition at 15 K (light blue dashed trace) and after heating to 135 K (solid red trace). Bottom: NH$_{4}$$^{+}$ $\nu_{4}$ mode after re-cooling from 135 K to 15 K (blue dashed trace) and after re-heating the annealed ice from 15 K to 135 K (solid dark red trace). Peak positions are marked with translucent vertical lines.}
\label{fig:annealing}
\end{figure}

\begin{table*}[h]
\caption{Peak positions, FWHMs, and integrated band areas of the NH$_{4}$$^{+}$ $\nu_{4}$ mode in H$_{2}$O:NH$_{3}$:H$_{2}$S ices with NH$_{3}$:H$_{2}$S mixing ratios of 2:1 at temperatures $\sim$15-170 K. The uncertainties of the peak positions and FWHMs are 0.5 cm$^{-1}$. Reported band areas are provided relative to the maximum amorphous value recorded during the warm-up. The reported band areas at different temperatures should not be used to calculate relative band strengths, as the salt abundance in the laboratory ice varies during warm-up as the acid-base reaction proceeds, and the salt begins desorbing at temperatures >135 K.}
\begin{center}
\begin{tabular}{|c|c|cc|cc|c|}
\hline
    H$_{2}$O:NH$_{3}$:H$_{2}$S & T & \multicolumn{2}{c|}{Peak} & \multicolumn{2}{c|}{FWHM} & \multirow{2}{*}{Rel. band area} \\
    mixing ratio & (K) & (cm$^{-1}$) & ($\mu$m) & (cm$^{-1}$) & ($\mu$m) & \\
    \hline 
    \multirow{10}{*}{10:2:1} & 15 & 1484.9 & 6.734 & 69.9 & 0.320 & 0.15 \\
     & 34 & 1485.2 & 6.733 & 69.9 & 0.320 & 0.17 \\
     & 56 & 1483.5 & 6.741 & 69.4 & 0.319 & 0.27 \\
     & 77 & 1480.3 & 6.755 & 70.9 & 0.327 & 0.46 \\
     & 99 & 1478.9 & 6.762 & 73.3 & 0.339 & 0.71 \\
     & 120 & 1478.4 & 6.764 & 75.2 & 0.348 & 0.94 \\
     & 133 & 1477.9 & 6.766 & 74.0 & 0.343 & 1 \\
     & 144 & 1475.5 & 6.777 & 74.7 & 0.347 & 0.84 \\
     & 154 & 1422.5 & 7.030 & 11.1 & 0.055 & 0.65 \\
     & 176 & 1422.7 & 7.029 & 7.7 & 0.038 & 0.18 \\
    \hline
    \multirow{10}{*}{8:2:1} & 15 & 1484.4 & 6.737 & 71.4 & 0.327 & 0.21 \\
    & 34 & 1484.7 & 6.735 & 68.9 & 0.315 & 0.22 \\
    & 56 & 1483.2 & 6.742 & 67.5 & 0.310 & 0.30 \\
    & 77 & 1480.1 & 6.756 & 71.1 & 0.328 & 0.51 \\
    & 99 & 1478.2 & 6.765 & 74.5 & 0.345 & 0.74 \\
    & 120 & 1477.0 & 6.771 & 75.7 & 0.351 & 0.92 \\
    & 134 & 1477.2 & 6.770 & 74.2 & 0.344 & 1 \\
    & 142 & 1474.1 & 6.784 & 75.5 & 0.351 & 0.89 \\
    & 156 & 1422.5 & 7.030 & 11.1 & 0.055 & 0.65 \\
    & 170 & 1422.7 & 7.029 & 9.2 & 0.045 & 0.22 \\
    \hline
    \multirow{10}{*}{6:2:1} & 15 & 1485.2 & 6.733 & 68.5 & 0.313 & 0.24 \\
    & 34 & 1484.4 & 6.737 & 68.5 & 0.313 & 0.26 \\
    & 56 & 1482.8 & 6.744 & 69.7 & 0.320 & 0.37 \\
    & 77 & 1479.4 & 6.760 & 71.8 & 0.331 & 0.54 \\
    & 99 & 1477.2 & 6.770 & 75.5 & 0.350 & 0.76 \\
    & 120 & 1476.2 & 6.774 & 76.2 & 0.353 & 0.89 \\
    & 135 & 1476.2 & 6.774 & 74.7 & 0.347 & 1 \\
    & 142 & 1473.1 & 6.788 & 76.7 & 0.357 & 0.96 \\
    & 156 & 1422.0 & 7.032 & 10.4 & 0.051 & 0.68 \\
    & 170 & 1421.8 & 7.034 & 8.9 & 0.044 & 0.26 \\
    \hline
    \multirow{10}{*}{5:2:1} & 15 & 1484.4 & 6.737 & 68.5 & 0.313 & 0.25 \\
    & 34 & 1483.7 & 6.740 & 68.7 & 0.315 & 0.28 \\
    & 55 & 1482.0 & 6.748 & 70.1 & 0.322 & 0.38 \\
    & 77 & 1478.4 & 6.764 & 72.6 & 0.335 & 0.55 \\
    & 98 & 1475.8 & 6.776 & 75.7 & 0.351 & 0.73 \\
    & 120 & 1475.0 & 6.779 & 76.7 & 0.356 & 0.85 \\
    & 134 & 1474.8 & 6.781 & 74.7 & 0.347 & 0.91 \\
    & 141 & 1471.9 & 9.794 & 76.4 & 0.356 & 1 \\
    & 156 & 1421.5 & 7.035 & 10.6 & 0.052 & 0.73 \\
    & 170 & 1421.3 & 7.036 & 8.9 & 0.044 & 0.33 \\
    \hline
\end{tabular}
\end{center}
\label{tab:nh4_h2o_char_2_1}
\end{table*}

\begin{table*}[h]
\caption{Peak positions, FWHMs, and integrated band areas of the NH$_{4}$$^{+}$ $\nu_{4}$ mode in H$_{2}$O:NH$_{3}$:H$_{2}$S ices with NH$_{3}$:H$_{2}$S mixing ratios of 3:2 at temperatures $\sim$15-170 K. The uncertainties of the peak positions and FWHMs are 0.5 cm$^{-1}$. Reported band areas are provided relative to the maximum amorphous value recorded during the warm-up. The reported band areas at different temperatures should not be used to calculate relative band strengths, as the salt abundance in the laboratory ice varies during warm-up as the acid-base reaction proceeds, and the salt begins desorbing at temperatures >135 K.}
\begin{center}
\begin{tabular}{|c|c|cc|cc|c|}
\hline
    H$_{2}$O:NH$_{3}$:H$_{2}$S & T & \multicolumn{2}{c|}{Peak} & \multicolumn{2}{c|}{FWHM} & \multirow{2}{*}{Rel. band area} \\
    mixing ratio & (K) & (cm$^{-1}$) & ($\mu$m) & (cm$^{-1}$) & ($\mu$m) & \\
    \hline 
    \multirow{10}{*}{12:3:2} & 15 & 1483.2 & 6.742 & 69.9 & 0.321 & 0.17 \\
     & 34 & 1483.0 & 6.743 & 72.1 & 0.331 & 0.20 \\
     & 56 & 1482.0 & 6.748 & 69.4 & 0.319 & 0.28 \\
     & 77 & 1477.4 & 6.768 & 72.6 & 0.336 & 0.47 \\
     & 99 & 1475.5 & 6.777 & 74.2 & 0.345 & 0.68 \\
     & 120 & 1473.6 & 6.786 & 75.5 & 0.351 & 0.88 \\
     & 135 & 1472.9 & 6.789 & 74.2 & 0.346 & 0.99 \\
     & 142 & 1471.2 & 6.797 & 74.7 & 0.349 & 0.95 \\
     & 156 & 1422.7 & 7.029 & 11.1 & 0.055 & 0.63 \\
     & 170 & 1422.7 & 7.029 & 8.7 & 0.043 & 0.30 \\
    \hline
    \multirow{9}{*}{10:3:2} & 15 & 1482.5 & 6.745 & 71.1 & 0.327 & 0.21 \\
    & 34 & 1482.5 & 6.745 & 71.4 & 0.328 & 0.23 \\
    & 56 & 1480.6 & 6.754 & 70.4 & 0.324 & 0.32 \\
    & 77 & 1477.0 & 6.771 & 72.6 & 0.336 & 0.51 \\
    & 99 & 1474.3 & 6.783 & 74.7 & 0.348 & 0.72 \\
    & 120 & 1472.4 & 6.792 & 75.7 & 0.353 & 0.91 \\
    & 136 & 1472.1 & 6.793 & 73.5 & 0.343 & 1 \\
    & 154 & 1422.5 & 7.030 & 11.3 & 0.056 & 0.66 \\
    & 172 & 1422.2 & 7.031 & 8.9 & 0.044 & 0.31 \\
    \hline
    \multirow{10}{*}{6:3:2} & 15 & 1482.0 & 6.748 & 69.9 & 0.321 & 0.27 \\
    & 34 & 1481.1 & 6.752 & 68.9 & 0.317 & 0.28 \\
    & 56 & 1478.7 & 6.763 & 70.6 & 0.326 & 0.38 \\
    & 77 & 1473.8 & 6.785 & 73.5 & 0.341 & 0.55 \\
    & 99 & 1470.2 & 6.802 & 75.2 & 0.351 & 0.70 \\
    & 120 & 1468.5 & 6.810 & 76.7 & 0.359 & 0.83 \\
    & 134 & 1468.0 & 6.812 & 75.0 & 0.351 & 0.90 \\
    & 145 & 1465.4 & 6.824 & 75.0 & 0.352 & 0.96 \\
    & 156 & 1421.5 & 7.035 & 11.1 & 0.055 & 0.77 \\
    & 170 & 1420.3 & 7.041 & 8.9 & 0.044 & 0.47 \\
    \hline
    \multirow{10}{*}{4:3:2} & 15 & 1478.7 & 6.763 & 71.1 & 0.327 & 0.41 \\
    & 34 & 1478.7 & 6.763 & 71.8 & 0.330 & 0.44 \\
    & 56 & 1476.5 & 6.773 & 71.8 & 0.332 & 0.52 \\
    & 77 & 1471.4 & 6.796 & 75.0 & 0.349 & 0.69 \\
    & 99 & 1467.1 & 6.816 & 76.4 & 0.358 & 0.83 \\
    & 120 & 1465.4 & 6.824 & 77.4 & 0.363 & 0.92 \\
    & 135 & 1465.2 & 6.825 & 76.2 & 0.358 & 0.97 \\
    & 145 & 1463.2 & 6.834 & 75.9 & 0.358 & 1 \\
    & 160 & 1421.0 & 7.037 & 10.1 & 0.050 & 0.98 \\
    & 170 & 1419.8 & 7.043 & 8.4 & 0.042 & 0.60 \\
    \hline
    \multirow{9}{*}{3:3:2} & 15 & 1477.4 & 6.768 & 72.3 & 0.333 & 0.51 \\
    & 33 & 1476.7 & 6.772 & 73.3 & 0.338 & 0.55 \\
    & 54 & 1474.6 & 6.782 & 74.0 & 0.342 & 0.63 \\
    & 76 & 1468.8 & 6.808 & 74.5 & 0.347 & 0.75 \\
    & 97 & 1464.4 & 6.829 & 76.4 & 0.358 & 0.89 \\
    & 119 & 1463.0 & 6.835 & 77.4 & 0.364 & 0.95 \\
    & 140 & 1461.5 & 6.842 & 76.7 & 0.362 & 1 \\
    & 158 & 1420.8 & 7.038 & 10.1 & 0.050 & 1.03 \\
    & 169 & 1420.3 & 7.041 & 9.4 & 0.047 & 0.67 \\
    \hline
\end{tabular}
\end{center}
\label{tab:nh4_h2o_char_3_2}
\end{table*}

\begin{table*}[h]
\caption{Peak positions, FWHMs, and integrated band areas of the NH$_{4}$$^{+}$ $\nu_{4}$ mode in H$_{2}$O:NH$_{3}$:H$_{2}$S ice with a mixing ratio of 8:5:4 at temperatures $\sim$15-170 K. The uncertainties of the peak positions and FWHMs are 0.5 cm$^{-1}$. Reported band areas are provided relative to the maximum amorphous value recorded during the warm-up. The reported band areas at different temperatures should not be used to calculate relative band strengths, as the salt abundance in the laboratory ice varies during warm-up as the acid-base reaction proceeds, and the salt begins desorbing at temperatures >135 K.}
\begin{center}
\begin{tabular}{|c|c|cc|cc|c|}
\hline
    H$_{2}$O:NH$_{3}$:H$_{2}$S & T & \multicolumn{2}{c|}{Peak} & \multicolumn{2}{c|}{FWHM} & \multirow{2}{*}{Rel. band area} \\
    mixing ratio & (K) & (cm$^{-1}$) & ($\mu$m) & (cm$^{-1}$) & ($\mu$m) & \\
    \hline 
    \multirow{9}{*}{8:5:4} & 15 & 1478.2 & 6.765 & 73.0 & 0.337 & 0.36 \\
    & 33 & 1477.4 & 6.768 & 73.3 & 0.338 & 0.40 \\
    & 55 & 1474.3 & 6.783 & 73.0 & 0.339 & 0.49 \\
    & 76 & 1467.8 & 6.813 & 75.0 & 0.351 & 0.69 \\
    & 98 & 1463.5 & 6.833 & 75.7 & 0.356 & 0.84 \\
    & 120 & 1460.6 & 6.847 & 76.4 & 0.362 & 0.96 \\
    & 135 & 1459.6 & 6.851 & 75.9 & 0.360 & 1 \\
    & 156 & 1421.0 & 7.037 & 11.1 & 0.055 & 0.97 \\
    & 174 & 1419.8 & 7.043 & 8.9 & 0.044 & 0.21 \\
    \hline
\end{tabular}
\end{center}
\label{tab:nh4_h2o_char_5_4}
\end{table*}

\begin{table*}[h]
\caption{Peak positions, FWHMs, and integrated band areas of the NH$_{4}$$^{+}$ $\nu_{4}$ mode in H$_{2}$O:NH$_{3}$:H$_{2}$S ices with NH$_{3}$:H$_{2}$S mixing ratios of 1:1 at temperatures $\sim$15-170 K. The uncertainties of the peak positions and FWHMs are 0.5 cm$^{-1}$. Reported band areas are provided relative to the maximum amorphous value recorded during the warm-up. The reported band areas at different temperatures should not be used to calculate relative band strengths, as the salt abundance in the laboratory ice varies during warm-up as the acid-base reaction proceeds, and the salt begins desorbing at temperatures >135 K.}
\begin{center}
\begin{tabular}{|c|c|cc|cc|c|}
\hline
    H$_{2}$O:NH$_{3}$:H$_{2}$S & T & \multicolumn{2}{c|}{Peak} & \multicolumn{2}{c|}{FWHM} & \multirow{2}{*}{Rel. band area} \\
    mixing ratio & (K) & (cm$^{-1}$) & ($\mu$m) & (cm$^{-1}$) & ($\mu$m) & \\
    \hline 
    \multirow{10}{*}{5:1:1} & 15 & 1482.8 & 6.744 & 71.4 & 0.328 & 0.12 \\
    & 34 & 1482.0 & 6.748 & 71.1 & 0.328 & 0.14 \\
    & 56 & 1480.6 & 6.754 & 70.9 & 0.327 & 0.22 \\
    & 77 & 1475.8 & 6.776 & 72.3 & 0.336 & 0.39 \\
    & 99 & 1473.8 & 6.785 & 72.6 & 0.338 & 0.59 \\
    & 120 & 1471.9 & 6.794 & 73.3 & 0.342 & 0.80 \\
    & 131 & 1470.0 & 6.803 & 73.3 & 0.343 & 0.92 \\
    & 142 & 1467.8 & 6.813 & 71.6 & 0.335 & 1 \\
    & 160 & 1422.7 & 7.029 & 11.6 & 0.057 & 0.62 \\
    & 170 & 1422.7 & 7.029 & 8.7 & 0.043 & 0.27 \\
    \hline
    \multirow{10}{*}{2:1:1} & 15 & 1478.9 & 6.762 & 72.6 & 0.335 & 0.27 \\
    & 34 & 1478.4 & 6.764 & 72.1 & 0.333 & 0.29 \\
    & 56 & 1475.5 & 6.777 & 72.8 & 0.337 & 0.39 \\
    & 77 & 1470.0 & 6.803 & 73.5 & 0.343 & 0.57 \\
    & 99 & 1465.2 & 6.825 & 74.5 & 0.350 & 0.75 \\
    & 120 & 1462.3 & 6.839 & 74.7 & 0.353 & 0.89 \\
    & 131 & 1460.6 & 6.847 & 74.7 & 0.353 & 0.96 \\
    & 142 & 1458.9 & 6.855 & 74.2 & 0.352 & 1 \\
    & 160 & 1421.5 & 7.035 & 11.6 & 0.057 & 0.97 \\
    & 170 & 1420.6 & 7.039 & 9.6 & 0.048 & 0.70 \\
    \hline
    \multirow{10}{*}{1:1:1} & 15 & 1473.8 & 6.785 & 73.3 & 0.339 & 0.41 \\
    & 34 & 1473.1 & 6.788 & 74.5 & 0.345 & 0.46 \\
    & 56 & 1470.2 & 6.802 & 73.3 & 0.341 & 0.53 \\
    & 77 & 1462.5 & 6.838 & 73.0 & 0.343 & 0.68 \\
    & 99 & 1457.7 & 6.860 & 73.8 & 0.350 & 0.84 \\
    & 120 & 1455.0 & 6.873 & 75.7 & 0.361 & 0.93 \\
    & 135 & 1453.1 & 6.882 & 76.7 & 0.367 & 0.98 \\
    & 145 & 1451.7 & 6.889 & 76.7 & 0.368 & 1 \\
    & 160 & 1421.5 & 7.035 & 11.1 & 0.055 & 0.90 \\
    & 170 & 1421.5 & 7.035 & 10.4 & 0.051 & 0.36 \\
    \hline
\end{tabular}
\end{center}
\label{tab:nh4_h2o_char_1_1}
\end{table*}

\begin{table*}[h]
\caption{Peak positions, FWHMs, and integrated band areas of the NH$_{4}$$^{+}$ $\nu_{4}$ mode in H$_{2}$O:NH$_{3}$:H$_{2}$S ice with a mixing ratio of 10:2:3 at temperatures $\sim$15-170 K. The uncertainties of the peak positions and FWHMs are 0.5 cm$^{-1}$. Reported band areas are provided relative to the maximum amorphous value recorded during the warm-up. The reported band areas at different temperatures should not be used to calculate relative band strengths, as the salt abundance in the laboratory ice varies during warm-up as the acid-base reaction proceeds, and the salt begins desorbing at temperatures >135 K.}
\begin{center}
\begin{tabular}{|c|c|cc|cc|c|}
\hline
    H$_{2}$O:NH$_{3}$:H$_{2}$S & T & \multicolumn{2}{c|}{Peak} & \multicolumn{2}{c|}{FWHM} & \multirow{2}{*}{Rel. band area} \\
    mixing ratio & (K) & (cm$^{-1}$) & ($\mu$m) & (cm$^{-1}$) & ($\mu$m) & \\
    \hline 
    \multirow{9}{*}{10:2:3} & 15 & 1479.4 & 6.760 & 69.9 & 0.322 & 0.14 \\
    & 34 & 1478.4 & 6.764 & 72.6 & 0.336 & 0.17 \\
    & 56 & 1477.4 & 6.768 & 70.4 & 0.326 & 0.26 \\
    & 77 & 1473.1 & 6.788 & 70.9 & 0.330 & 0.45 \\
    & 99 & 1470.9 & 6.798 & 70.9 & 0.331 & 0.67 \\
    & 120 & 1469.0 & 6.807 & 70.4 & 0.329 & 0.85 \\
    & 135 & 1467.3 & 6.815 & 69.7 & 0.326 & 0.97 \\
    & 142 & 1466.8 & 6.817 & 68.9 & 0.323 & 1 \\
    & 160 & 1422.2 & 7.031 & 11.6 & 0.057 & 0.68 \\
    & 170 & 1422.2 & 7.031 & 9.9 & 0.049 & 0.34 \\
    \hline
\end{tabular}
\end{center}
\label{tab:nh4_h2o_char_2_3}
\end{table*}

\begin{table*}[h]
\caption{Peak positions and FWHMs of the SH$^{-}$ stretching mode in H$_{2}$O:NH$_{3}$:H$_{2}$S ices at 135 K. The uncertainties of the peak positions and FWHMs are 0.5 cm$^{-1}$.}
\begin{center}
\begin{tabular}{|c|cc|cc|}
\hline
    H$_{2}$O:NH$_{3}$:H$_{2}$S & \multicolumn{2}{c|}{Peak} & \multicolumn{2}{c|}{FWHM} \\
    mixing ratio & (cm$^{-1}$) & ($\mu$m) & (cm$^{-1}$) & ($\mu$m) \\
    \hline 
    10:2:1 & 2565.7 & 3.898 & 22.4 & 0.034 \\
    8:2:1 & 2566.1 & 3.897 & 27.2 & 0.041 \\
    6:2:1 & 2566.5 & 3.896 & 26.2 & 0.040 \\
    5:2:1 & 2566.6 & 3.896 & 25.8 & 0.039 \\
    \hline
    12:3:2 & 2564.2 & 3.900 & 29.3 & 0.045 \\
    10:3:2 & 2564.5 & 3.899 & 27.3 & 0.042 \\
    6:3:2 & 2561.2 & 3.904 & 32.2 & 0.049 \\
    4:3:2 & 2560.8 & 3.905 & 29.2 & 0.045 \\
    3:3:2 & 2559.8 & 3.907 & 31.1 & 0.047 \\
    \hline
    8:5:4 & 2558.2 & 3.909 & 34.6 & 0.053 \\
    \hline
    5:1:1 & 2561.3 & 3.904 & 32.3 & 0.049 \\
    2:1:1 & 2557.2 & 3.910 & 35.9 & 0.055 \\
    1:1:1 & 2555.9 & 3.913 & 30.4 & 0.047 \\
    \hline
\end{tabular}
\end{center}
\label{tab:sh_h2o_char}
\end{table*}

\clearpage

\section{Observational supplementary data}
\label{app:obs}

\subsection{Continua}
\label{app:obs_cont}
This appendix contains figures showing the local and global continua used to extract the observed features of interest for further analysis as well as fits to the anion features in B1-c, L1527, and W33A and the silicate and H$_{2}$O libration features in L1527.

\begin{figure}[ht!]
\centering
\includegraphics[width=\linewidth]{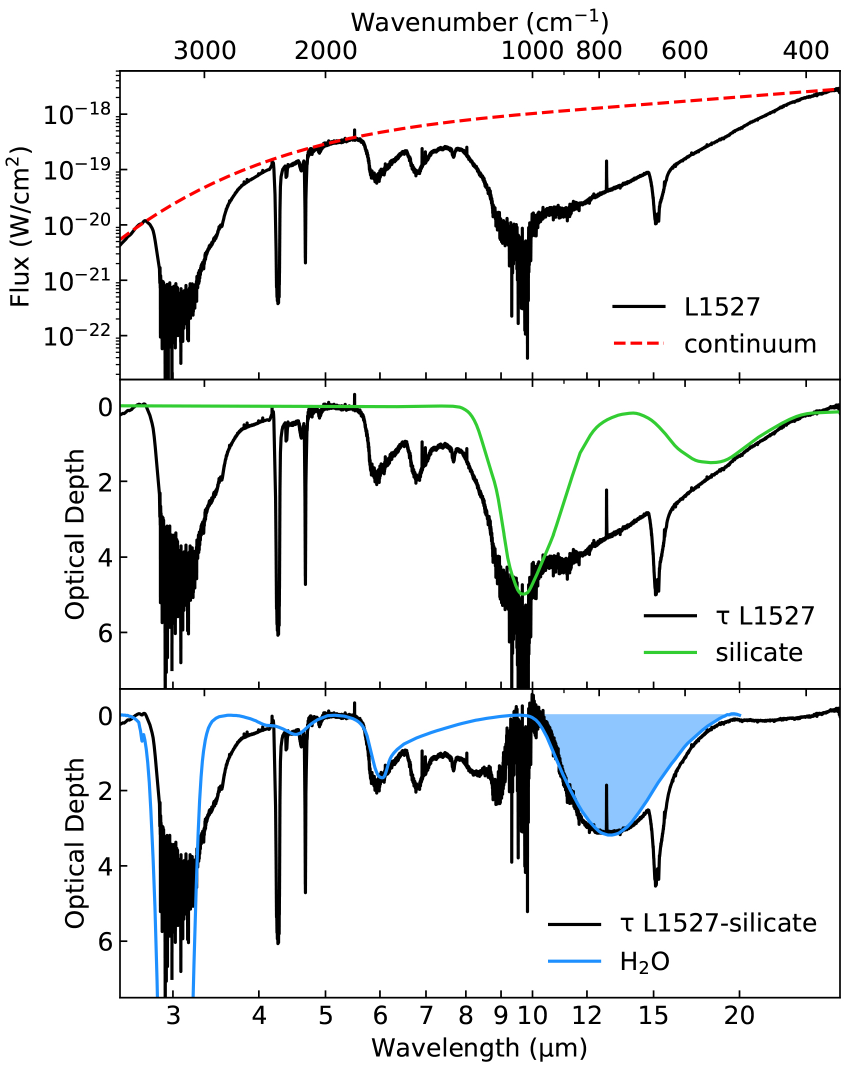}
\caption{Fits used to extract the optical depth of the silicate feature and H$_{2}$O ice column density from the spectrum of L1527. The fitting procedure is analogous to that described in \cite{boogert2008c2d}. Top: global continuum used to derive the spectrum of L1527 on the optical depth scale. Middle: subtraction of the silicate template spectrum GCS 3 \citep{kemper2004absence} from the spectrum of L1527 on the optical depth scale. Bottom: fit of a laboratory 15 K H$_{2}$O ice spectrum to the 12 $\mu$m H$_{2}$O libration mode in the silicate subtracted spectrum of L1527 on the optical depth scale.}
\label{fig:l1527_cont_sil_h2o}
\end{figure}
\begin{figure*}[ht!]
\centering
\includegraphics[width=0.8\linewidth]{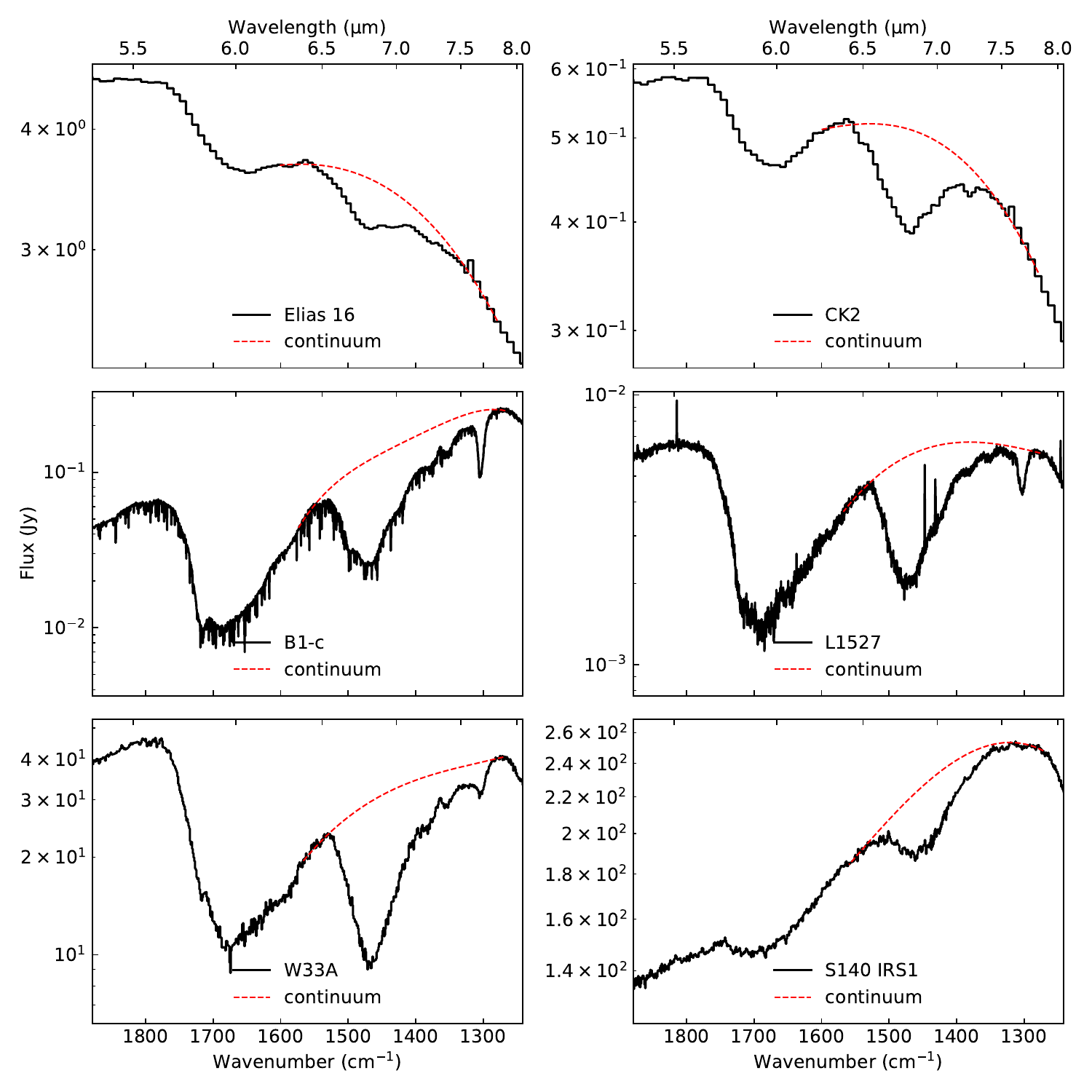}
\caption{Local continua used to extract the 6.85 $\mu$m feature for fitting with laboratory data.}
\label{fig:nh4_obs_cont}
\end{figure*}

\begin{figure*}[ht!]
\centering
\includegraphics[width=0.8\linewidth]{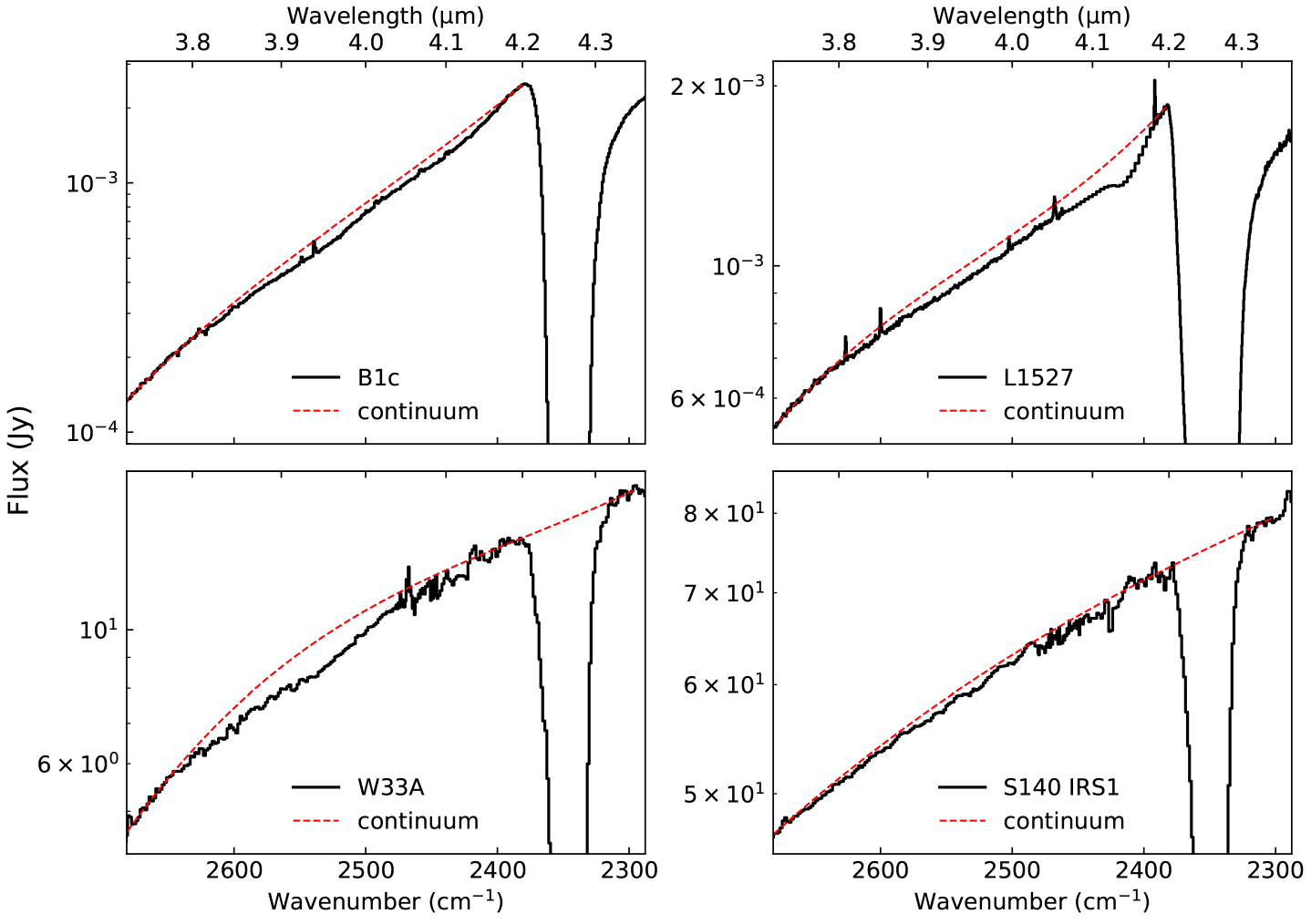}
\caption{Local continua used to extract the 3.7-4.2 $\mu$m spectral region for fitting with laboratory data.}
\label{fig:sh_obs_cont}
\end{figure*}

\begin{figure*}[ht!]
\centering
\includegraphics[width=0.8\linewidth]{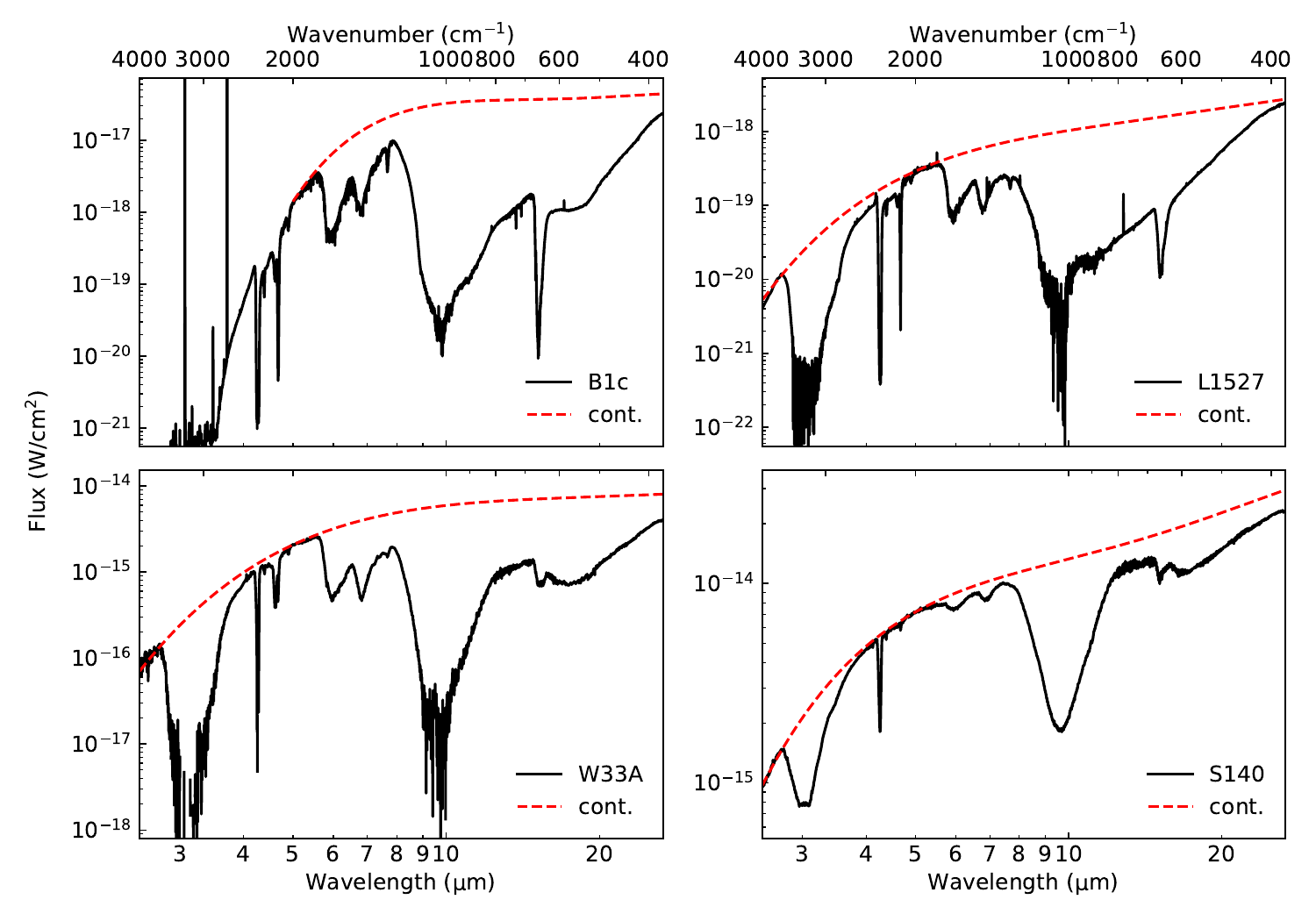}
\caption{Global continua used to extract the 5.3 $\mu$m feature for fitting with laboratory data.}
\label{fig:global_conts}
\end{figure*}

\begin{figure*}[ht!]
\centering
\includegraphics[width=0.9\linewidth]{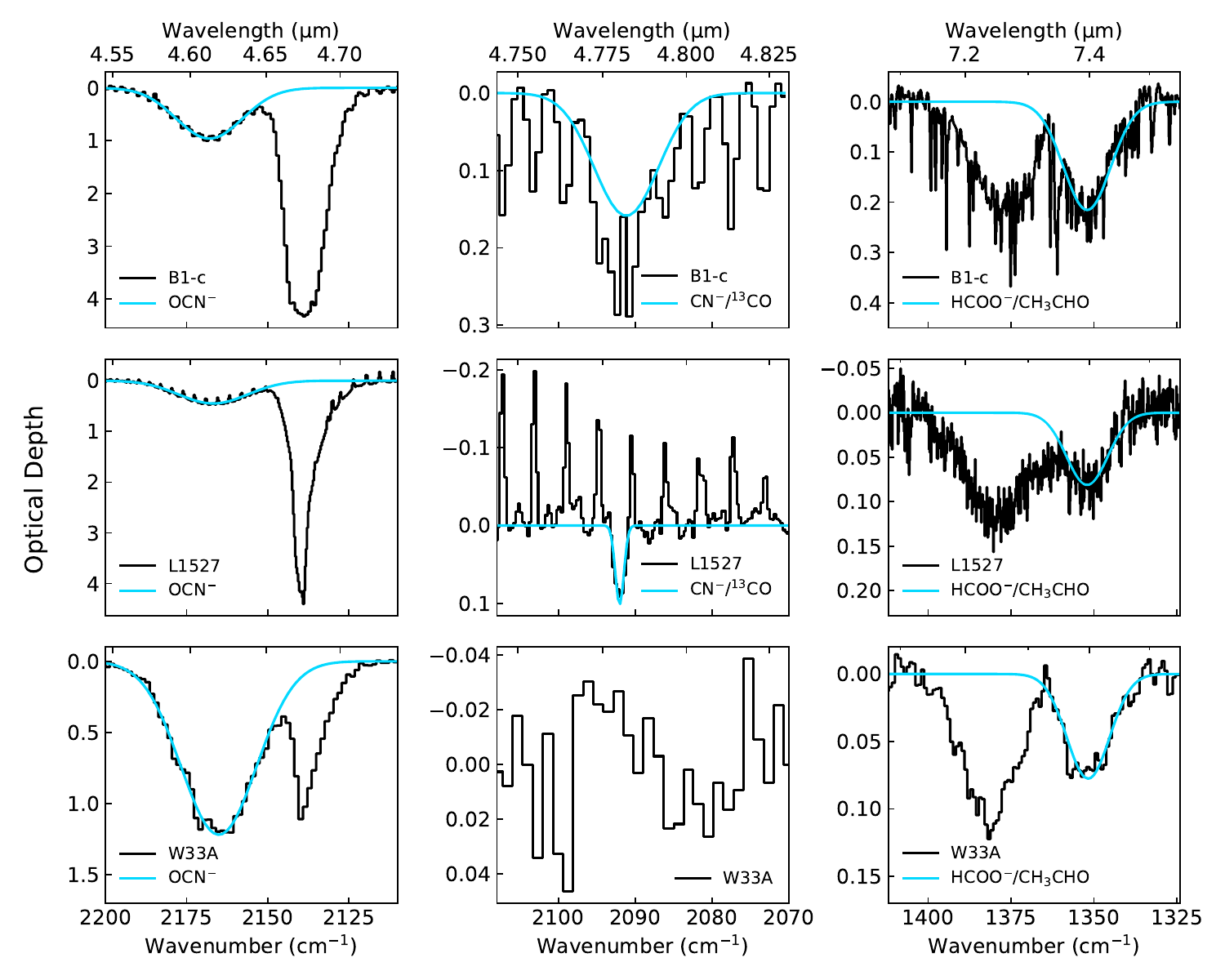}
\caption{Gaussian fits used to estimate abundances and upper limits of the OCN$^{-}$, CN$^{-}$, and HCOO$^{-}$ anions toward three protostellar sources. The selected IR features of CN$^{-}$ and HCOO$^{-}$ overlap with those of $^{13}$CO and CH$_{3}$CHO, respectively, so their column densities resulting from these fits are only considered upper limits. No feature is detected in the CN$^{-}$/$^{13}$CO spectral region in the W33A spectrum.}
\label{fig:anions}
\end{figure*}

\subsection{N$_{\rm{H}}$ from $\tau_{9.7}$}
\label{app:obs_nh}

The hydrogen column densities of the investigated sources were estimated in all of the investigated sources except L1527 by \cite{boogert2013infrared} using the relation:

\begin{equation}
    \indent N_{H} = 3.08\times10^{22} \times A_{K} (cm^{-2}),
\end{equation}

where A$_{K}$ was estimated using the fitting parameters derived from the optical depths of the silicate features ($\tau_{9.7}$) and measured A$_{K}$ values in the dense cores of Lupus:

\begin{equation}
    \indent \tau_{9.7} = 0.26 \times A_{K}.
\end{equation}

Assuming that the relationships between $\tau_{9.7}$ and A$_{K}$ in the clouds and protostellar envelopes investigated here are consistent with those observed in the dense cores of Lupus, an uncertainty of $\sim$30\% applies to the derived N$_{\rm{H}}$ values \citep{boogert2013infrared}. Using fitting parameters derived from measurements of the diffuse medium would provide N$_{\rm{H}}$ values up to a factor of 2-4 lower \citep{bohlin1978survey,boogert2013infrared,whittet2018dust}, resulting in total \% S budgets that are a factor of 2-4 higher than our reported values, but the environments of the sources investigated here are more likely to be analogous to the dense clouds of Lupus rather than the diffuse medium.

\end{appendix}

\end{document}